\documentclass[11pt]{article}%
\usepackage{amsmath}
\usepackage{graphicx}%
\usepackage{amsfonts}%
\usepackage{amssymb}
\newcommand{\eqnb}{\begin{equation}}
\newcommand{\eqne}{\end{equation}}

\newtheorem{The}{Theorem}

\newtheorem{Def}{Definition}

\newtheorem{Lem}{Lemma}

\newtheorem{Rem}{Remark}

\begin{document}

\title{Block-Structured Supermarket Models\footnote{The main results of this paper will be published in "Discrete Event
Dynamic Systems" 2014. On the other hand, the three appendices are
the online supplementary material for this paper published in
"Discrete Event Dynamic Systems" 2014}}
\author{Quan-Lin Li\\School of Economics and Management Sciences \\Yanshan University, Qinhuangdao 066004, China \\
John C.S. Lui\\Department of Computer Science \& Engineering \\
The Chinese University of Hong Kong, Shatin, N.T, Hong Kong}

\date{March 25, 2014}
\maketitle

\begin{abstract}
Supermarket models are a class of parallel queueing networks with an adaptive
control scheme that play a key role in the study of resource management of,
such as, computer networks, manufacturing systems and transportation networks.
When the arrival processes are non-Poisson and the service times are
non-exponential, analysis of such a supermarket model is always limited,
interesting, and challenging.

This paper describes a supermarket model with non-Poisson inputs: Markovian
Arrival Processes (MAPs) and with non-exponential service times: Phase-type
(PH) distributions, and provides a generalized matrix-analytic method which is
first combined with the operator semigroup and the mean-field limit. When
discussing such a more general supermarket model, this paper makes some new
results and advances as follows: (1) Providing a detailed probability analysis
for setting up an infinite-dimensional system of differential vector equations
satisfied by the expected fraction vector, where \textit{the invariance of
environment factors} is given as an important result. (2) Introducing the
phase-type structure to the operator semigroup and to the mean-field limit,
and a Lipschitz condition can be obtained by means of a unified
matrix-differential algorithm. (3) The matrix-analytic method is used to
compute the fixed point which leads to performance computation of this system.
Finally, we use some numerical examples to illustrate how the performance
measures of this supermarket model depend on the non-Poisson inputs and on the
non-exponential service times. Thus the results of this paper give new
highlight on understanding influence of non-Poisson inputs and of
non-exponential service times on performance measures of more general
supermarket models.


\end{abstract}

\section{Introduction}

Supermarket models are a class of parallel queueing networks with an adaptive
control scheme that play a key role in the study of resource management of,
such as computer networks (e.g., see the dynamic randomized load balancing),
manufacturing systems and transportation networks. Since a simple supermarket
model was discussed by Mitzenmacher \cite{Mit:1996}, Vvedenskaya et al
\cite{Vve:1996} and Turner \cite{Tur:1996} through queueing theory as well as
Markov processes, subsequent papers have been published on this theme, among
which, see, Vvedenskaya and Suhov \cite{Vve:1997}, Jacquet and Vvedenskaya
\cite{Jac:1998}, Jacquet et al \cite{Jac:1999}, Mitzenmacher \cite{Mit:1999},
Graham \cite{Gra:2000a, Gra:2000b, Gra:2004}, Mitzenmacher et al
\cite{Mit:2001}, Vvedenskaya and Suhov \cite{Vve:2005}, Luczak and Norris
\cite{LucN:2005}, Luczak and McDiarmid \cite{Luc:2006, Luc:2007}, Bramson et
al \cite{Bra:2010, Bra:2012, Bra:2013}, Li et al \cite{LiLW:2011}, Li
\cite{Li:2014} and Li et al \cite{Li:2013a}. For the fast Jackson networks (or
the supermarket networks), readers may refer to Martin and Suhov
\cite{Mar:1999}, Martin \cite{Mar:2001} and Suhov and Vvedenskaya
\cite{Suh:2002}.

The available results of the supermarket models with non-exponential service
times are still few in the literature. Important examples include an
approximate method of integral equations by Vvedenskaya and Suhov
\cite{Vve:1997}, the Erlang service times by Mitzenmacher \cite{Mit:1999} and
Mitzenmacher et al \cite{Mit:2001}, the PH service times by Li et al
\cite{LiLW:2011} and Li and Lui \cite{LiL:2010}, and the ansatz-based
modularized program for the general service times by Bramson et al
\cite{Bra:2010, Bra:2012, Bra:2013}.

Little work has been done on the analysis of the supermarket models with
non-Poisson inputs, which are more difficult and challenging due to the higher
complexity of that $N$ arrival processes are superposed. Li and Lui
\cite{LiL:2010} and Li \cite{Li:2011} used the superposition of $N$ MAP inputs
to study the infinite-dimensional Markov processes of supermarket modeling
type. Comparing with the results given in Li and Lui \cite{LiL:2010} and Li
\cite{Li:2011}, this paper provides more necessary phase-level probability
analysis in setting up the infinite-dimensional system of differential vector
equations, which leads some new results and methodologies in the study of
block-structured supermarket models. Note that the PH distributions constitute
a versatile class of distributions that can approximate arbitrarily closely
any probability distribution defined on the nonnegative real line, and the
MAPs are a broad class of renewal or non-renewal point processes that can
approximate arbitrarily closely any stochastic counting process (e.g., see
Neuts \cite{Neu:1981, Neu:1989} and Li \cite{Li:2010} for more details), thus
the results of this paper are a key advance of those given in Mitzenmacher
\cite{Mit:1996} and Vvedenskaya et al \cite{Vve:1996} under the Poisson and
exponential setting.

The main contributions of this paper are threefold. The first one is to use
the MAP inputs and the PH service times to describe a more general supermarket
model with non-Poisson inputs and with non-exponential service times. Based on
the phase structure, we define the random fraction vector and construct an
infinite-dimensional Markov process, which expresses the state of this
supermarket model by means of an infinite-dimensional Markov process.
Furthermore, we set up an infinite-dimensional system of differential vector
equations satisfied by the expected fraction vector through a detailed
probability analysis. To that end, we obtain an important result: The
invariance of environment factors, which is a key for being able to simplify
the differential equations in a vector form. Based on the differential vector
equations, we can provide a generalized matrix-analytic method to investigate
more general supermarket models with non-Poisson inputs and with
non-exponential service times. The second contribution of this paper is to
provide phase-structured expression for the operator semigroup with respect to
the MAP inputs and to the PH service times, and use the operator semigroup to
provide the mean-field limit for the sequence of Markov processes who
asymptotically approaches a single trajectory identified by the unique and
global solution to the infinite-dimensional system of limiting differential
vector equations. To prove the existence and uniqueness of solution through
the Picard approximation, we provide a unified computational method for
establishing a Lipschitz condition, which is crucial in all the rigor proofs
involved. The third contribution of this paper is to provide an effective
matrix-analytic method both for computing the fixed point and for analyzing
performance measures of this supermarket model. Furthermore, we use some
numerical examples to indicate how the performance measures of this
supermarket model depend on the non-Poisson MAP inputs and on the
non-exponential PH service times. Therefore, the results of this paper gives
new highlight on understanding performance analysis and nonlinear Markov
processes for more general supermarket models with non-Poisson inputs and
non-exponential service times.

The remainder of this paper is organized as follows. In Section 2, we first
introduce a new MAP whose transition rates are controlled by the number of
servers in the system. Then we describe a more general supermarket model of
$N$ identical servers with MAP inputs and PH service times. In Section 3, we
define a random fraction vector and construct an infinite-dimensional Markov
process, which expresses the state of this supermarket model. In Section 4, we
set up an infinite-dimensional system of differential vector equations
satisfied by the expected fraction vector through a detailed probability
analysis, and establish an important result: The invariance of environment
factors. In Section 5, we show that the mean-field limit for the sequence of
Markov processes who asymptotically approaches a single trajectory identified
by the unique and global solution to the infinite-dimensional system of
limiting differential vector equations. To prove the existence and uniqueness
of the solution, we provide a unified matrix-differential algorithm for
establishing the Lipschitz condition. In Section 6, we first discuss the
stability of this supermarket model in terms of a coupling method. Then we
provide a generalized matrix-analytic method for computing the fixed point
whose doubly exponential solution and phase-structured tail are obtained.
Finally, we discuss some useful limits of the fraction vector $\mathbf{u}%
^{\left(  N\right)  }\left(  t\right)  $ as $N\rightarrow\infty$ and
$t\rightarrow+\infty$. In Section 7, we provide two performance
measures of this supermarket model, and use some numerical examples
to indicate how the performance measures of this system depend on
the non-Poisson MAP inputs and on the non-exponential PH service
times. Some concluding remarks are given in Section 8. Finally,
Appendices A and C are respectively designed for the proofs of
Theorems 1 and 3, and Appendix B contains the proof of Theorem 2,
where the mean-field limit of the sequence of Markov processes in
this supermarket model is given a detailed analysis through the
operator semigroup.

\section{Supermarket Model Description}

In this section, we first introduce a new MAP whose transition rates are
controlled by the number of servers in the system. Then we describe a more
general supermarket model of $N$ identical servers with MAP inputs and PH
service times.

\subsection{A new Markovian arrival process}

Based on Chapter 5 in Neuts \cite{Neu:1989}, the MAP is a bivariate Markov
process $\left\{  \left(  N(t),J(t)\right)  :t\geq0\right\}  $ with state
space $S=\left\{  1,2,3,\ldots\right\}  \times\left\{  1,2,\ldots
,m_{A}\right\}  $, where $\left\{  N(t):t\geq0\right\}  $ is a counting
process of arrivals and $\left\{  J(t):t\geq0\right\}  $ is a Markov
environment process. When $J(t)=i$, if the random environment shall go to
state $j$ in the next time, then the counting process $\left\{  N(t):t\geq
0\right\}  $ is a Poisson process with arrival rate $d_{i,j}$ for $1\leq
i,j\leq m_{A}$. The matrix $D$ with elements $d_{i,j}$ satisfies
$D\gvertneqq0$. The matrix $C$ with elements $c_{i,j}$ has negative diagonal
elements and nonnegative off-diagonal elements, and the matrix $C$ is
invertible, where $c_{i,j}$ is a state transition rate of the Markov chain
$\left\{  J(t):t\geq0\right\}  $ from state $i$ to state $j$ for $i\neq j$.
The matrix $Q=C+D$ is the infinitesimal generator of an irreducible Markov
chain. We assume that $Qe=0$, where $e$ is a column vector of ones with a
suitable size. Hence, we have%
\[
c_{i,i}=-\left[  \sum\limits_{j=1}^{m_{A}}d_{i,j}+\sum\limits_{j\neq i}%
^{m_{A}}c_{i,j}\right]  .
\]

Let%
\[
\mathbb{C}=\left(
\begin{array}
[c]{cccc}%
-\sum\limits_{j\neq1}^{m_{A}}c_{1,j} & c_{1,2} & \cdots & c_{1,m_{A}}\\
c_{2,1} & -\sum\limits_{j\neq2}^{m_{A}}c_{2,j} & \cdots & c_{2,m_{A}}\\
\vdots & \vdots & \ddots & \vdots\\
c_{m_{A},1} & c_{m_{A},2} & \cdots & -\sum\limits_{j\neq m_{A}}^{m_{A}%
}c_{m_{A},j}%
\end{array}
\right)  ,
\]%
\[
C(N)=\mathbb{C}-N\text{diag}(De),
\]%
\[
D(N)=ND,
\]
where%
\[
\text{diag}(De)=\text{diag}\left(  \sum\limits_{j=1}^{m_{A}}d_{1,j}%
,\sum\limits_{j=1}^{m_{A}}d_{2,j},\ldots,\sum\limits_{j=1}^{m_{A}}d_{m_{A}%
,j}\right)  .
\]
Then%
\[
Q\left(  N\right)  =C(N)+D(N)=\left[  \mathbb{C}-N\text{diag}(De)\right]  +ND
\]
is obviously the infinitesimal generator of an irreducible Markov chain with
$m_{A}$ states. Thus $\left(  C(N),D(N)\right)  $ is the irreducible matrix
descriptor of a new MAP of order $m_{A}$. Note that the new MAP is non-Poisson
and may also be non-renewal, and its arrival rate at each environment state is
controlled by the number $N$ of servers in the system.

Note that%
\[
Q\left(  N\right)  e=\left[  \mathbb{C}-N\text{diag}(De)\right]  e+NDe=0,
\]
the Markov chain $Q\left(  N\right)  $ with $m_{A}$ states is irreducible and
positive recurrent. Let $\omega_{N}$ be the stationary probability vector of
the Markov chain $Q\left(  N\right)  $. Then $\omega_{N}$ depends on the
number $N\geq1$, and the stationary arrival rate of the MAP is given by
$N\lambda_{N}=N\omega_{N}De$.

\subsection{Model description}

Based on the new MAP, we describe a more general supermarket model of $N$
identical servers with MAP inputs and PH service times as follows:

\textbf{Non-Poisson inputs:} Customers arrive at this system as the MAP of
irreducible matrix descriptor $\left(  C\left(  N\right)  ,D\left(  N\right)
\right)  $ of size $m_{A}$, whose stationary arrival rate is given by
$N\lambda_{N}=N\omega_{N}De$.

\textbf{Non-exponential service times:} The service times of each server are
i.i.d. and are of phase type with an irreducible representation $\left(
\alpha,T\right)  $ of order $m_{B}$, where the row vector $\alpha$ is a
probability vector whose $j$th entry is the probability that a service begins
in phase $j$ for $1\leq j\leq m_{B}$; $T$ is a matrix of size $m_{B}$\ whose
$\left(  i,j\right)  ^{\text{th}}$ entry is denoted by $t_{i,j}$ with
$t_{i,i}<0$ for $1\leq i\leq m_{B}$, and $t_{i,j}\geq0$ for $i\neq j$. Let
$T^{0}=-Te=\left(  t_{1}^{0},t_{2}^{0},\ldots,t_{m_{B}}^{0}\right)
^{\text{T}}\gvertneqq0$, where \textquotedblleft$A^{T}$\textquotedblright
\ denotes the transpose of matrix (or vector) $A$. When a PH service time is
in phase $i$, the transition rate from phase $i$ to phase $j$ is $t_{i,j}$,
the service completion rate is $t_{i}^{0}$, and the output rate from phase $i$
is $\mu_{i}=-t_{i,i}$. At the same time, the mean of the PH service time is
given by $1/\mu=-\alpha T^{-1}e$.

\textbf{Arrival and service disciplines: }Each arriving customer chooses
$d\geq1$ servers independently and uniformly at random from the $N$ identical
servers, and waits for its service at the server which currently contains the
fewest number of customers. If there is a tie, servers with the fewest number
of customers will be chosen randomly. All customers in any server will be
served in the FCFS manner. Figure 1 gives a physical interpretation for this
supermarket model.

\begin{figure}[ptb]
%Requires \usepackage{graphicx}
\centering          \includegraphics[width=12cm]{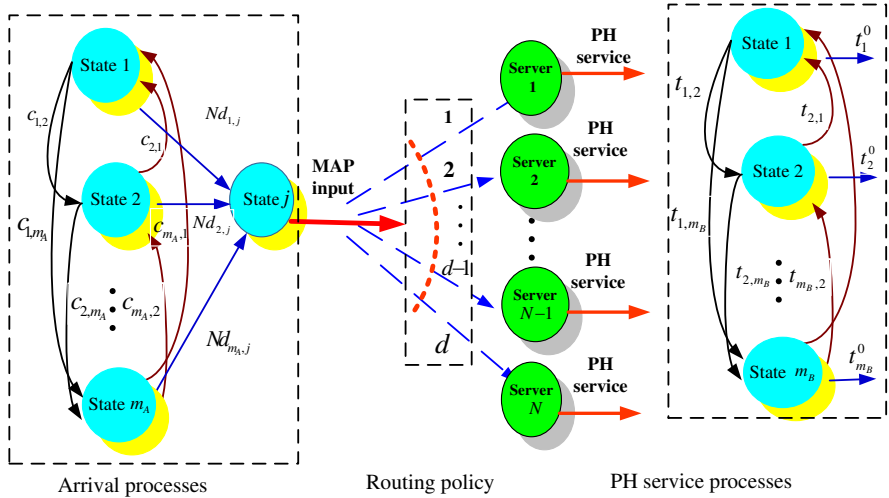}  \newline
\caption{The supermarket model with MAP inputs and PH service times}%
\label{figure:fig-1}%
\end{figure}

\begin{Rem}
The block-structured supermarket models can have many practical
applications to, such as, computer networks and manufacturing
system, where it is a key to introduce the PH service times and the
MAP inputs to such a practical model, because the PH distributions
contain many useful distributions such as exponential,
hyper-exponential and Erlang distributions; while the MAPs include,
for example, Poisson process, PH-renewal processes, and Markovian
modulated Poisson processes (MMPPs). Note that the probability
distributions and stochastic point processes have extensively been
used in most practical stochastic modeling. On the other hand, in
many practical applications, the block-structured supermarket model
is an important queueing model to analyze the relation between the
system performance and the job routing rule, and it can also help to
design reasonable architecture to improve the performance and to
balance the load.
\end{Rem}

\section{An Infinite-Dimensional Markov Process}

In this section, we first define the random fraction vector of this
supermarket model. Then we use the the random fraction vector to construct an
infinite-dimensional Markov process, which describes the state of this
supermarket model.

For this supermarket model, let $n_{k;i,j}^{\left(  N\right)  }\left(
t\right)  $ be the number of servers with at least $k$ customers (note that
the serving customer is also taken into account), and with the MAP be in phase
$i$ and the PH service time be in phase $j$ at time $t\geq0$. Clearly, $0\leq
n_{0;i}^{\left(  N\right)  }\left(  t\right)  \leq N$ and $0\leq
n_{k;i,j}^{\left(  N\right)  }\left(  t\right)  \leq N$ for $k\geq1$, $1\leq
i\leq m_{A}$ and $1\leq j\leq m_{B}$. Let%
\[
U_{0;i}^{\left(  N\right)  }\left(  t\right)  =\frac{n_{0;i}^{\left(
N\right)  }\left(  t\right)  }{N},\text{ \ }1\leq i\leq m_{A},
\]
and for $k\geq1$%
\[
U_{k;i,j}^{\left(  N\right)  }\left(  t\right)  =\frac{n_{k;i,j}^{\left(
N\right)  }\left(  t\right)  }{N},\text{ \ }1\leq i\leq m_{A},1\leq j\leq
m_{B}.
\]
Then $U_{k;i,j}^{\left(  N\right)  }\left(  t\right)  $ is the fraction of
servers with at least $k$ customers, and with the MAP be in phase $i$ and the
PH service time be in phase $j$ at time $t$. Using the lexicographic order we
write%
\[
U_{0}^{\left(  N\right)  }\left(  t\right)  =\left(  U_{0;1}^{\left(
N\right)  }\left(  t\right)  ,U_{0;2}^{\left(  N\right)  }\left(  t\right)
,\ldots,U_{0;m_{A}}^{\left(  N\right)  }\left(  t\right)  \right)  \text{ }%
\]
for $k\geq1$%
\begin{align*}
U_{k}^{\left(  N\right)  }\left(  t\right)  =  &  \left(  U_{k;1,1}^{\left(
N\right)  }\left(  t\right)  ,U_{k;1,2}^{\left(  N\right)  }\left(  t\right)
,\ldots,U_{k;1,m_{B}}^{\left(  N\right)  }\left(  t\right)  ;\ldots;\right. \\
&  \left.  U_{k;m_{A},1}^{\left(  N\right)  }\left(  t\right)  ,U_{k;m_{A}%
,2}^{\left(  N\right)  }\left(  t\right)  ,\ldots,U_{k;m_{A},m_{B}}^{\left(
N\right)  }\left(  t\right)  \right)  ,
\end{align*}
and%
\begin{equation}
U^{\left(  N\right)  }\left(  t\right)  =\left(  U_{0}^{\left(  N\right)
}\left(  t\right)  ,U_{1}^{\left(  N\right)  }\left(  t\right)  ,U_{2}%
^{\left(  N\right)  }\left(  t\right)  ,\ldots\right)  . \label{Equ1}%
\end{equation}

Let $a=\left(  a_{1},a_{2},a_{3},\ldots\right)  $ and $b=\left(  b_{1}%
,b_{2},b_{3},\ldots\right)  $. We write $a<b$ if $a_{k}<b_{k}$ for some
$k\geq1$; $a\leq b$ if $a_{k}\leq b_{k}$ for every $k\geq1$.

For a fixed quaternary array $\left(  t,N,i,j\right)  $ with $t\geq
0,N\in\left\{  1,2,3,\ldots\right\}  ,i\in\left\{  1,2,\ldots,m_{A}\right\}  $
and $j\in\left\{  1,2,\ldots,m_{B}\right\}  $, it is easy to see from the
stochastic order that $n_{k;i,j}^{\left(  N\right)  }\left(  t\right)  \geq
n_{k+1;i,j}^{\left(  N\right)  }\left(  t\right)  $ for $k\geq1$. This gives
\begin{equation}
U_{1}^{\left(  N\right)  }\left(  t\right)  \geq U_{2}^{\left(  N\right)
}\left(  t\right)  \geq U_{3}^{\left(  N\right)  }\left(  t\right)  \cdots
\geq0 \label{Equ1-1}%
\end{equation}
and%
\begin{equation}
1=U_{0}^{\left(  N\right)  }\left(  t\right)  e\geq U_{1}^{\left(  N\right)
}\left(  t\right)  e\geq U_{2}^{\left(  N\right)  }\left(  t\right)  e\geq
U_{3}^{\left(  N\right)  }\left(  t\right)  e\geq\cdots\geq0. \label{Equ1-2}%
\end{equation}

Note that the state of this supermarket model is described as the random
fraction vector $U^{\left(  N\right)  }\left(  t\right)  $ for $t\geq0$, and
$\left\{  U^{\left(  N\right)  }\left(  t\right)  ,t\geq0\right\}  $ is a
stochastic vector process for each $N=1,2,\ldots$. Since the arrival process
to this supermarket model is the MAP and the service times in each server are
of phase type, $\left\{  U^{\left(  N\right)  }\left(  t\right)
,t\geq0\right\}  $ is an infinite-dimensional Markov process whose state space
is given by%
\begin{align}
\widetilde{\Omega}_{N}=  &  \left\{  \left(  h_{0}^{\left(  N\right)  }%
,h_{1}^{\left(  N\right)  },h_{2}^{\left(  N\right)  }\ldots\right)
:h_{0}^{\left(  N\right)  }\text{ is a probability vector of size }%
m_{A},\right. \nonumber\\
&  \left.  h_{1}^{\left(  N\right)  }\geq h_{2}^{\left(  N\right)  }\geq
h_{3}^{\left(  N\right)  }\geq\cdots\geq0,h_{k}^{\left(  N\right)  }\text{ is
a row vector of size }m_{A}m_{B}\text{ for }k\geq1,\right. \nonumber\\
&  \left.  1=h_{0}^{\left(  N\right)  }e\geq h_{1}^{\left(  N\right)  }e\geq
h_{2}^{\left(  N\right)  }e\geq\cdots\geq0,\right. \nonumber\\
&  \left.  \text{and \ }Nh_{k}^{\left(  N\right)  }\text{ \ is a row vector of
nonnegative integers for }k\geq0\right\}  , \label{Equ2}%
\end{align}

We write%
\[
u_{0;i}^{\left(  N\right)  }\left(  t\right)  =E\left[  U_{0;i}^{\left(
N\right)  }\left(  t\right)  \right]
\]
and for $k\geq1$%
\[
u_{k;i,j}^{\left(  N\right)  }\left(  t\right)  =E\left[  U_{k;i,j}^{\left(
N\right)  }\left(  t\right)  \right]  .
\]
Using the lexicographic order we write%
\[
u_{0}^{\left(  N\right)  }\left(  t\right)  =\left(  u_{0;1}^{\left(
N\right)  }\left(  t\right)  ,u_{0;2}^{\left(  N\right)  }\left(  t\right)
,\ldots,u_{0;m_{A}}^{\left(  N\right)  }\left(  t\right)  \right)
\]
and for $k\geq1$%
\begin{align*}
u_{k}^{\left(  N\right)  }\left(  t\right)  =  &  \left(  u_{k;1,1}^{\left(
N\right)  }\left(  t\right)  ,u_{k;1,2}^{\left(  N\right)  }\left(  t\right)
,\ldots,u_{k;1,m_{B}}^{\left(  N\right)  }\left(  t\right)  ;\ldots;\right. \\
&  \left.  u_{k;m_{A},1}^{\left(  N\right)  }\left(  t\right)  ,u_{k;m_{A}%
,2}^{\left(  N\right)  }\left(  t\right)  ,\ldots,u_{k;m_{A},m_{B}}^{\left(
N\right)  }\left(  t\right)  \right)  ,
\end{align*}%
\[
\mathbf{u}^{\left(  N\right)  }\left(  t\right)  =\left(  u_{0}^{\left(
N\right)  }\left(  t\right)  ,u_{1}^{\left(  N\right)  }\left(  t\right)
,u_{2}^{\left(  N\right)  }\left(  t\right)  ,\ldots\right)  .
\]
It is easy to see from Equations (\ref{Equ1-1}) and (\ref{Equ1-2}) that%
\begin{equation}
u_{1}^{\left(  N\right)  }\left(  t\right)  \geq u_{2}^{\left(  N\right)
}\left(  t\right)  \geq u_{3}^{\left(  N\right)  }\left(  t\right)  \cdots
\geq0 \label{Equ2-1}%
\end{equation}
and%
\begin{equation}
1=u_{0}^{\left(  N\right)  }\left(  t\right)  e\geq u_{1}^{\left(  N\right)
}\left(  t\right)  e\geq u_{2}^{\left(  N\right)  }\left(  t\right)
e\geq\cdots\geq0. \label{Equ2-2}%
\end{equation}

In the remainder of this section, for convenience of readers, it is necessary
to explain the structure of this long paper which is outlined as follows.
\textit{Part one: The limit of the sequence of Markov processes.} It is seen
from (\ref{Equ1}) and (\ref{Equ2}) that we need to deal with the limit of the
sequence $\left\{  U^{\left(  N\right)  }\left(  t\right)  \right\}  $ of
infinite-dimensional Markov processes. This is organized in Appendix B by
means of the convergence theorems of operator semigroups, e.g., see Ethier and
Kurtz \cite{Eth:1986} for more details. \textit{Part two: The existence and
uniqueness of the solution.} As seen from Theorem \ref{The:Conver} and
(\ref{EquK-16}), we need to study the two means $E\left[  U^{\left(  N\right)
}\left(  t\right)  \right]  $ and $E\left[  U\left(  t\right)  \right]
=\lim_{N\rightarrow\infty}E\left[  U^{\left(  N\right)  }\left(  t\right)
\right]  $, or $\mathbf{u}^{\left(  N\right)  }\left(  t\right)  $ and
$\mathbf{u}\left(  t\right)  =\lim_{N\rightarrow\infty}\mathbf{u}^{\left(
N\right)  }\left(  t\right)  $. To that end, Section 4 sets up the system of
differential vector equations satisfied by $\mathbf{u}^{\left(  N\right)
}\left(  t\right)  $, while Section 5 provides a unified matrix-differential
algorithm for establishing the Lipschitz condition, which is a key in proving
the existence and uniqueness of the solution to the limiting system of
differential vector equations satisfied by $\mathbf{u}\left(  t\right)
$\ through the Picard approximation. \textit{Part three: Computation of the
fixed point and performance analysis.} Section 6 discusses the stability of
this supermarket model in terms of a coupling method, and provide an effective
matrix-analytic method for computing the fixed point. Section 7 analyzes the
performance of this supermarket model by means of some numerical examples.

\section{The System of Differential Vector Equations}

In this section, we set up an infinite-dimensional system of differential
vector equations satisfied by the expected fraction vector through a detailed
probability analysis. Specifically, we obtain an important result: The
invariance of environment factors, which is a key to rewriting the
differential equations as a simple vector form.

To derive the system of differential vector equations, we first discuss an
example with the number $k\geq2$ of customers through the following three steps:

\textbf{Step one: Analysis of the Arrival Processes}

In this supermarket model of $N$ identical servers, we need to determine the
change in the expected number of servers with at least $k$ customers over a
small time period $\left[  0,\text{d}t\right)  $. When the MAP environment
process $\left\{  J\left(  t\right)  :t\geq0\right\}  $ jumps form state $l$
to state $i$ for $1\leq l,i\leq m_{A}$ and the PH service environment process
$\left\{  I\left(  t\right)  :t\geq0\right\}  $ sojourns at state $j$ for
$1\leq j\leq m_{B}$, one arrival occurs in a small time period $\left[
0,\text{d}t\right)  $. In this case, the rate that any arriving customer
selects $d$ servers with at least $k-1$ customers at random and joins the
shortest one with $k-1$ customers, is given by%
\begin{align}
&  \sum_{l=1}^{m_{A}}\left[  u_{k-1;l,j}^{\left(  N\right)  }\left(  t\right)
d_{l,i}-u_{k;i,j}^{\left(  N\right)  }\left(  t\right)  \left(  d_{i,1}%
,d_{i,2},\ldots,d_{i,m_{A}}\right)  e\right] \nonumber\\
&  \times L_{k;l}^{\left(  N\right)  }\left(  u_{k-1}\left(  t\right)
,u_{k}\left(  t\right)  \right)  N\text{d}t, \label{Equ3-0}%
\end{align}
where%
\begin{align*}
&  L_{k;l}^{\left(  N\right)  }\left(  u_{k-1}\left(  t\right)  ,u_{k}\left(
t\right)  \right)  =\sum_{m=1}^{d}C_{d}^{m}\left\{  \sum_{j=1}^{m_{B}}\left[
u_{k-1;l,j}^{(N)}\left(  t\right)  -u_{k;l,j}^{(N)}\left(  t\right)  \right]
\right\}  ^{m-1}\left\{  \sum_{j=1}^{m_{B}}\left[  u_{k;l,j}^{(N)}\left(
t\right)  \right]  \right\}  ^{d-m}\\
&  +\sum_{m=1}^{d-1}C_{d}^{m}\left\{  \sum_{j=1}^{m_{B}}\left[  u_{k-1;l,j}%
^{(N)}\left(  t\right)  -u_{k;l,j}^{(N)}\left(  t\right)  \right]  \right\}
^{m-1}\sum_{\substack{r_{1}+r_{2}+\cdots+r_{m_{A}}=d-m\\\sum_{i\neq l}^{m_{A}%
}r_{i}\geq1\\0\leq r_{j}\leq d-m,1\leq j\leq m_{A}}}\left(
\begin{array}
[c]{c}%
d-m\\
r_{1},r_{2},\ldots,r_{m_{A}}%
\end{array}
\right) \\
&  \times\prod_{i=1}^{m_{A}}\left\{  \sum_{j=1}^{m_{B}}\left[  u_{k;i,j}%
^{(N)}\left(  t\right)  \right]  \right\}  ^{r_{i}}+\sum_{m=2}^{d}C_{d}%
^{m}\sum_{m_{1}=1}^{m-1}\frac{m_{1}}{m}C_{m}^{m_{1}}\left\{  \sum_{j=1}%
^{m_{B}}\left[  u_{k-1;l,j}^{(N)}\left(  t\right)  -u_{k;l,j}^{(N)}\left(
t\right)  \right]  \right\}  ^{m_{1}-1}\\
\end{align*}%
\begin{align}
&  \times\sum_{\substack{n_{1}+n_{2}+\cdots+n_{m_{A}}=m-m_{1}\\\sum_{i\neq
l}^{m_{A}}n_{i}\geq1\\0\leq n_{j}\leq m-m_{1},1\leq j\leq m_{A}}}\left(
\begin{array}
[c]{c}%
m-m_{1}\\
n_{1},n_{2},\ldots,n_{m_{A}}%
\end{array}
\right)  \prod_{i=1}^{m_{A}}\left\{  \sum_{j=1}^{m_{B}}\left[  u_{k-1;i,j}%
^{(N)}\left(  t\right)  -u_{k;i,j}^{(N)}\left(  t\right)  \right]  \right\}
^{n_{i}}\nonumber\\
&  \times\sum_{\substack{r_{1}+r_{2}+\cdots+r_{m_{A}}=d-m\\0\leq r_{j}\leq
d-m,1\leq j\leq m_{A}}}\left(
\begin{array}
[c]{c}%
d-m\\
r_{1},r_{2},\ldots,r_{m_{A}}%
\end{array}
\right)  \prod_{i=1}^{m_{A}}\left\{  \sum_{j=1}^{m_{B}}\left[  u_{k;i,j}%
^{(N)}\left(  t\right)  \right]  \right\}  ^{r_{i}}. \label{Equ3}%
\end{align}

Note that $\left[  u_{k-1;l,j}^{\left(  N\right)  }\left(  t\right)
d_{l,i}-u_{k;i,j}^{\left(  N\right)  }\left(  t\right)  \left(  d_{i,1}%
,d_{i,2},\ldots,d_{i,m_{A}}\right)  e\right]  $ is the rate that any arriving
customer joins one server with the shortest queue length $k-1$, where the MAP
goes to phase $i$ from phase $l$, and the PH service time is in phase $j$.

Now, we provide a detailed interpretation for how to derive (\ref{Equ3})
through a set decomposition of all possible events given in Figure 2, where
each of the $d$ selected servers has at least $k-1$ customers, the MAP arrival
environment is in phase $i$ or $l$, and the PH service environment is in phase
$j$. Hence, the probability that any arriving customer selects $d$ servers
with at least $k-1$ customers at random and joins a server with the shortest
queue length $k-1$ and with the MAP phase $i$ or $l$ is determined by means of
Figure 2 through the following three parts:

\begin{figure}[ptb]
\centering  \includegraphics[width=12cm]{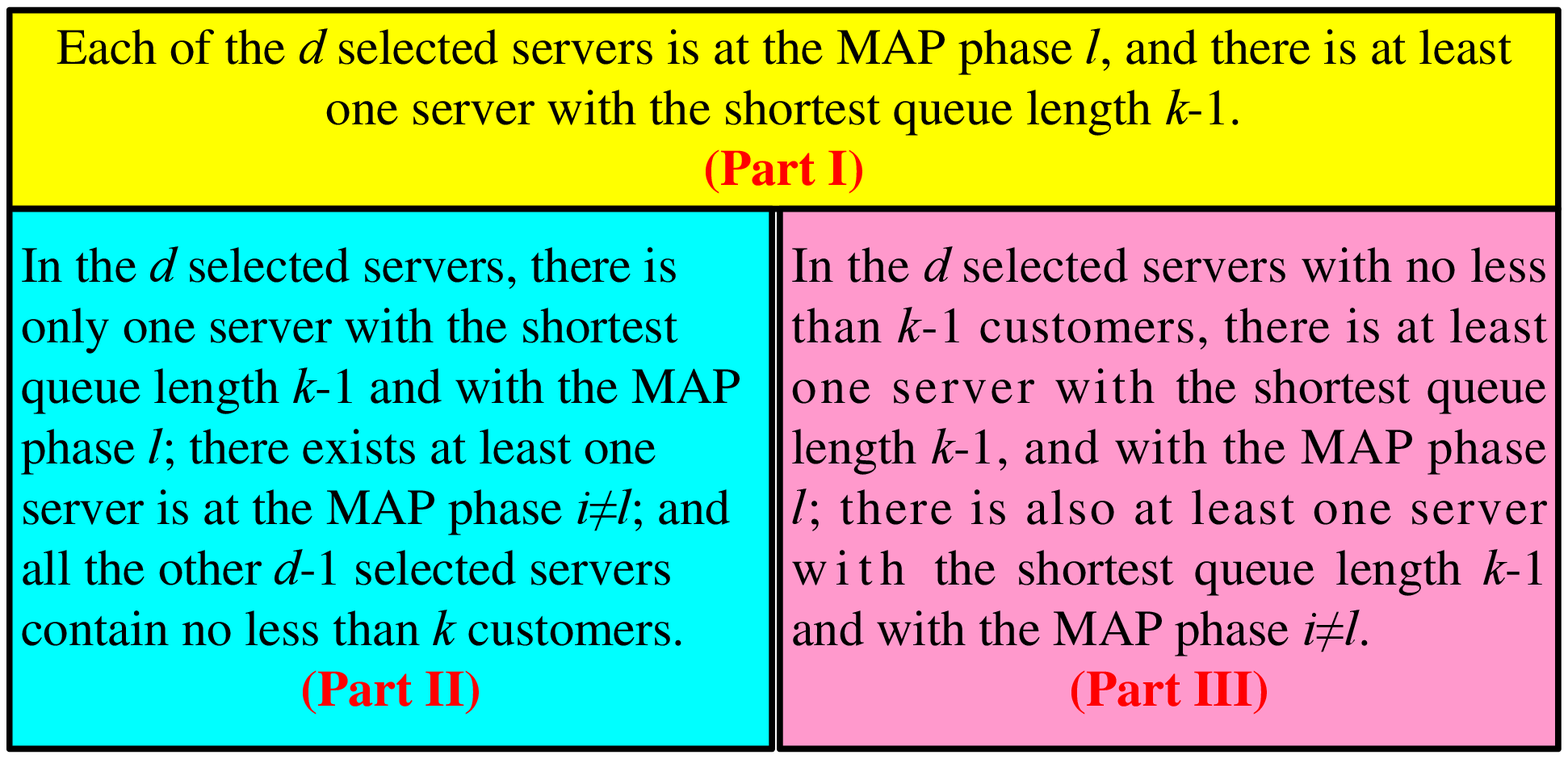}  \newline \caption{A set
decomposition of all possible events}%
\label{figure:fig-2}%
\end{figure}

\textbf{Part I: }The probability that any arriving customer joins a server
with the shortest queue length $k-1$ and with the MAP phase $l$, and the queue
lengths of the other selected $d-1$ servers are not shorter than $k-1$, is
given by%
\[
\sum_{m=1}^{d}C_{d}^{m}\left\{  \sum_{j=1}^{m_{B}}\left[  u_{k-1;l,j}%
^{(N)}\left(  t\right)  -u_{k;l,j}^{(N)}\left(  t\right)  \right]  \right\}
^{m-1}\left\{  \sum_{j=1}^{m_{B}}\left[  u_{k;l,j}^{(N)}\left(  t\right)
\right]  \right\}  ^{d-m},
\]
where $C_{d}^{m}=d!/\left[  m!\left(  d-m\right)  !\right]  $ is a binomial
coefficient, and%
\[
\left\{  \sum_{j=1}^{m_{B}}\left[  u_{k-1;l,j}^{(N)}\left(  t\right)
-u_{k;l,j}^{(N)}\left(  t\right)  \right]  \right\}  ^{m-1}%
\]
is the probability that any arriving customer who can only choose one server
makes $m-1$ independent selections during the $m-1$ servers with the queue
length $k-1$ and with the MAP phase $l$ at time $t$; while $\left\{
\sum_{j=1}^{m_{B}}\left[  u_{k;l,j}^{(N)}\left(  t\right)  \right]  \right\}
^{d-m}$ is the probability that there are $d-m$ servers whose queue lengths
are not shorter than $k$ and with the MAP phase $l$.

\textbf{Part I\negthinspace I:} The probability that any arriving customer
joins a server with the shortest queue length $k-1$ and with the MAP phase
$l$; and the queue lengths of the other selected $d-1$ servers are not shorter
than $k-1$, and there exist at least one server with no less than $k$
customers and with the MAP phase $i\neq l$, is given by
\begin{align*}
&  \sum_{m=1}^{d-1}C_{d}^{m}\left\{  \sum_{j=1}^{m_{B}}\left[  u_{k-1;l,j}%
^{(N)}\left(  t\right)  -u_{k;l,j}^{(N)}\left(  t\right)  \right]  \right\}
^{m-1}\\
&  \times\sum_{\substack{r_{1}+r_{2}+\cdots+r_{m_{A}}=d-m\\\sum_{i\neq
l}^{m_{A}}r_{i}\geq1\\0\leq r_{j}\leq d-m,1\leq j\leq m_{A}}}\left(
\begin{array}
[c]{c}%
d-m\\
r_{1},r_{2},\ldots,r_{m_{A}}%
\end{array}
\right)  \prod_{i=1}^{m_{A}}\left\{  \sum_{j=1}^{m_{B}}\left[  u_{k;i,j}%
^{(N)}\left(  t\right)  \right]  \right\}  ^{r_{i}},
\end{align*}
where when $r_{1}+r_{2}+\cdots+r_{m_{A}}=n$, $\left(
\begin{array}
[c]{c}%
n\\
r_{1},r_{2},\ldots,r_{m_{A}}%
\end{array}
\right)  =\dfrac{n}{\prod\nolimits_{i=1}^{m_{A}}r_{i}!}$ is a multinomial coefficient.

\textbf{Part I\negthinspace I\negthinspace I:} If there are $m$ selected
servers with the shortest queue length $k-1$ where there are $m_{1}\ $servers
with the MAP phase $l$ and $m-m_{1}$ servers with the MAP phases $i\neq l$,
then the probability that any arriving customer joins a server with the
shortest queue length $k-1$ and with the MAP phase $l$ is equal to $m_{1}/m$.
In this case, the probability that any arriving customer joins a server with
the shortest queue length $k-1$ and with the MAP phase $l$, the queue lengths
of the other selected $d-1$ servers are not shorter than $k-1$, is given by%
\begin{align*}
&  \sum_{m=2}^{d}C_{d}^{m}\sum_{m_{1}=1}^{m-1}\frac{m_{1}}{m}C_{m}^{m_{1}%
}\left\{  \sum_{j=1}^{m_{B}}\left[  u_{k-1;l,j}^{(N)}\left(  t\right)
-u_{k;l,j}^{(N)}\left(  t\right)  \right]  \right\}  ^{m_{1}-1}\\
&  \times\sum_{\substack{n_{1}+n_{2}+\cdots+n_{m_{A}}=m-m_{1}\\\sum_{i\neq
l}^{m_{A}}n_{i}\geq1\\0\leq n_{j}\leq m-m_{1},1\leq j\leq m_{A}}}\left(
\begin{array}
[c]{c}%
m-m_{1}\\
n_{1},n_{2},\ldots,n_{m_{A}}%
\end{array}
\right)  \prod_{i=1}^{m_{A}}\left\{  \sum_{j=1}^{m_{B}}\left[  u_{k-1;i,j}%
^{(N)}\left(  t\right)  -u_{k;i,j}^{(N)}\left(  t\right)  \right]  \right\}
^{n_{i}}\\
&  \times\sum_{_{\substack{r_{1}+r_{2}+\cdots+r_{m_{A}}=d-m\\0\leq r_{j}\leq
d-m,1\leq j\leq m_{A}}}}\left(
\begin{array}
[c]{c}%
d-m\\
r_{1},r_{2},\ldots,r_{m_{A}}%
\end{array}
\right)  \prod_{i=1}^{m_{A}}\left\{  \sum_{j=1}^{m_{B}}\left[  u_{k;i,j}%
^{(N)}\left(  t\right)  \right]  \right\}  ^{r_{i}}.
\end{align*}

Using the above three parts, (\ref{Equ3-0}) and (\ref{Equ3}) can be obtained immediately.

For any two matrices $A=\left(  a_{i,j}\right)  $ and $B=\left(
b_{i,j}\right)  $, their Kronecker product is defined as $A\otimes B=\left(
a_{i,j}B\right)  $, and their Kronecker sum is given by $A\oplus B=A\otimes
I+I\otimes B$.

The following theorem gives an important result, called \textit{the invariance
of environment factors}, which will play an important role in setting up the
infinite-dimensional system of differential vector equations. This enables us
to apply the matrix-analytic method to the study of more general supermarket
models with non-Poisson inputs and non-exponential service times.

\begin{The}
\label{The:InvEnF}%
\begin{align}
L_{1;l}^{\left(  N\right)  }\left(  u_{0}^{\left(  N\right)  }\left(
t\right)  \otimes\alpha,u_{1}^{\left(  N\right)  }\left(  t\right)  \right)
=  &  \sum_{m=1}^{d}C_{d}^{m}\left[  \sum_{l=1}^{m_{A}}\sum_{j=1}^{m_{B}%
}\left(  u_{0;l}^{\left(  N\right)  }\left(  t\right)  \alpha_{j}%
-u_{1;l,j}^{\left(  N\right)  }\left(  t\right)  \right)  \right]
^{m-1}\nonumber\\
&  \times\left[  \sum_{l=1}^{m_{A}}\sum_{j=1}^{m_{B}}u_{1;l,j}^{\left(
N\right)  }\left(  t\right)  \right]  ^{d-m} \label{Equ4-1}%
\end{align}
and for $k\geq2$%
\begin{align}
L_{k;l}^{\left(  N\right)  }\left(  u_{k-1}^{\left(  N\right)  }\left(
t\right)  ,u_{k}^{\left(  N\right)  }\left(  t\right)  \right)  =  &
\sum_{m=1}^{d}C_{d}^{m}\left[  \sum_{l=1}^{m_{A}}\sum_{j=1}^{m_{B}}\left(
u_{k-1;l,j}^{\left(  N\right)  }\left(  t\right)  -u_{k;l,j}^{\left(
N\right)  }\left(  t\right)  \right)  \right]  ^{m-1}\nonumber\\
&  \times\left[  \sum_{l=1}^{m_{A}}\sum_{j=1}^{m_{B}}u_{k;l,j}^{\left(
N\right)  }\left(  t\right)  \right]  ^{d-m}. \label{Equ4-2}%
\end{align}
Thus $L_{1;l}^{\left(  N\right)  }\left(  u_{0}^{\left(  N\right)  }\left(
t\right)  \otimes\alpha,u_{1}^{\left(  N\right)  }\left(  t\right)  \right)
$\ and $L_{k;l}^{\left(  N\right)  }\left(  u_{k-1}^{\left(  N\right)
}\left(  t\right)  ,u_{k}^{\left(  N\right)  }\left(  t\right)  \right)  $ for
$k\geq2$\ are independent of the MAP phase $l\in\left\{  1,2,\ldots
,m_{A}\right\}  $. In this case, we have%
\begin{equation}
L_{1;l}^{\left(  N\right)  }\left(  u_{0}^{\left(  N\right)  }\left(
t\right)  \otimes\alpha,u_{1}^{\left(  N\right)  }\left(  t\right)  \right)
\overset{\text{def}}{=}L_{1}^{\left(  N\right)  }\left(  u_{0}^{\left(
N\right)  }\left(  t\right)  \otimes\alpha,u_{1}^{\left(  N\right)  }\left(
t\right)  \right)  \label{Equ4-3}%
\end{equation}
and for $k\geq2$%
\begin{equation}
L_{k;l}^{\left(  N\right)  }\left(  u_{k-1}^{\left(  N\right)  }\left(
t\right)  ,u_{k}^{\left(  N\right)  }\left(  t\right)  \right)  \overset
{\text{def}}{=}L_{k}^{\left(  N\right)  }\left(  u_{k-1}^{\left(  N\right)
}\left(  t\right)  ,u_{k}^{\left(  N\right)  }\left(  t\right)  \right)  .
\label{Equ4-4}%
\end{equation}
\end{The}

\textbf{Proof:} See Appendix A. \textbf{{\rule{0.08in}{0.08in}}}

It is seen from the invariance of environment factors in Theorem 1 that
Equation (\ref{Equ3-0}) is rewritten as, in a vector form,%
\begin{align}
&  \left\{  u_{k-1}^{\left(  N\right)  }\left(  t\right)  \left(  D\otimes
I\right)  -u_{k}^{\left(  N\right)  }\left(  t\right)  \left[  \text{diag}%
\left(  De\right)  \otimes I\right]  \right\} \nonumber\\
&  \times L_{k}^{\left(  N\right)  }\left(  u_{k-1}^{\left(  N\right)
}\left(  t\right)  ,u_{k}^{\left(  N\right)  }\left(  t\right)  \right)
N\text{d}t. \label{Equ3-1}%
\end{align}
Note that $L_{1}^{\left(  N\right)  }\left(  u_{0}^{\left(  N\right)  }\left(
t\right)  \otimes\alpha,u_{1}^{\left(  N\right)  }\left(  t\right)  \right)  $
and $L_{k}^{\left(  N\right)  }\left(  u_{k-1}^{\left(  N\right)  }\left(
t\right)  ,u_{k}^{\left(  N\right)  }\left(  t\right)  \right)  $ are scale
for $k\geq2$.

\textbf{Step two: Analysis of the Environment State Transitions in the MAP}

When there are at least $k$ customers in the server, the rate that the MAP
environment process jumps from state $l$ to state $i$ with rate $c_{l,j}$, and
no arrival of the MAP occurs during a small time period $\left[
0,\text{d}t\right)  $, is given by
\[
\left[  \sum\limits_{l=1}^{m_{A}}u_{k;l,j}^{(N)}(t)c_{l,i}+u_{k,i,j}%
^{(N)}(t)\left(  d_{i,1},d_{i,2},\ldots,d_{i,m_{A}}\right)  e\right]
N\text{d}t.
\]
This gives, in a vector form,%
\begin{equation}
u_{k}^{\left(  N\right)  }\left(  t\right)  \left(  \left[  C+\text{diag}%
\left(  De\right)  \right]  \otimes I\right)  N\text{d}t. \label{Equ3-2}%
\end{equation}

\textbf{Step three: Analysis of the Service Processes}

To analyze the PH service process, we need to consider the following two cases:

Case one: One service completion occurs with rate $t_{l}^{0}$ during a small
time period $\left[  0,\text{d}t\right)  $. In this case, when there are at
least $k+1$ customers in the server, the rate that a customer is completed its
service with entering PH phase $j$ and the MAP is in phase $i$\ is given by%
\[
\left[  u_{k+1;i,1}^{(N)}(t)t_{1}^{0}\alpha_{j}+u_{k+1;i,2}^{(N)}(t)t_{2}%
^{0}\alpha_{j}+\cdots+u_{k+1;i,m_{B}}^{(N)}(t)t_{m_{B}}^{0}\alpha_{j}\right]
N\text{d}t.
\]

Case two: No service completion occurs during a small time period $\left[
0,\text{d}t\right)  $, but the MAP is in phase $i$ and the PH service
environment process goes to phase $j$. Thus, when there are at least $k$
customers in the server, the rate of this case is given by%
\[
\left[  u_{k;i,1}^{(N)}(t)t_{1,j}+u_{k;i,2}^{(N)}(t)t_{2,j}+u_{k;i,3}%
^{(N)}(t)t_{3,j}+\cdots+u_{k;i,m_{B}}^{(N)}(t)t_{m_{B},j}\right]  N\text{d}t.
\]
Thus, for the PH service process, we obtain that in a vector form,%
\begin{equation}
\left[  u_{k}^{\left(  N\right)  }\left(  t\right)  \left(  I\otimes T\right)
+u_{k+1}^{\left(  N\right)  }\left(  t\right)  \left(  I\otimes T^{0}%
\alpha\right)  \right]  N\text{d}t \label{Equ3-3}%
\end{equation}

Let%
\begin{align*}
n_{k}^{(N)}(t) =  &  \left(  n_{k;1,1}^{(N)}(t),n_{k;1,2}^{(N)}(t),\ldots
,n_{k;1,m_{B}}^{(N)}(t);\ldots;\right. \\
&  \left.  n_{k;m_{A},1}^{(N)}(t),n_{k;m_{A},2}^{(N)}(t),\ldots,n_{k;m_{A}%
,m_{B}}^{(N)}(t)\right)  .
\end{align*}
Then it follows from Equation (\ref{Equ3-1}) to (\ref{Equ3-3}) that%
\begin{align*}
\text{d}E\left[  n_{k}^{(N)}(t)\right]  =  &  \left\{  \left\{  u_{k-1}%
^{\left(  N\right)  }\left(  t\right)  \left(  D\otimes I\right)
-u_{k}^{\left(  N\right)  }\left(  t\right)  \left[  \text{diag}\left(
De\right)  \otimes I\right]  \right\}  L_{k}^{\left(  N\right)  }\left(
u_{k-1}^{\left(  N\right)  }\left(  t\right)  ,u_{k}^{\left(  N\right)
}\left(  t\right)  \right)  \right. \\
&  \left.  +u_{k}^{\left(  N\right)  }\left(  t\right)  \left\{  \left[
C+\text{diag}\left(  De\right)  \right]  \oplus T\right\}  +u_{k+1}^{\left(
N\right)  }\left(  t\right)  \left(  I\otimes T^{0}\alpha\right)  \right\}
N\text{d}t.
\end{align*}
Since $E\left[  n_{k}^{(N)}(t)/N\right]  =u_{k}^{(N)}(t)$ and $A\otimes
I+I\otimes B=A\oplus B$, we obtain%
\begin{align}
\frac{\text{d}u_{k}^{(N)}(t)}{\text{d}t}=  &  \left\{  u_{k-1}^{\left(
N\right)  }\left(  t\right)  \left(  D\otimes I\right)  -u_{k}^{\left(
N\right)  }\left(  t\right)  \left(  t\right)  \left[  \text{diag}\left(
De\right)  \otimes I\right]  \right\}  L_{k}^{\left(  N\right)  }\left(
u_{k-1}^{\left(  N\right)  }\left(  t\right)  ,u_{k}^{\left(  N\right)
}\left(  t\right)  \right) \nonumber\\
&  +u_{k}^{\left(  N\right)  }\left(  t\right)  \left\{  \left[
C+\text{diag}\left(  De\right)  \right]  \oplus T\right\}  +u_{k+1}^{\left(
N\right)  }\left(  t\right)  \left(  I\otimes T^{0}\alpha\right)  .
\label{Equ3-4}%
\end{align}
Using a similar analysis to Equation (\ref{Equ3-4}), we obtain an
infinite-dimensional system of differential vector equations satisfied by the
expected fraction vector $\mathbf{u}^{(N)}\left(  t\right)  $ as follows:%
\begin{align}
\frac{\text{d}u_{1}^{(N)}(t)}{\text{d}t}=  &  \left\{  \left[  u_{0}^{\left(
N\right)  }\left(  t\right)  \otimes\alpha\right]  \left(  D\otimes I\right)
-u_{1}^{\left(  N\right)  }\left(  t\right)  \left[  \text{diag}\left(
De\right)  \otimes I\right]  \right\}  L_{1}^{\left(  N\right)  }\left(
u_{0}^{\left(  N\right)  }\left(  t\right)  \otimes\alpha,u_{1}^{\left(
N\right)  }\left(  t\right)  \right) \nonumber\\
&  +u_{1}^{(N)}(t)\left\{  \left[  C+\text{diag}\left(  De\right)  \right]
\oplus T\right\}  +u_{2}^{\left(  N\right)  }\left(  t\right)  \left(
I\otimes T^{0}\alpha\right)  , \label{Equ6}%
\end{align}
and for $k\geq2$%
\begin{align}
\frac{\text{d}u_{k}^{(N)}(t)}{\text{d}t}=  &  \left\{  u_{k-1}^{\left(
N\right)  }\left(  t\right)  \left(  D\otimes I\right)  -u_{k}^{\left(
N\right)  }\left(  t\right)  \left[  \text{diag}\left(  De\right)  \otimes
I\right]  \right\}  L_{k}^{\left(  N\right)  }\left(  u_{k-1}^{\left(
N\right)  }\left(  t\right)  ,u_{k}^{\left(  N\right)  }\left(  t\right)
\right) \nonumber\\
&  +u_{k}^{\left(  N\right)  }\left(  t\right)  \left\{  \left[
C+\text{diag}\left(  De\right)  \right]  \oplus T\right\}  +u_{k+1}^{\left(
N\right)  }\left(  t\right)  \left(  I\otimes T^{0}\alpha\right)  ,
\label{Equ7}%
\end{align}
with the boundary condition%
\begin{equation}
\frac{\text{d}u_{0}^{(N)}(t)}{\text{d}t}=u_{0}^{(N)}(t)\left(  C+D\right)  ,
\label{Equ7-1}%
\end{equation}%
\begin{equation}
u_{0}^{(N)}(t)e=1; \label{Equ7-2}%
\end{equation}
and with the initial condition%
\begin{equation}
u_{k}^{(N)}(0)=g_{k},\text{ }k\geq1, \label{Equ8}%
\end{equation}
where%
\[
g_{1}\geq g_{2}\geq g_{3}\geq\cdots\geq0
\]
and%
\[
1=g_{0}e\geq g_{1}e\geq g_{2}e\geq\cdots\geq0.
\]

\begin{Rem}
It is necessary to explain some probability setting for the invariance of
environment factors. It follows from Theorem 1 that%
\[
L_{1}^{\left(  N\right)  }\left(  u_{0}^{\left(  N\right)  }\left(  t\right)
\otimes\alpha,u_{1}^{\left(  N\right)  }\left(  t\right)  \right)
=\frac{\left[  u_{0}^{\left(  N\right)  }\left(  t\right)  e\right]
^{d}-\left[  u_{1}^{\left(  N\right)  }\left(  t\right)  e\right]  ^{d}}%
{u_{0}^{\left(  N\right)  }\left(  t\right)  e-u_{1}^{\left(  N\right)
}\left(  t\right)  e}%
\]
and for $k\geq2$%
\[
L_{k}^{\left(  N\right)  }\left(  u_{k-1}^{\left(  N\right)  }\left(
t\right)  ,u_{k}^{\left(  N\right)  }\left(  t\right)  \right)  =\frac{\left[
u_{k-1}^{\left(  N\right)  }\left(  t\right)  e\right]  ^{d}-\left[
u_{k}^{\left(  N\right)  }\left(  t\right)  e\right]  ^{d}}{u_{k-1}^{\left(
N\right)  }\left(  t\right)  e-u_{k}^{\left(  N\right)  }\left(  t\right)
e}.
\]
Note that the two expressions will be useful in our later study, for example,
establishing the Lipschitz condition, and computing the fixed point.
Specifically, for $d=1$ we have%
\[
L_{1}^{\left(  N\right)  }\left(  u_{0}^{\left(  N\right)  }\left(  t\right)
\otimes\alpha,u_{1}^{\left(  N\right)  }\left(  t\right)  \right)  =1
\]
and for $k\geq2$%
\[
L_{k}^{\left(  N\right)  }\left(  u_{k-1}^{\left(  N\right)  }\left(
t\right)  ,u_{k}^{\left(  N\right)  }\left(  t\right)  \right)  =1.
\]
For $d=2$ we have%
\[
L_{1}^{\left(  N\right)  }\left(  u_{0}^{\left(  N\right)  }\left(  t\right)
\otimes\alpha,u_{1}^{\left(  N\right)  }\left(  t\right)  \right)
=u_{0}^{\left(  N\right)  }\left(  t\right)  e+u_{1}^{\left(  N\right)
}\left(  t\right)  e>1
\]
and for $k\geq2$%
\[
L_{k}^{\left(  N\right)  }\left(  u_{k-1}^{\left(  N\right)  }\left(
t\right)  ,u_{k}^{\left(  N\right)  }\left(  t\right)  \right)  =u_{k-1}%
^{\left(  N\right)  }\left(  t\right)  e+u_{k}^{\left(  N\right)  }\left(
t\right)  e.
\]
This shows that $\left(  L_{1}^{\left(  N\right)  }\left(  u_{0}^{\left(
N\right)  }\left(  t\right)  \otimes\alpha,u_{1}^{\left(  N\right)  }\left(
t\right)  \right)  ,L_{2}^{\left(  N\right)  }\left(  u_{1}^{\left(  N\right)
}\left(  t\right)  ,u_{2}^{\left(  N\right)  }\left(  t\right)  \right)
,\ldots\right)  $ is not a probability vector.
\end{Rem}

\section{The Lipschitz Condition}

In this section, we show that the mean-field limit of the sequence of Markov
processes asymptotically approaches a single trajectory identified by the
unique and global solution to the infinite-dimensional system of limiting
differential vector equations. To that end, we provide a unified
matrix-differential algorithm for establishing the Lipschitz condition, which
is a key in proving the existence and uniqueness of the solution by means of
the Picard approximation according to the basic results of the Banach space.

Let $\mathbf{T}_{N}(t)$ be the operator semigroup of the Markov process
$\left\{  \mathbf{U}^{(N)}(t),t\geq0\right\}  $. If $f:\Omega_{N}%
\rightarrow\mathbf{C}^{1}$, where $\Omega_{N}=\left\{  \mathbf{g}\in
\widetilde{\Omega}_{N}:\mathbf{g}e<+\infty\right\}  $, then for $\mathbf{g}%
\in\Omega_{N}$ and $t\geq0$%
\[
\mathbf{T}_{N}(t)f(\mathbf{g})=E\left[  f(\mathbf{U}_{N}(t)\text{ }|\text{
}\mathbf{U}_{N}(0)=\mathbf{g}\right]  .
\]
We denote by $\mathbf{A}_{N}$ the generating operator of the operator
semigroup $\mathbf{T}_{N}(t)$, it is easy to see that $\mathbf{T}_{N}%
(t)=\exp\left\{  \mathbf{A}_{N}t\right\}  $ for $t\geq0$. In \textbf{Appendix
B}, we will provide a detailed analysis for the limiting behavior of the
sequence $\{(\mathbf{U}^{(N)}(t),t\geq0\}$ of Markov processes for
$N=1,2,3,\ldots$, where two formal limits for the sequence $\left\{
\mathbf{A}_{N}\right\}  $ of generating operators and for the sequence
$\left\{  \mathbf{T}_{N}(t)\right\}  $ of operator semigroups are expressed as
$\mathbf{A}=\lim_{N\rightarrow\infty}\mathbf{A}_{N}$ and $\mathbf{T}\left(
t\right)  =\lim_{N\rightarrow\infty}\mathbf{T}_{N}(t)$ for $t\geq0$, respectively.

We write%
\[
L_{1}\left(  u_{0}\left(  t\right)  \otimes\alpha,u_{1}\left(  t\right)
\right)  =\sum_{m=1}^{d}C_{d}^{m}\left[  \sum_{l=1}^{m_{A}}\sum_{j=1}^{m_{B}%
}\left(  u_{0,l}\left(  t\right)  \alpha_{j}-u_{1;l,j}\left(  t\right)
\right)  \right]  ^{m-1}\left[  \sum_{l=1}^{m_{A}}\sum_{j=1}^{m_{B}}%
u_{1;l,j}\left(  t\right)  \right]  ^{d-m},
\]
for $k\geq2$%
\begin{align*}
L_{k}\left(  u_{k-1}\left(  t\right)  ,u_{k}\left(  t\right)  \right)   &
=\sum_{m=1}^{d}C_{d}^{m}\left[  \sum_{l=1}^{m_{A}}\sum_{j=1}^{m_{B}}\left(
u_{k-1;l,j}\left(  t\right)  -u_{k;l,j}\left(  t\right)  \right)  \right]
^{m-1}\\
&  \times\left[  \sum_{l=1}^{m_{A}}\sum_{j=1}^{m_{B}}u_{k;l,j}\left(
t\right)  \right]  ^{d-m}.
\end{align*}

Let $\mathbf{u}(t)=\lim_{N\rightarrow\infty}\mathbf{u}^{(N)}(t)$ where
$u_{k}\left(  t\right)  =\lim_{N\rightarrow\infty}u_{k}^{(N)}(t)$ for $k\geq0$
and $t\geq0$. Based on the limiting operator semigroup $\mathbf{T}\left(
t\right)  $ or the limiting generating operator $\mathbf{A}$, as
$N\rightarrow\infty$ it follows from Equations (\ref{Equ6}) to (\ref{Equ8})
that $\mathbf{u}(t)$ is a solution to the system of differential vector
equations as follows:%
\begin{align}
\frac{\text{d}u_{1}(t)}{\text{d}t}=  &  \left\{  \left[  u_{0}(t)\otimes
\alpha\right]  \left(  D\otimes I\right)  -u_{1}\left(  t\right)  \left[
\text{diag}\left(  De\right)  \otimes I\right]  \right\}  L_{1}\left(
u_{0}(t)\otimes\alpha,u_{1}\left(  t\right)  \right) \nonumber\\
&  +u_{1}(t)\left\{  \left[  C+\text{diag}\left(  De\right)  \right]  \oplus
T\right\}  +u_{2}\left(  t\right)  \left(  I\otimes T^{0}\alpha\right)  ,
\label{EqS1}%
\end{align}
and for $k\geq2$%
\begin{align}
\frac{\text{d}u_{k}(t)}{\text{d}t}=  &  \left\{  u_{k-1}\left(  t\right)
\left(  D\otimes I\right)  -u_{k}\left(  t\right)  \left[  \text{diag}\left(
De\right)  \otimes I\right]  \right\}  L_{k}\left(  u_{k-1}\left(  t\right)
,u_{k}\left(  t\right)  \right) \nonumber\\
&  +u_{k}\left(  t\right)  \left\{  \left[  C+\text{diag}\left(  De\right)
\right]  \oplus T\right\}  +u_{k+1}\left(  t\right)  \left(  I\otimes
T^{0}\alpha\right)  , \label{EqS2}%
\end{align}
with the boundary condition%
\begin{equation}
u_{0}^{\left(  N\right)  }\left(  t\right)  =u_{0}^{\left(  N\right)  }\left(
0\right)  \exp\left\{  \left(  C+D\right)  t\right\}  , \label{EqS3}%
\end{equation}%
\begin{equation}
u_{0}^{\left(  N\right)  }\left(  t\right)  e=1, \label{EqS4}%
\end{equation}
and with initial condition%
\begin{equation}
u_{k}\left(  0\right)  =g_{k},\text{ \ \ }k\geq0. \label{EqS6}%
\end{equation}

Based on the solution $\mathbf{u}(t,\mathbf{g)}$ to the system of differential
vector equations (\ref{EqS1}) to (\ref{EqS6}), we define a mapping:
$\mathbf{g}\rightarrow\mathbf{u}(t,\mathbf{g})$. Note that the operator
semigroup $\mathbf{T}(t)$ acts in the space $L$, where $L=C(\widetilde{\Omega
})$ is the Banach space of continuous functions $f:\widetilde{\Omega
}\rightarrow\mathbf{R}$ with uniform metric $\left\|  f\right\|
=\underset{u\in\widetilde{\Omega}}{\max}\left|  f(u)\right|  $, and%
\[
\widetilde{\Omega}=\{\mathbf{u}:u_{1}\geq u_{2}\geq u_{3}\geq\cdots
\geq0;\ \ 1=u_{0}^{\left(  N\right)  }e\geq u_{1}^{\left(  N\right)  }e\geq
u_{2}^{\left(  N\right)  }e\geq\cdots\geq0\}
\]
for the vector $\mathbf{u}=\left(  u_{0},u_{1},u_{2},\ldots\right)  $ with
$u_{0}$ be a probability vector of size $m_{A}$ and the size of the row vector
$u_{k}$ be $m_{A}m_{B}$ for $k\geq1$. If $f\in L$ and $\mathbf{g}\in
\widetilde{\Omega}$, then%
\[
\mathbf{T}(t)f(\mathbf{g})=f\left(  \mathbf{u}(t,\mathbf{g})\right)  .
\]

The following theorem uses the operator semigroup to provide the mean-field
limit in this supermarket model. Note that the mean-field limit shows that
there always exists the limiting process $\{U\left(  t\right)  ,t\geq0\}$ of
the sequence $\{U^{\left(  N\right)  }\left(  t\right)  ,t\geq0\}$ of Markov
processes, and also indicates the asymptotic independence of the
block-structured queueing processes in this supermarket model.

\begin{The}
\label{The:Conver}For any continuous function $f:$ $\Omega\rightarrow
\mathbf{R}$ and $t>0$,%
\[
\underset{N\rightarrow\infty}{\lim}\underset{\mathbf{g}\in\Omega}{\sup}\left|
\mathbf{T}_{N}(t)f(\mathbf{g})-f(\mathbf{u}(t;\mathbf{g}))\right|  =0,
\]
and the convergence is uniform in $t$ with any bounded interval.
\end{The}

\textbf{Proof:} See Appendix B. \textbf{{\rule{0.08in}{0.08in}}}

Finally, we provide some interpretation on Theorem \ref{The:Conver}. If
$\lim_{N\rightarrow\infty}U^{\left(  N\right)  }\left(  0\right)
=\mathbf{u}(0)=\mathbf{g}\in$ $\Omega$ in probability, then Theorem
\ref{The:Conver} shows that $U\left(  t\right)  =\lim_{N\rightarrow\infty
}U^{\left(  N\right)  }\left(  t\right)  $ is concentrated on the trajectory
$\Gamma_{\mathbf{g}}=\left\{  \mathbf{u}(t,\mathbf{g}):t\geq0\right\}  $. This
indicates the functional strong law of large numbers for the time evolution of
the fraction of each state of this supermarket model, thus the sequence
$\left\{  U^{\left(  N\right)  }\left(  t\right)  ,t\geq0\right\}  $ of Markov
processes converges weakly to the expected fraction vector $\mathbf{u}%
(t,\mathbf{g})$ as $N\rightarrow\infty$, that is, for any $T>0$%
\begin{equation}
\lim_{N\rightarrow\infty}\sup_{0\leq s\leq T}\left\Vert U^{\left(  N\right)
}\left(  s\right)  -\mathbf{u}(s,\mathbf{g})\right\Vert =0\text{ \ in
probability}. \label{EquK-16}%
\end{equation}

In the remainder of this section, we provide a unified matrix-differential
algorithm for establishing a Lipschitz condition for the expected fraction
vector $f:\mathbf{R}_{+}^{\infty}\rightarrow\mathbf{C}^{1}\left(
\mathbf{R}_{+}^{\infty}\right)  $. The Lipschitz condition is a key for
proving the existence and uniqueness of solution to the infinite-dimensional
system of limiting differential vector equations (\ref{EqS1}) to (\ref{EqS6}).
On the other hand, the proof of the existence and uniqueness of solution is
standard by means of the Picard approximation according to the basic results
of the Banach space. Readers may refer to Li, Dai, Lui and Wang
\cite{Li:2013a} for more details.

To provide the Lipschitz condition, we need to use the derivative of the
infinite-dimensional vector $G:\mathbf{R}_{+}^{\infty}\rightarrow
\mathbf{C}^{1}\left(  \mathbf{R}_{+}^{\infty}\right)  $. Thus we first provide
some definitions and preliminaries for such derivatives as follows.

For the infinite-dimensional vector $G:\mathbf{R}_{+}^{\infty}\rightarrow
\mathbf{C}^{1}\left(  \mathbf{R}_{+}^{\infty}\right)  $, we write
$x=(x_{1},x_{2},x_{3},\ldots)$ and $G(x)=(G_{1}(x),G_{2}(x),G_{3}(x),\ldots)$,
where $x_{k}$ and $G_{k}(x)$ are scalar for $k\geq1$. Then the matrix of
partial derivatives of the infinite-dimensional vector $G(x)$ is defined as%
\begin{equation}
\mathcal{D}G(x)=\dfrac{\partial G(x)}{\partial x}=\left(
\begin{array}
[c]{cccc}%
\dfrac{\partial G_{1}(x)}{\partial x_{1}} & \dfrac{\partial G_{2}(x)}{\partial
x_{1}} & \dfrac{\partial G_{3}(x)}{\partial x_{1}} & \cdots\\
\dfrac{\partial G_{1}(x)}{\partial x_{2}} & \dfrac{\partial G_{2}(x)}{\partial
x_{2}} & \dfrac{\partial G_{3}(x)}{\partial x_{2}} & \cdots\\
\dfrac{\partial G_{1}(x)}{\partial x_{3}} & \dfrac{\partial G_{2}(x)}{\partial
x_{3}} & \dfrac{\partial G_{3}(x)}{\partial x_{3}} & \cdots\\
\vdots & \vdots & \vdots &
\end{array}
\right)  , \label{Eq7.3}%
\end{equation}
if each of the partial derivatives exists.

For the infinite-dimensional vector $G:\mathbf{R}_{+}^{\infty}\rightarrow
\mathbf{C}^{1}\left(  \mathbf{R}_{+}^{\infty}\right)  $, if there exists a
linear operator $A:\mathbf{R}_{+}^{\infty}\rightarrow\mathbf{C}^{1}\left(
\mathbf{R}_{+}^{\infty}\right)  $ such that for any vector $h\in
\mathbf{R}^{\infty}$ and a scalar $t\in\mathbf{R}$%
\[
\lim_{t\rightarrow0}\frac{||G\left(  x+th\right)  -G\left(  x\right)
-thA||}{t}=0,
\]
then the function $G\left(  x\right)  $ is called to be Gateaux differentiable
at $x\in\mathbf{R}_{+}^{\infty}$. In this case, we write the Gateaux
derivative $A=\mathcal{D}G(x)=\dfrac{\partial G(x)}{\partial x}$.

Let $\boldsymbol{t}  =\left(  t_{1},t_{2},t_{3},\ldots\right)  $ with $0\leq
t_{k}\leq1$ for $k\geq1$. Then we write%
\[
\mathcal{D}G(x+\boldsymbol{t\oslash}  \left(  y-x\right)  )=\left(
\begin{array}
[c]{cccc}%
\dfrac{\partial G_{1}(x+t_{1}\left(  y-x\right)  )}{\partial x_{1}} &
\dfrac{\partial G_{2}(x+t_{2}\left(  y-x\right)  )}{\partial x_{1}} &
\dfrac{\partial G_{3}(x+t_{3}\left(  y-x\right)  )}{\partial x_{1}} & \cdots\\
\dfrac{\partial G_{1}(x+t_{1}\left(  y-x\right)  )}{\partial x_{2}} &
\dfrac{\partial G_{2}(x+t_{2}\left(  y-x\right)  )}{\partial x_{2}} &
\dfrac{\partial G_{3}(x+t_{3}\left(  y-x\right)  )}{\partial x_{2}} & \cdots\\
\dfrac{\partial G_{1}(x+t_{1}\left(  y-x\right)  )}{\partial x_{3}} &
\dfrac{\partial G_{2}(x+t_{2}\left(  y-x\right)  )}{\partial x_{3}} &
\dfrac{\partial G_{3}(x+t_{3}\left(  y-x\right)  )}{\partial x_{3}} & \cdots\\
\vdots & \vdots & \vdots &
\end{array}
\right)  .
\]
If the infinite-dimensional vector $G:\mathbf{R}_{+}^{\infty}\rightarrow
\mathbf{C}^{1}\left(  \mathbf{R}_{+}^{\infty}\right)  $ is Gateaux
differentiable, then there exists a vector $\boldsymbol{t}  =\left(
t_{1},t_{2},t_{3},\ldots\right)  $ with $0\leq t_{k}\leq1$ for $k\geq1$ such
that%
\begin{equation}
G\left(  y\right)  -G\left(  x\right)  =\left(  y-x\right)  \mathcal{D}%
G(x+\boldsymbol{t\oslash}  \left(  y-x\right)  ). \label{Eq7.3.2}%
\end{equation}
Furthermore, we have%
\begin{equation}
||G\left(  y\right)  -G\left(  x\right)  ||\text{ }\leq\sup_{0\leq t\leq
1}||\mathcal{D}G(x+t\left(  y-x\right)  )||\text{ }||y-x||. \label{Eq7.3.3}%
\end{equation}

For convenience of description, Equations (\ref{EqS1}) to (\ref{EqS6}) are
rewritten as an initial value problem as follows:%
\begin{align}
\frac{\text{d}}{\text{d}t}u_{1}=  &  \left\{  \left(  u_{0}\otimes
\alpha\right)  \left(  D\otimes I\right)  -u_{1}\left[  \text{diag}\left(
De\right)  \otimes I\right]  \right\}  L_{1}\left(  u_{0}\otimes\alpha
,u_{1}\right) \nonumber\\
&  +u_{1}\left\{  \left[  C+\text{diag}\left(  De\right)  \right]  \oplus
T\right\}  +u_{2}\left(  I\otimes T^{0}\alpha\right)  \label{Equ5.1}%
\end{align}
and for $k\geq2$,%
\begin{align}
\frac{\text{d}}{\text{d}t}u_{k}=  &  \left\{  u_{k-1}\left(  D\otimes
I\right)  -u_{k}\left[  \text{diag}\left(  De\right)  \otimes I\right]
\right\}  L_{k}\left(  u_{k-1},u_{k}\right) \nonumber\\
&  +u_{k}\left\{  \left[  C+\text{diag}\left(  De\right)  \right]  \oplus
T\right\}  +u_{k+1}\left(  I\otimes T^{0}\alpha\right)  , \label{Equ5.2}%
\end{align}
with the initial condition%
\begin{equation}
u_{k}\left(  0\right)  =g_{k},\text{ \ }k\geq0, \label{Equ5.4}%
\end{equation}
where for $t\geq0$%
\[
u_{0}\left(  t\right)  =u_{0}\left(  0\right)  \exp\left\{  \left(
C+D\right)  t\right\}
\]
and%
\[
u_{0}\left(  t\right)  e=1.
\]

Let $x=\left(  x_{1},x_{2},x_{3},\ldots\right)  =\left(  u_{1},u_{2}%
,u_{3},\ldots\right)  $ and $F(x)=(F_{1}(x),F_{2}(x),F_{3}(x),\ldots)$, where%
\begin{align}
F_{1}(x)=  &  \left\{  \left(  u_{0}\otimes\alpha\right)  \left(  D\otimes
I\right)  -x_{1}\left[  \text{diag}\left(  De\right)  \otimes I\right]
\right\}  L_{1}\left(  u_{0}\otimes\alpha,x_{1}\right) \nonumber\\
&  +x_{1}\left\{  \left[  C+\text{diag}\left(  De\right)  \right]  \oplus
T\right\}  +x_{2}\left(  I\otimes T^{0}\alpha\right)  \label{Eq7.2-0}%
\end{align}
and for $k\geq2$%
\begin{align}
F_{k}(x)=  &  \left\{  x_{k-1}\left(  D\otimes I\right)  -x_{k}\left[
\text{diag}\left(  De\right)  \otimes I\right]  \right\}  L_{k}\left(
x_{k-1},x_{k}\right) \nonumber\\
&  +x_{k}\left\{  \left[  C+\text{diag}\left(  De\right)  \right]  \oplus
T\right\}  +x_{k+1}\left(  I\otimes T^{0}\alpha\right)  . \label{Eq7.2}%
\end{align}
Note that $u_{0}=g_{0}\exp\left\{  \left(  C+D\right)  t\right\}  $ may be
regarded as a given vector. Thus $F(x)$ is in $\mathbf{C}^{2}\left(
\mathbf{R}_{+}^{\infty}\right)  $, and the system of differential vector
equations (\ref{Equ5.1}) to (\ref{Equ5.4}) is rewritten as%
\begin{equation}
\frac{\text{d}}{\text{d}t}x=F(x) \label{Eq4.6}%
\end{equation}
with the initial condition%
\begin{equation}
x\left(  0\right)  =\widetilde{\mathbf{g}}=\left(  g_{1},g_{2},g_{3}%
,\ldots\right)  . \label{Eq4.7}%
\end{equation}

In what follows we show that the expected fraction vector $F(x)$ is Lipschitz.

Based on the definition of the Gateaux derivative, it follows from
(\ref{Eq7.2-0}) and (\ref{Eq7.2}) that%
\[
\dfrac{\partial F(x)}{\partial x}=\left(
\begin{array}
[c]{ccccc}%
\dfrac{\partial F_{1}(x)}{\partial x_{1}} & \dfrac{\partial F_{2}(x)}{\partial
x_{1}} &  &  & \\
\dfrac{\partial F_{1}(x)}{\partial x_{2}} & \dfrac{\partial F_{2}(x)}{\partial
x_{2}} & \dfrac{\partial F_{3}(x)}{\partial x_{2}} &  & \\
& \dfrac{\partial F_{2}(x)}{\partial x_{3}} & \dfrac{\partial F_{3}%
(x)}{\partial x_{3}} & \dfrac{\partial F_{4}(x)}{\partial x_{3}} & \\
&  & \ddots & \ddots & \ddots
\end{array}
\right)  .
\]
We write%
\begin{equation}
\mathcal{D}F(x)=\left(
\begin{array}
[c]{ccccc}%
A_{1}(x) & B_{1}(x) &  &  & \\
C_{2}(x) & A_{2}(x) & B_{2}(x) &  & \\
& C_{3}(x) & A_{3}(x) & B_{3}(x) & \\
&  & \ddots & \ddots & \ddots
\end{array}
\right)  =\dfrac{\partial F(x)}{\partial x}. \label{Eq7.4}%
\end{equation}
where $A_{k}\left(  x\right)  $, $B_{k}\left(  x\right)  $ and $C_{j}\left(
x\right)  $ are the matrices of size $m_{A}m_{B}$ for $k\geq1$ and $j\geq2$.

To compute the matrix $\mathcal{D}F(x)$, we need to use two basic properties
of the Gateaux derivative as follows:

\textbf{Property one}%
\[
\frac{\partial x_{k}}{\partial x_{k}}=I,\text{ \ }\frac{\partial x_{k}%
S}{\partial x_{k}}=S,
\]
where $S$ is a matrix of size $m_{A}m_{B}$.

Note that%
\[
L_{1}\left(  u_{0}\otimes\alpha,x_{1}\right)  =\frac{\left(  u_{0}e\right)
^{d}-\left(  x_{1}e\right)  ^{d}}{u_{0}e-x_{1}e}=\frac{1-\left(
x_{1}e\right)  ^{d}}{1-x_{1}e}%
\]
and for $k\geq2$%
\[
L_{k}\left(  x_{k-1},x_{k}\right)  =\frac{\left(  x_{k-1}e\right)
^{d}-\left(  x_{k}e\right)  ^{d}}{x_{k-1}e-x_{k}e}.
\]
Let $y_{1}=x_{1}e$. Then%
\begin{align*}
\frac{\partial L_{1}\left(  u_{0}\otimes\alpha,x_{1}\right)  }{\partial
x_{1}}  &  =\frac{\partial y_{1}}{\partial x_{1}}\frac{\partial L_{1}\left(
u_{0}\otimes\alpha,x_{1}\right)  }{\partial y_{1}}\\
&  =e\frac{\left[  \left(  u_{0}e\right)  ^{d}-\left(  x_{1}e\right)
^{d}\right]  -d\left(  x_{1}e\right)  ^{d-1}\left(  u_{0}e-x_{1}e\right)
}{\left(  u_{0}e-x_{1}e\right)  ^{2}}.
\end{align*}
Similarly, for $k\geq2$\ we can obtain%
\[
\frac{\partial L_{k}\left(  x_{k-1},x_{k}\right)  }{\partial x_{k-1}%
}=e\frac{d\left(  x_{k-1}e\right)  ^{d-1}\left(  x_{k-1}e-x_{k}e\right)
-\left[  \left(  x_{k-1}e\right)  ^{d}-\left(  x_{k}e\right)  ^{d}\right]
}{\left(  x_{k-1}e-x_{k}e\right)  ^{2}}%
\]
and%
\[
\frac{\partial L_{k}\left(  x_{k-1},x_{k}\right)  }{\partial x_{k}%
}=e\frac{\left[  \left(  x_{k-1}e\right)  ^{d}-\left(  x_{k}e\right)
^{d}\right]  -d\left(  x_{k}e\right)  ^{d-1}\left(  x_{k-1}e-x_{k}e\right)
}{\left(  x_{k-1}e-x_{k}e\right)  ^{2}}.
\]

It is easy to check that%
\begin{align}
A_{1}(x)=  &  \left[  C+\text{diag}\left(  De\right)  \right]  \oplus
T+\left[  \text{diag}\left(  De\right)  \otimes I\right]  \frac{\left(
u_{0}e\right)  ^{d}-\left(  x_{1}e\right)  ^{d}}{u_{0}e-x_{1}e}\nonumber\\
&  +ex_{1}\left[  \text{diag}\left(  De\right)  \otimes I\right]
\frac{\left[  \left(  u_{0}e\right)  ^{d}-\left(  x_{1}e\right)  ^{d}\right]
-d\left(  x_{1}e\right)  ^{d-1}\left(  u_{0}e-x_{1}e\right)  }{\left(
u_{0}e-x_{1}e\right)  ^{2}}, \label{Norm1}%
\end{align}%
\begin{align}
B_{1}(x)=  &  \left(  D\otimes I\right)  \frac{\left(  x_{1}e\right)
^{d}-\left(  x_{2}e\right)  ^{d}}{x_{1}e-x_{2}e}+e\left\{  x_{1}\left(
D\otimes I\right)  -x_{2}\left[  \text{diag}\left(  De\right)  \otimes
I\right]  \right\} \nonumber\\
&  \times\frac{d\left(  x_{1}e\right)  ^{d-1}\left(  x_{1}e-x_{2}e\right)
-\left[  \left(  x_{1}e\right)  ^{d}-\left(  x_{2}e\right)  ^{d}\right]
}{\left(  x_{1}e-x_{2}e\right)  ^{2}}; \label{Norm2}%
\end{align}
and for $k\geq2$%
\begin{equation}
C_{k}(x)=I\otimes T^{0}\alpha, \label{Norm3}%
\end{equation}%
\begin{align}
B_{k}(x)=  &  \left(  D\otimes I\right)  \frac{\left(  x_{k}e\right)
^{d}-\left(  x_{k+1}e\right)  ^{d}}{x_{k}e-x_{k+1}e}+e\left\{  x_{k}\left(
D\otimes I\right)  -x_{k+1}\left[  \text{diag}\left(  De\right)  \otimes
I\right]  \right\} \nonumber\\
&  \times\frac{d\left(  x_{k}e\right)  ^{d-1}\left(  x_{k}e-x_{k+1}e\right)
-\left[  \left(  x_{k}e\right)  ^{d}-\left(  x_{k+1}e\right)  ^{d}\right]
}{\left(  x_{k}e-x_{k+1}e\right)  ^{2}}, \label{Norm3-1}%
\end{align}%
\begin{align}
A_{k}(x)=  &  \left[  C+\text{diag}\left(  De\right)  \right]  \oplus
T+\left[  \text{diag}\left(  De\right)  \otimes I\right]  \frac{\left(
x_{k-1}e\right)  ^{d}-\left(  x_{k}e\right)  ^{d}}{x_{k-1}e-x_{k}e}\nonumber\\
&  +e\left\{  x_{k-1}\left(  D\otimes I\right)  -x_{k}\left[  \text{diag}%
\left(  De\right)  \otimes I\right]  \right\} \nonumber\\
&  \times\frac{\left[  \left(  x_{k-1}e\right)  ^{d}-\left(  x_{k}e\right)
^{d}\right]  -d\left(  x_{k}e\right)  ^{d-1}\left(  x_{k-1}e-x_{k}e\right)
}{\left(  x_{k-1}e-x_{k}e\right)  ^{2}}. \label{Norm4}%
\end{align}

Note that $\left\Vert \mathbf{A}\right\Vert =\max_{i}\left\{  \sum
\limits_{j}\left\vert a_{i,j}\right\vert \right\}  $, it follows from
(\ref{Eq7.4}) that%
\begin{equation}
||\mathcal{D}F\left(  x\right)  ||=\max\left\{  \left\Vert A_{1}\left(
x\right)  \right\Vert +\left\Vert B_{2}\left(  x\right)  \right\Vert
,\sup_{k\geq2}\left\{  ||A_{k}\left(  x\right)  ||+||B_{k}\left(  x\right)
||+||C_{k}\left(  x\right)  ||\right\}  \right\}  . \label{EquK-11}%
\end{equation}

Since $u_{0}e\leq1$ and $x_{1}e\leq1$, we obtain%
\[
\frac{\left(  u_{0}e\right)  ^{d}-\left(  x_{1}e\right)  ^{d}}{u_{0}e-x_{1}%
e}=\sum_{j=0}^{d-1}\left(  u_{0}e\right)  ^{j}\left(  x_{1}e\right)
^{d-1-j}\leq d,
\]%
\[
\frac{\left[  \left(  u_{0}e\right)  ^{d}-\left(  x_{1}e\right)  ^{d}\right]
-d\left(  x_{1}e\right)  ^{d-1}\left(  u_{0}e-x_{1}e\right)  }{\left(
u_{0}e-x_{1}e\right)  ^{2}}=\sum\limits_{k=0}^{d-2}\sum\limits_{j=0}%
^{k}\left(  u_{0}e\right)  ^{j}\left(  x_{1}e\right)  ^{k-j}\leq\frac{\left(
d-1\right)  \left(  d-2\right)  }{2};
\]%
\[
\frac{\left(  x_{k-1}e\right)  ^{d}-\left(  x_{k}e\right)  ^{d}}%
{x_{k-1}e-x_{k}e}\leq d,
\]%
\[
\frac{\left[  \left(  x_{k-1}e\right)  ^{d}-\left(  x_{k}e\right)
^{d}\right]  -d\left(  x_{k}e\right)  ^{d-1}\left(  x_{k-1}e-x_{k}e\right)
}{\left(  x_{k-1}e-x_{k}e\right)  ^{2}}\leq\frac{\left(  d-1\right)  \left(
d-2\right)  }{2}.
\]
Thus it follows from (\ref{Norm1}) and (\ref{Norm2}) that%
\[
\left\Vert A_{1}(x)\right\Vert \leq\left\Vert C+\text{diag}\left(  De\right)
\right\Vert +\frac{2d+\left(  d-1\right)  \left(  d-2\right)  }{2}\left\Vert
D\right\Vert +\left\Vert T\right\Vert ,
\]%
\[
\left\Vert B_{1}(x)\right\Vert \leq\left[  d+\left(  d-1\right)  \left(
d-2\right)  \right]  \left\Vert D\right\Vert ,
\]%
\[
\left\Vert A_{1}(x)\right\Vert +\left\Vert B_{1}(x)\right\Vert \leq\left\Vert
C+\text{diag}\left(  De\right)  \right\Vert +\left[  2d+\frac{3\left(
d-1\right)  \left(  d-2\right)  }{2}\right]  \left\Vert D\right\Vert
+\left\Vert T\right\Vert .
\]
It follows from (\ref{Norm3}) to (\ref{Norm4}) that for $k\geq2$%
\[
\left\Vert A_{k}(x)\right\Vert \leq\left\Vert C+\text{diag}\left(  De\right)
\right\Vert +\left[  d+\left(  d-1\right)  \left(  d-2\right)  \right]
\left\Vert D\right\Vert +\left\Vert T\right\Vert ,
\]%
\[
\left\Vert B_{k}(x)\right\Vert \leq\left[  d+\left(  d-1\right)  \left(
d-2\right)  \right]  \left\Vert D\right\Vert ,
\]%
\[
\left\Vert C_{k}(x)\right\Vert =\left\Vert T^{0}\alpha\right\Vert ,
\]
hence we have%
\begin{align*}
&  \left\Vert A_{k}(x)\right\Vert +\left\Vert B_{k}(x)\right\Vert +\left\Vert
C_{k}(x)\right\Vert \\
&  \leq\left\Vert C+\text{diag}\left(  De\right)  \right\Vert +2\left[
d+\left(  d-1\right)  \left(  d-2\right)  \right]  \left\Vert D\right\Vert
+\left\Vert T\right\Vert +\left\Vert T^{0}\alpha\right\Vert .
\end{align*}
Let%
\[
M=\max\left\{  \left\Vert C+\text{diag}\left(  De\right)  \right\Vert
+2\left[  d+\left(  d-1\right)  \left(  d-2\right)  \right]  \left\Vert
D\right\Vert +\left\Vert T\right\Vert +\left\Vert T^{0}\alpha\right\Vert
\right\}  .
\]
Then%
\[
\left\Vert A_{1}(x)\right\Vert +\left\Vert B_{1}(x)\right\Vert \leq M
\]
and for $k\geq2$%
\[
\left\Vert A_{k}(x)\right\Vert +\left\Vert B_{k}(x)\right\Vert +\left\Vert
C_{k}(x)\right\Vert \leq M.
\]
Hence, it follows from Equation (\ref{EquK-11}) that%
\[
||\mathcal{D}F\left(  x\right)  ||\leq M.
\]
Note that $x=\mathbf{u}$, this gives that for $\mathbf{u}\in\widetilde{\Omega
}$%
\begin{equation}
\left\Vert \mathcal{D}F\left(  \mathbf{u}\right)  \right\Vert \leq M.
\label{EquK-13-1}%
\end{equation}
For $\mathbf{u},\mathbf{v}\in\widetilde{\Omega}$,%
\begin{align}
||F\left(  \mathbf{u}\right)  -F\left(  \mathbf{v}\right)  ||\text{ }  &
\leq\sup_{0\leq t\leq1}||\mathcal{D}F(\mathbf{u}+t\left(  \mathbf{v}%
-\mathbf{u}\right)  )||\text{ }||\mathbf{u}-\mathbf{v}||\nonumber\\
&  \leq M||\mathbf{u}-\mathbf{v}||. \label{EquK-14}%
\end{align}
This indicates that the function $F\left(  \mathbf{u}\right)  $ is Lipschitz
for $\mathbf{u}\in\widetilde{\Omega}$.

Note that $x=\mathbf{u}$, it follows from Equations (\ref{Equ5.1}) and
(\ref{Equ5.4}) that for $\mathbf{u}\in\widetilde{\Omega}$%
\[
\mathbf{u}\left(  t\right)  =\mathbf{u}\left(  0\right)  +\int_{0}^{t}F\left(
\mathbf{u}\left(  \xi\right)  \right)  \text{d}\xi,
\]
this gives%
\begin{equation}
\mathbf{u}\left(  t\right)  =\widetilde{\mathbf{g}}+\int_{0}^{t}F\left(
\mathbf{u}\left(  \xi\right)  \right)  \text{d}\xi. \label{EquK-15}%
\end{equation}

Using the Picard approximation as well as the Lipschitz condition, it is easy
to prove that there exists the unique solution to the integral equation
(\ref{EquK-15}) according to the basic results of the Banach space. Therefore,
there exists the unique solution to the system of differential vector
equations (\ref{Equ5.1}) to (\ref{Equ5.4}) (that is, (\ref{EqS1}) to
(\ref{EqS6})).

\section{A Matrix-Analytic Solution}

In this section, we first discuss the stability of this supermarket model in
terms of a coupling method. Then we provide a generalized matrix-analytic
method for computing the fixed point whose doubly exponential solution and
phase-structured tail are obtained. Finally, we discuss some useful limits of
the fraction vector $\mathbf{u}^{\left(  N\right)  }\left(  t\right)  $ as
$N\rightarrow\infty$ and $t\rightarrow+\infty$.

\subsection{Stability of this supermarket model}

In this subsection, we provide a coupling method to study the stability of
this supermarket model of $N$ identical servers with MAP inputs and PH service
times, and give a sufficient condition under which this supermarket model is stable.

Let $Q$ and $R$ denote two supermarket models with MAP inputs and PH service
times, both of which have the same parameters $N,d,m_{A},C,D,m_{B},\alpha,T$,
and the same initial state at $t=0$. Let $d\left(  Q\right)  $ and $d\left(
R\right)  $ be two choice numbers in the two supermarket models $Q$ and $R$,
respectively. We assume $d\left(  Q\right)  =1$ and $d\left(  R\right)  \geq
2$. Thus, the only difference between the two supermarket models $Q$ and $R$
is the two different choice numbers: $d\left(  Q\right)  =1$ and $d\left(
R\right)  \geq2$.

For the two supermarket models $Q$ and $R$, we define two infinite-dimensional
Markov processes $\left\{  U_{N}^{\left(  Q\right)  }\left(  t\right)
:t\geq0\right\}  $ and $\left\{  U_{N}^{\left(  R\right)  }\left(  t\right)
:t\geq0\right\}  $, respectively. The following theorem sets up a coupling
between the two processes $\left\{  U_{N}^{\left(  Q\right)  }\left(
t\right)  :t\geq0\right\}  $ and $\left\{  U_{N}^{\left(  R\right)  }\left(
t\right)  :t\geq0\right\}  $.

\begin{The}
\label{The:Coupling}For the two supermarket models $Q$ and $R$, there is a
coupling between the two processes $\left\{  U_{N}^{\left(  Q\right)  }\left(
t\right)  :t\geq0\right\}  $ and $\left\{  U_{N}^{\left(  R\right)  }\left(
t\right)  :t\geq0\right\}  $ such that the total number of customers in the
supermarket model $R$ is no greater than the total number of customers in the
supermarket model $Q$ at time $t\geq0$.
\end{The}

\textbf{Proof:} See Appendix C. \textbf{{\rule{0.08in}{0.08in}}}

\begin{Rem}
\label{Rem:QLC}Note that the $N$ queueing processes in this supermarket model
is symmetric, it is easy to see from Theorem \ref{The:Coupling} that the queue
length of each server in the supermarket model $R$ is no greater than that in
the supermarket model $Q$ at time $t\geq0$.
\end{Rem}

Since this supermarket model with MAP inputs and PH service times is more
general, it is necessary to extend the coupling method given in Turner
\cite{Tur:1996} and Martin and Suhov \cite{Mar:1999} through a detailed
probability analysis given in Appendix C. We show that such a coupling method
can be applied to discussing stability of more general supermarket models.

Note that the stationary arrival rate of the MAP of irreducible matrix
descriptor $\left(  C,D\right)  $ is given by $\lambda=\omega De$, and the
mean of the PH service time is given by $1/\mu=-\alpha T^{-1}e$. The following
theorem provides a sufficient condition under which this supermarket model is stable.

\begin{The}
This supermarket model of $N$ identical servers with MAP inputs and PH service
times is stable if $\rho=\lambda/\mu<1$.
\end{The}

\textbf{Proof:} From the two different choice numbers: $d\left(  Q\right)  =1$
and $d\left(  R\right)  \geq2$, we set up two different supermarket models $Q$
and $R$, respectively. Note that the supermarket model $Q$ is the set of $N$
parallel and independent MAP/PH/1 queues. Obviously, the MAP/PH/1 queue is
described as a QBD process whose infinitesimal generator is given by%
\[
\mathbf{Q}=\left(
\begin{array}
[c]{ccccc}%
C & D\otimes\alpha &  &  & \\
I\otimes T^{0} & C\oplus T & D\otimes I &  & \\
& I\otimes\left(  T^{0}\alpha\right)  & C\oplus T & D\otimes I & \\
&  & \ddots & \ddots & \ddots
\end{array}
\right)  .
\]
Note that%
\[
A=A_{-1}+A_{0}+A_{1}=\left(  C+D\right)  \oplus\left(  T+T^{0}\alpha\right)
,
\]
where%
\[
A_{-1}=I\otimes\left(  T^{0}\alpha\right)  ,\text{ \ }A_{0}=C\oplus T,\text{
\ }A_{1}=D\otimes I,
\]
thus it is easy to check that $\omega\otimes\theta$ is the stationary
probability vector of the Markov chain $A$, where $\theta$ is the stationary
probability vector of the Markov chain $T+T^{0}\alpha$. Using Chapter 3 of Li
\cite{Li:2010}, it is clear that the QBD process $\mathbf{Q}$ is stable if
$\left(  \omega\otimes\theta\right)  A_{-1}e>\left(  \omega\otimes
\theta\right)  A_{2}e$, that is, $\rho=\lambda/\mu<1$. Hence, the supermarket
model $Q$ is stable if $\rho<1$. It is seen from Theorem \ref{The:Coupling}
and Remark \ref{Rem:QLC} that the queue length of each server in the
supermarket model $R$ is no greater than that in the supermarket model $Q$ at
time $t\geq0$, this shows that the supermarket model $R$ is stable if the
supermarket model $Q$ is stable. Thus the supermarket model $R$ is stable if
$\rho=\lambda/\mu<1$. This completes the proof. \textbf{{\rule{0.08in}{0.08in}%
}}

\subsection{Computation of the fixed point}

A row vector $\pi=\left(  \pi_{0},\pi_{1},\pi_{2},\ldots\right)  $ is called a
fixed point of the infinite-dimensional system of differential vector
equations (\ref{EqS1}) to (\ref{EqS6}) satisfied by the limiting fraction
vector $\mathbf{u}\left(  t\right)  $ if $\pi=\lim_{t\rightarrow+\infty
}\mathbf{u}\left(  t\right)  $, or $\pi_{k}=\lim_{t\rightarrow+\infty}%
u_{k}\left(  t\right)  $ for $k\geq0$.

It is well-known that if $\pi$ is a fixed point of the vector $\mathbf{u}%
\left(  t\right)  $, then%
\[
\lim_{t\rightarrow+\infty}\left[  \frac{\text{d}}{\text{d}t}\mathbf{u}\left(
t\right)  \right]  =0.
\]

Let
\[
L_{1}\left(  \pi_{0}\otimes\alpha,\pi_{1}\right)  =\sum_{m=1}^{d}C_{d}%
^{m}\left[  \sum_{l=1}^{m_{A}}\sum_{j=1}^{m_{B}}\left(  \pi_{0;l}\alpha
_{j}-\pi_{1;l,j}\right)  \right]  ^{m-1}\left[  \sum_{l=1}^{m_{A}}\sum
_{j=1}^{m_{B}}\pi_{1;l,j}\right]  ^{d-m}%
\]
for $k\geq2$%
\[
L_{k}\left(  \pi_{k-1},\pi_{k}\right)  =\sum_{m=1}^{d}C_{d}^{m}\left[
\sum_{l=1}^{m_{A}}\sum_{j=1}^{m_{B}}\left(  \pi_{k-1;l,j}-\pi_{k;l,j}\right)
\right]  ^{m-1}\left[  \sum_{l=1}^{m_{A}}\sum_{j=1}^{m_{B}}\pi_{k;l,j}\right]
^{d-m}.
\]
Then%
\[
L_{1}\left(  \pi_{0}\otimes\alpha,\pi_{1}\right)  =\frac{1-\left(  \pi
_{1}e\right)  ^{d}}{1-\pi_{1}e}%
\]
and for $k\geq2$%
\[
L_{k}\left(  \pi_{k-1},\pi_{k}\right)  =\frac{\left(  \pi_{k-1}e\right)
^{d}-\left(  \pi_{k}e\right)  ^{d}}{\pi_{k-1}e-\pi_{k}e}.
\]

To determine the fixed point $\pi=\left(  \pi_{0},\pi_{1},\pi_{2}%
,\ldots\right)  $, as $t\rightarrow+\infty$ taking limits on both sides of
Equations (\ref{EqS1}) to (\ref{EqS6}) we obtain the system of nonlinear
vector equations as follows:%
\begin{equation}
\pi_{0}\left(  C+D\right)  =0,\text{ }\pi_{0}e=1, \label{Equ14}%
\end{equation}%
\begin{align}
&  \left\{  \left(  \pi_{0}\otimes\alpha\right)  \left(  D\otimes I\right)
-\pi_{1}\left[  \text{diag}\left(  De\right)  \otimes I\right]  \right\}
L_{1}\left(  \pi_{0}\otimes\alpha,\pi_{1}\right) \nonumber\\
&  +\pi_{1}\left\{  \left[  C+\text{diag}\left(  De\right)  \right]  \oplus
T\right\}  +\pi_{2}\left(  I\otimes T^{0}\alpha\right)  =0, \label{Equ14-5}%
\end{align}
for $k\geq2$%
\begin{align}
&  \left\{  \pi_{k-1}\left(  D\otimes I\right)  -\pi_{k}\left[  \text{diag}%
\left(  De\right)  \otimes I\right]  \right\}  L_{k}\left(  \pi_{k-1},\pi
_{k}\right) \nonumber\\
&  +\pi_{k}\left\{  \left[  C+\text{diag}\left(  De\right)  \right]  \oplus
T\right\}  +\pi_{k+1}\left(  I\otimes T^{0}\alpha\right)  =0. \label{Equ15}%
\end{align}
Since $\omega$ is the stationary probability vector of the Markov chain $C+D$,
then it follows from (\ref{Equ14}) that%
\begin{equation}
\pi_{0}=\omega. \label{Equ13-0}%
\end{equation}

For the fixed point $\pi=\left(  \pi_{0},\pi_{1},\pi_{2},\ldots\right)  $,
$\left(  \pi_{0}e,\pi_{1}e,\pi_{2}e,\cdots\right)  $ is the tail vector of the
stationary queue length distribution. The following theorem shows that the
tail vector $\left(  \pi_{0}e,\pi_{1}e,\pi_{2}e,\cdots\right)  $ of the
stationary queue length distribution is doubly exponential.

\begin{The}
\label{The:TDE} If $\rho=\lambda/\mu<1$, then the tail vector $\left(  \pi
_{0}e,\pi_{1}e,\pi_{2}e,\cdots\right)  $ of the stationary queue length
distribution is doubly exponential, that is, for $k\geq0$%
\begin{equation}
\pi_{k}e=\rho^{\frac{d^{k}-1}{d-1}}. \label{Equ15-1}%
\end{equation}
\end{The}

\textbf{Proof: }Multiplying both sides of the equation (\ref{Equ15}) by the
vector $e$, and noting that $\left[  C+\text{diag}\left(  De\right)  \right]
e=0$ and $Te=-T^{0}$, we obtain that%
\begin{equation}
\left[  \left(  \pi_{0}\otimes\alpha\right)  \left(  De\otimes e\right)
-\pi_{1}\left(  De\otimes e\right)  \right]  L_{k}\left(  \pi_{0}\otimes
,\pi_{1}\right)  -\mu\left[  \pi_{1}\left(  e\otimes T^{0}\right)  -\pi
_{2}\left(  e\otimes T^{0}\right)  \right]  =0 \label{Equ15-2}%
\end{equation}
for $k\geq2$,%
\begin{equation}
\left[  \pi_{k-1}\left(  De\otimes e\right)  -\pi_{k}\left(  De\otimes
e\right)  \right]  L_{k}\left(  \pi_{k-1},\pi_{k}\right)  -\mu\left[  \pi
_{k}\left(  e\otimes T^{0}\right)  -\pi_{k+1}\left(  e\otimes T^{0}\right)
\right]  =0. \label{Equ15-3}%
\end{equation}
Let $\pi_{k}=\eta_{k}\left(  \omega\otimes\theta\right)  $ for $k\geq1$, and
$\zeta_{1}=L_{1}\left(  \pi_{0}\otimes\alpha,\pi_{1}\right)  $ and $\zeta
_{k}=L_{k}\left(  \pi_{k-1},\pi_{k}\right)  $ for $k\geq2$. Note that
$\lambda=\omega De$, $\mu=\theta T^{0}$ and $\rho=\lambda/\mu$, it follows
from (\ref{Equ15-3}) that%
\[
\rho\left(  1-\eta_{1}^{d}\right)  -\left(  \eta_{1}-\eta_{2}\right)  =0
\]
and%
\[
\rho\left(  \eta_{k-1}^{d}-\eta_{k}^{d}\right)  -\left(  \eta_{k}-\eta
_{k+1}\right)  =0.
\]
This gives%
\[
\pi_{k}e=\eta_{k}=\rho^{\frac{d^{k}-1}{d-1}}.
\]
This completes the proof. \rule{1.8mm}{2.5mm}

Note that%
\[
\zeta_{k}=\frac{\rho^{\frac{d^{k}-d}{d-1}}-\rho^{\frac{d^{k+1}-d}{d-1}}}%
{\rho^{\frac{d^{k-1}-1}{d-1}}-\rho^{\frac{d^{k}-1}{d-1}}},\text{ \ }k\geq1,
\]
we obtain%
\[
B_{k}=\left[  C+\left(  1-\zeta_{k}\right)  \text{ diag}\left(  De\right)
\right]  \oplus T
\]
and%
\[
Q=\left(
\begin{array}
[c]{ccccc}%
B_{1} & \zeta_{2}\left(  D\otimes I\right)  &  &  & \\
I\otimes\left(  T^{0}\alpha\right)  & B_{2} & \zeta_{3}\left(  D\otimes
I\right)  &  & \\
& I\otimes\left(  T^{0}\alpha\right)  & B_{3} & \zeta_{4}\left(  D\otimes
I\right)  & \\
&  & \ddots & \ddots & \ddots
\end{array}
\right)  .
\]
Then the level-dependent QBD process is irreducible and transient, since%
\[
\zeta_{1}>\zeta_{2}>\zeta_{3}>\cdots>0,
\]%
\[
\left[  B_{1}+\zeta_{2}\left(  D\otimes I\right)  \right]  e=-\left(
\zeta_{1}-\zeta_{2}\right)  \left[  \left(  De\right)  \otimes e\right]
-e\otimes T^{0}\lvertneqq0
\]
and%
\[
\left[  I\otimes\left(  T^{0}\alpha\right)  +B_{k}+\zeta_{k}\left(  D\otimes
I\right)  \right]  e=-\left(  \zeta_{k}-\zeta_{k+1}\right)  \left[  \left(
De\right)  \otimes e\right]  \lvertneqq0.
\]

In what follows we provide the UL-type of $RG$-factorization of the QBD
process $Q$ according to Chapter 1 in Li \cite{Li:2010} or Li and Cao
\cite{Li:2004}. Applying the UL-type of $RG$-Factorization, we can give the
maximal non-positive inverse of matrix $Q$, which leads to the matrix-product
solution of the fixed point $\left(  \pi_{0},\pi_{1},\pi_{2},\cdots\right)  $
by means of the $R$- and $U$-measures.

Let the matrix sequence $\left\{  R_{k},k\geq1\right\}  $ be the minimal
nonnegative solution to the nonlinear matrix equations%
\[
\xi_{k+1}\left(  D\otimes I\right)  +R_{k}B_{k+1}+R_{k}R_{k+1}\left[
I\otimes\left(  T^{0}\alpha\right)  \right]  =0,
\]
and the matrix sequence $\left\{  G_{k},k\geq2\right\}  $ be the minimal
nonnegative solution to the nonlinear matrix equations%
\[
I\otimes\left(  T^{0}\alpha\right)  +B_{k}G_{k}+\xi_{k+1}\left(  D\otimes
I\right)  G_{k+1}G_{k}=0.
\]
Let the matrix sequence $\left\{  U_{k},k\geq0\right\}  $ be%
\begin{align*}
U_{k}  &  =B_{k+1}+\left[  \zeta_{k+2}\left(  D\otimes I\right)  \right]
\left[  -U_{k+1}\right]  ^{-1}\left[  I\otimes\left(  T^{0}\alpha\right)
\right] \\
&  =B_{k+1}+R_{k+1}\left[  I\otimes\left(  T^{0}\alpha\right)  \right] \\
&  =B_{k+1}+\left[  \zeta_{k+2}\left(  D\otimes I\right)  \right]  G_{k+1.}%
\end{align*}
Hence we obtain%
\[
R_{0}=\zeta_{1}\left(  D\otimes I\right)  \left(  -U_{1}\right)  ^{-1}%
\]
and%
\[
G_{1}=\left(  -U_{0}\right)  ^{-1}\left[  I\otimes\left(  T^{0}\alpha\right)
\right]  .
\]

Based on the $R$-measure $\left\{  R_{k},k\geq0\right\}  $, $G$-measure
$\left\{  G_{k},k\geq1\right\}  $ and $U$-measure $\left\{  U_{k}%
,k\geq0\right\}  $, we can get the UL-type of $RG$-factorization of the matrix
$Q$ as follows%
\[
Q=\left(  I-R_{U}\right)  U_{D}\left(  I-G_{L}\right)  ,
\]
where%
\[
R_{U}=\left(
\begin{array}
[c]{ccccc}%
0 & R_{0} &  &  & \\
& 0 & R_{1} &  & \\
&  & 0 & R_{2} & \\
&  &  & \ddots & \ddots
\end{array}
\right)  ,
\]%
\[
U_{D}=\text{diag}\left(  U_{0},U_{1},U_{2},\ldots\right)
\]
and%
\[
G_{L}=\left(
\begin{array}
[c]{ccccc}%
I &  &  &  & \\
G_{1} & I &  &  & \\
& G_{2} & I &  & \\
&  & \ddots & \ddots & \ddots
\end{array}
\right)  .
\]

Using the $RG$-factorization, we obtain the maximal non-positive inverse of
the matrix $Q$ as follows%
\begin{equation}
Q^{-1}=\left(  I-G_{L}\right)  ^{-1}U_{D}^{-1}\left(  I-R_{U}\right)  ^{-1},
\label{Equ15-4}%
\end{equation}
where%
\[
\left(  I-R_{U}\right)  ^{-1}=\left(
\begin{array}
[c]{ccccc}%
I & X_{1}^{\left(  0\right)  } & X_{2}^{\left(  0\right)  } & X_{3}^{\left(
0\right)  } & \cdots\\
& I & X_{1}^{\left(  1\right)  } & X_{2}^{\left(  1\right)  } & \cdots\\
&  & I & X_{1}^{\left(  2\right)  } & \cdots\\
&  &  & I & \cdots\\
&  &  &  & \ddots
\end{array}
\right)  ,
\]%
\[
X_{k}^{\left(  l\right)  }=R_{l}R_{l+1}R_{l+2}\cdots R_{l+k-1},\hbox{ \ }
k\geq1,l\geq0;
\]%
\[
U_{D}^{-1}=\text{diag}\left(  U_{0}^{-1},U_{1}^{-1},U_{2}^{-1},\ldots\right)
;
\]%
\[
\left(  I-G_{L}\right)  ^{-1}=\left(
\begin{array}
[c]{ccccc}%
I &  &  &  & \\
Y_{1}^{\left(  1\right)  } & I &  &  & \\
Y_{2}^{\left(  2\right)  } & Y_{1}^{\left(  2\right)  } & I &  & \\
Y_{3}^{\left(  3\right)  } & Y_{2}^{\left(  3\right)  } & Y_{1}^{\left(
3\right)  } & I & \\
\vdots & \vdots & \vdots & \vdots & \ddots
\end{array}
\right)  ,
\]%
\[
Y_{k}^{\left(  l\right)  }=G_{l}G_{l-1}G_{l-2}\cdots G_{l-k+1},\hbox{ \ }
l\geq k\geq1.
\]

The following theorem illustrates that the fixed point $\left(  \pi_{0}%
,\pi_{1},\pi_{2},\cdots\right)  $ is matrix-product.

\begin{The}
\label{The:Fix} If $\rho<1$, then the fixed point $\pi=\left(  \pi_{0},\pi
_{1},\pi_{2},\ldots\right)  $ is given by%
\[
\pi_{0}=\omega,
\]%
\begin{equation}
\pi_{1}=\zeta_{1}\left(  \omega\otimes\alpha\right)  \left(  D\otimes
I\right)  \left(  -U_{0}\right)  ^{-1} \label{FixE-1}%
\end{equation}
and for $k\geq2$%
\begin{equation}
\pi_{k}=\zeta_{1}\left(  \omega\otimes\alpha\right)  \left(  D\otimes
I\right)  \left(  -U_{0}\right)  ^{-1}R_{0}R_{1}\cdots R_{k-2}. \label{FixE-2}%
\end{equation}
\end{The}

\textbf{Proof: }It follows from (\ref{Equ15-3}) that%
\begin{align*}
&  \left(  \pi_{1},\pi_{2},\pi_{3},\ldots\right)  \left(
\begin{array}
[c]{ccccc}%
B_{1} & \zeta_{2}\left(  D\otimes I\right)  &  &  & \\
I\otimes\left(  T^{0}\alpha\right)  & B_{2} & \zeta_{3}\left(  D\otimes
I\right)  &  & \\
& I\otimes\left(  T^{0}\alpha\right)  & B_{3} & \zeta_{4}\left(  D\otimes
I\right)  & \\
&  & \ddots & \ddots & \ddots
\end{array}
\right) \\
&  =-\left(  \zeta_{1}\left(  \omega\otimes\alpha\right)  \left(  D\otimes
I\right)  ,0,0,\ldots\right)  .
\end{align*}
This gives%
\[
\left(  \pi_{1},\pi_{2},\pi_{3},\ldots\right)  =-\left(  \zeta_{1}\left(
\omega\otimes\alpha\right)  \left(  D\otimes I\right)  ,0,0,\ldots\right)
\left(  I-G_{L}\right)  ^{-1}U_{D}^{-1}\left(  I-R_{U}\right)  ^{-1}.
\]
Thus we obtain%
\[
\pi_{1}=\zeta_{1}\left(  \omega\otimes\alpha\right)  \left(  D\otimes
I\right)  \left(  -U_{0}\right)  ^{-1}%
\]
and for $k\geq2$%
\[
\pi_{k}=\zeta_{1}\left(  \omega\otimes\alpha\right)  \left(  D\otimes
I\right)  \left(  -U_{0}\right)  ^{-1}R_{0}R_{1}\cdots R_{k-2}.
\]
This completes the proof. \rule{1.8mm}{2.5mm}

In what follows we consider the block-structured supermarket model with
Poisson inputs and PH service times. In this case, we can give an interesting
explicit expression of the fixed point.

Note that $C=-\lambda$, $D=\lambda$, it is clear that $\omega=1$ and $\pi
_{0}=1$. It follows from Equations (\ref{Equ14-5}) and (\ref{Equ15}) that%
\[
\lambda\left(  \theta-\pi_{1}\right)  \frac{1-\left(  \pi_{1}e\right)  ^{d}%
}{1-\left(  \pi_{1}e\right)  }+\pi_{1}T+\pi_{2}T^{0}\alpha=0
\]
and for $k\geq2$%
\[
\lambda\left(  \pi_{k-1}-\pi_{k}\right)  \frac{\left(  \pi_{k-1}e\right)
^{d}-\left(  \pi_{k}e\right)  ^{d}}{\left(  \pi_{k-1}e\right)  -\left(
\pi_{k}e\right)  }+\pi_{k}T+\pi_{k+1}T^{0}\alpha=0.
\]
Thus we obtain%
\begin{equation}
\left(  \pi_{1},\pi_{2},\pi_{3},\ldots\right)  \Theta=\lambda\left(  \left(
\theta-\pi_{1}\right)  \frac{1-\left(  \pi_{1}e\right)  ^{d}}{1-\left(
\pi_{1}e\right)  },\left(  \pi_{1}-\pi_{2}\right)  \frac{\left(  \pi
_{1}e\right)  ^{d}-\left(  \pi_{2}e\right)  ^{d}}{\left(  \pi_{1}e\right)
-\left(  \pi_{2}e\right)  },\ldots\right)  , \label{Sp5}%
\end{equation}
where%
\[
\Theta=\left(
\begin{array}
[c]{cccc}%
-T &  &  & \\
-T^{0}\alpha & -T &  & \\
& -T^{0}\alpha & -T & \\
&  & \ddots & \ddots
\end{array}
\right)  .
\]
Since%
\[
\Theta^{-1}=\left(
\begin{array}
[c]{ccccc}%
\left(  -T\right)  ^{-1} &  &  &  & \\
\left(  e\alpha\right)  \left(  -T\right)  ^{-1} & \left(  -T\right)  ^{-1} &
&  & \\
\left(  e\alpha\right)  \left(  -T\right)  ^{-1} & \left(  e\alpha\right)
\left(  -T\right)  ^{-1} & \left(  -T\right)  ^{-1} &  & \\
\left(  e\alpha\right)  \left(  -T\right)  ^{-1} & \left(  e\alpha\right)
\left(  -T\right)  ^{-1} & \left(  e\alpha\right)  \left(  -T\right)  ^{-1} &
\left(  -T\right)  ^{-1} & \\
\vdots & \vdots & \vdots & \vdots & \ddots
\end{array}
\right)  .
\]
It follows from (\ref{Sp5}) that%
\begin{equation}
\pi_{1}\left[  I+\lambda\frac{1-\left(  \pi_{1}e\right)  ^{d}}{1-\left(
\pi_{1}e\right)  }\left(  -T\right)  ^{-1}\right]  =\lambda\frac{1-\left(
\pi_{1}e\right)  ^{d}}{1-\left(  \pi_{1}e\right)  }\theta\left(  -T\right)
^{-1}+\lambda\alpha\left(  -T\right)  ^{-1}\left(  \pi_{1}e\right)  ^{d}
\label{Sp8}%
\end{equation}
and for $k\geq2$%
\begin{equation}
\pi_{k}\left[  I+\lambda\frac{\left(  \pi_{k-1}e\right)  ^{d}-\left(  \pi
_{k}e\right)  ^{d}}{\left(  \pi_{k-1}e\right)  -\left(  \pi_{k}e\right)
}\left(  -T\right)  ^{-1}\right]  =\lambda\frac{\left(  \pi_{k-1}e\right)
^{d}-\left(  \pi_{k}e\right)  ^{d}}{\left(  \pi_{k-1}e\right)  -\left(
\pi_{k}e\right)  }\theta\left(  -T\right)  ^{-1}+\lambda\alpha\left(
-T\right)  ^{-1}\left(  \pi_{k}e\right)  ^{d}. \label{Sp9}%
\end{equation}
Note that the matrices $I+\lambda\frac{1-\left(  \pi_{1}e\right)  ^{d}%
}{1-\left(  \pi_{1}e\right)  }\left(  -T\right)  ^{-1}$ and $I+\lambda
\frac{\left(  \pi_{k-1}e\right)  ^{d}-\left(  \pi_{k}e\right)  ^{d}}{\left(
\pi_{k-1}e\right)  -\left(  \pi_{k}e\right)  }\left(  -T\right)  ^{-1}$ for
$k\geq2$ are all invertible, it follows from (\ref{Sp8}) and (\ref{Sp9}) that%
\[
\pi_{1}=\left[  \lambda\frac{1-\left(  \pi_{1}e\right)  ^{d}}{1-\left(
\pi_{1}e\right)  }\omega\left(  -T\right)  ^{-1}+\lambda\alpha\left(
-T\right)  ^{-1}\left(  \pi_{1}e\right)  ^{d}\right]  \left[  I+\lambda
\frac{1-\left(  \pi_{1}e\right)  ^{d}}{1-\left(  \pi_{1}e\right)  }\left(
-T\right)  ^{-1}\right]  ^{-1}.
\]
and for $k\geq2$%
\begin{align*}
\pi_{k}=  &  \left[  \lambda\frac{\left(  \pi_{k-1}e\right)  ^{d}-\left(
\pi_{k}e\right)  ^{d}}{\left(  \pi_{k-1}e\right)  -\left(  \pi_{k}e\right)
}\omega\left(  -T\right)  ^{-1}+\lambda\alpha\left(  -T\right)  ^{-1}\left(
\pi_{k}e\right)  ^{d}\right] \\
&  \times\left[  I+\lambda\frac{\left(  \pi_{k-1}e\right)  ^{d}-\left(
\pi_{k}e\right)  ^{d}}{\left(  \pi_{k-1}e\right)  -\left(  \pi_{k}e\right)
}\left(  -T\right)  ^{-1}\right]  ^{-1}.
\end{align*}
Thus we obtain%
\begin{equation}
\pi_{1}=\left[  \lambda\zeta_{1}\omega\left(  -T\right)  ^{-1}+\lambda
\alpha\left(  -T\right)  ^{-1}\rho^{d}\right]  \left[  I+\lambda\zeta
_{1}\left(  -T\right)  ^{-1}\right]  ^{-1} \label{Sp12}%
\end{equation}
and for $k\geq2$%
\begin{equation}
\pi_{k}=\left[  \lambda\zeta_{k}\omega\left(  -T\right)  ^{-1}+\lambda
\alpha\left(  -T\right)  ^{-1}\rho^{\frac{d^{k+1}-d}{d-1}}\right]  \left[
I+\lambda\zeta_{k}\left(  -T\right)  ^{-1}\right]  ^{-1}. \label{Sp13}%
\end{equation}

\begin{Rem}
For this block-structured supermarket model, the fixed point is matrix-product
and depends on the R-measure $\left\{  R_{k},k\geq0\right\}  $, see
(\ref{FixE-1}) and (\ref{FixE-2}). However, when the input is a Poisson
process, we can give the explicit expression of the fixed point by
(\ref{Sp12}) and (\ref{Sp13}). This explains the reason why the MAP input
makes the study of block-structured supermarket models more difficult and challenging.
\end{Rem}

\subsection{The double limits}

In this subsection, we discuss some useful limits of the fraction vector
$\mathbf{u}^{\left(  N\right)  }\left(  t\right)  $ as $N\rightarrow\infty$
and $t\rightarrow+\infty$. Note that the limits are necessary for using the
stationary probabilities of the limiting process to give an effective
approximate performance of this supermarket model.

The following theorem gives the limit of the vector $\mathbf{u}(t,\mathbf{g})$
as $t\rightarrow+\infty$, that is,%
\[
\lim_{t\rightarrow+\infty}\mathbf{u}(t,\mathbf{g})=\lim_{t\rightarrow+\infty
}\lim_{N\rightarrow\infty}\mathbf{u}^{\left(  N\right)  }(t,\mathbf{g}).
\]

\begin{The}
\label{The:Limit1}If $\rho<1$, then for any $\mathbf{g}\in\Omega$%
\[
\lim_{t\rightarrow+\infty}\mathbf{u}(t,\mathbf{g})=\pi.
\]
Furthermore, there exists a unique probability measure $\varphi$ on $\Omega$,
which is invariant under the map $\mathbf{g}\longmapsto\mathbf{u}%
(t,\mathbf{g})$, that is, for any continuous function $f:$ $\Omega
\rightarrow\mathbf{R}$ and $t>0$%
\[
\int_{\Omega}f(\mathbf{g})\text{d}\varphi(\mathbf{g})=\int_{\Omega
}f(\mathbf{u}(t,\mathbf{g}))\text{d}\varphi(\mathbf{g}).
\]
Also, $\varphi=\delta_{\pi}$ is the probability measure concentrated at the
fixed point $\pi$.
\end{The}

\textbf{Proof:} It is seen from Theorem \ref{The:Fix} that the condition
$\rho<1$ guarantees the existence of solution in $\Omega$ to the system of
nonlinear equations (\ref{Equ14}) to (\ref{Equ15}). This indicates that if
$\rho<1$, then as $t\rightarrow+\infty$,\ the limit of $\mathbf{u}%
(t,\mathbf{g})$ exists in $\Omega$. Since $\mathbf{u}(t,\mathbf{g})$ is the
unique and global solution to the infinite-dimensional system of differential
vector equations (\ref{EqS1}) to (\ref{EqS6}) for $t\geq0$, the vector
$\lim_{t\rightarrow+\infty}\mathbf{u}(t,\mathbf{g})$ is also a solution to the
system of nonlinear equations (\ref{Equ14}) to (\ref{Equ15}). Note that $\pi$
is the unique solution to the system of nonlinear equations (\ref{Equ14}) to
(\ref{Equ15}), hence we obtain that $\lim_{t\rightarrow+\infty}\mathbf{u}%
(t,\mathbf{g})=\pi$. The second statement in this theorem can be immediately
given by the probability measure of the limiting process $\left\{
U(t),t\geq0\right\}  $ on state space $\Omega$. This completes the proof.
\textbf{{\rule{0.08in}{0.08in}}}

The following theorem indicates the weak convergence of the sequence $\left\{
\varphi_{N}\right\}  $ of stationary probability distributions for the
sequence $\left\{  U^{(N)}(t),t\geq0\right\}  $ of Markov processes to the
probability measure concentrated at the fixed point $\pi$.

\begin{The}
\label{The:Limit2}(1) If $\rho<1$, then for a fixed number $N=1,2,3,\ldots$,
the Markov process $\left\{  U^{(N)}(t),t\geq0\right\}  $ is positive
recurrent, and has a unique invariant distribution $\varphi_{N}$.

(2) $\left\{  \varphi_{N}\right\}  $ weakly converges to $\delta_{\pi}$, that
is, for any continuous function $f:$ $\Omega\rightarrow\mathbf{R}$%
\[
\lim_{N\rightarrow\infty}E_{\varphi_{N}}\left[  f(\mathbf{g})\right]
=f\left(  \pi\right)  .
\]
\end{The}

\textbf{Proof:} (1) From Theorem 3, this supermarket model of $N$ identical
servers is stable if $\rho<1$, hence this supermarket model has a unique
invariant distribution $\varphi_{N}$.

(2) Since $\widetilde{\Omega}$ is compact under the metric $\rho\left(
\mathbf{u},\mathbf{u}^{\prime}\right)  $ given in (\ref{Eq10.0}), so is the
set $\mathcal{P}\left(  \widetilde{\Omega}\right)  $ of probability measures.
Hence the sequence $\left\{  \varphi_{N}\right\}  $ of invariant distributions
has limiting points. A similar analysis to the proof of Theorem 5 in Martin
and Suhov \cite{Mar:1999} shows that $\left\{  \varphi_{N}\right\}  $ weakly
converges to $\delta_{\pi}$ and $\lim_{N\rightarrow\infty}E_{\varphi_{N}%
}\left[  f(\mathbf{g})\right]  =f\left(  \pi\right)  $. This completes the
proof. \textbf{{\rule{0.08in}{0.08in}}}

Based on Theorems \ref{The:Limit1} and \ref{The:Limit2}, we obtain a useful
relation as follows%
\[
\lim_{t\rightarrow+\infty}\lim_{N\rightarrow\infty}\mathbf{u}^{\left(
N\right)  }(t,\mathbf{g})=\lim_{N\rightarrow\infty}\lim_{t\rightarrow+\infty
}\mathbf{u}^{\left(  N\right)  }(t,\mathbf{g})=\pi.
\]
Therefore, we have%
\[
\lim_{\substack{N\rightarrow\infty\\t\rightarrow+\infty}}\mathbf{u}^{\left(
N\right)  }(t,\mathbf{g})=\pi,
\]
which justifies the interchange of the limits of $N\rightarrow\infty$ and
$t\rightarrow+\infty$. This is necessary in many practical applications when
using the stationary probabilities of the limiting process to give an
effective approximate performance of this supermarket model.

\section{Performance Computation}

In this section, we provide two performance measures of this supermarket
model, and use some numerical examples to show how the two performance
measures of this supermarket model depend on the non-Poisson MAP inputs and on
the non-exponential PH service times.

\subsection{Performance measures}

For this supermarket model, we provide two simple performance measures as follows:

\textbf{(1) The mean of the stationary queue length in any server}

The mean of the stationary queue length in any server is given by%
\begin{equation}
E\left[  Q_{d}\right]  =\sum_{k=1}^{\infty}\pi_{k}e=\sum_{k=1}^{\infty}%
\rho^{\frac{d^{k}-1}{d-1}}. \label{Perf1}%
\end{equation}

\textbf{(2) The expected sojourn time that any arriving customer spends in
this system}

Note that $u_{0}^{\left(  N\right)  }\left(  0\right)  \geq0$ and
$u_{0}^{\left(  N\right)  }\left(  0\right)  e=1$, it is clear that%
\[
\lim_{t\rightarrow+\infty}u_{0}^{\left(  N\right)  }\left(  t\right)
=\lim_{t\rightarrow+\infty}u_{0}^{\left(  N\right)  }\left(  0\right)
\exp\left\{  \left(  C+D\right)  t\right\}  =\omega.
\]

For the PH service times, any arriving customer finds $k$ customer in any
server whose probability is given by $\left(  \omega\otimes\alpha-\pi
_{1}\right)  L_{d}\left(  \omega\otimes\alpha,\pi_{1}\right)  e$ for $k=0$ and
$\left(  \pi_{k}-\pi_{k+1}\right)  L_{d}\left(  \pi_{k},\pi_{k+}\right)  e$
for $k\geq1$. When $k\geq1$, the head customer in the server has been served,
and so its service time is residual and is denoted as $X_{R}$. Let $X$ be of
phase type with irreducible representation $\left(  \alpha,T\right)  $. Then
$X_{R}$ is also of phase type with irreducible representation $\left(
\theta,T\right)  $, where $\theta$ is the stationary probability vector of the
Markov chain $T+T^{0}\alpha$. Clearly, we have%
\[
E\left[  X\right]  =\alpha\left(  -T\right)  ^{-1}e,\text{ \ }E\left[
X_{R}\right]  =\theta\left(  -T\right)  ^{-1}e.
\]
Thus it is easy to see that the expected sojourn time that any arriving
customer spends in this system is given by%
\begin{align}
E\left[  T_{d}\right]  =  &  \left(  \omega\otimes\alpha-\pi_{1}\right)
L_{d}\left(  \omega\otimes\alpha,\pi_{1}\right)  eE\left[  X\right]
\nonumber\\
&  +\sum_{k=1}^{\infty}\left(  \pi_{k}-\pi_{k+1}\right)  L_{d}\left(  \pi
_{k},\pi_{k+}\right)  e\left\{  E\left[  X_{R}\right]  +kE\left[  X\right]
\right\} \nonumber\\
=  &  \left(  1-\rho\right)  E\left[  X\right]  +\sum_{k=1}^{\infty}\left(
\rho^{\frac{d^{k}-1}{d-1}}-\rho^{\frac{d^{k+1}-1}{d-1}}\right)  \left\{
E\left[  X_{R}\right]  +kE\left[  X\right]  \right\}  .\nonumber\\
=  &  E\left[  X\right]  +\rho E\left[  X_{R}\right]  +E\left[  X\right]
\sum_{k=2}^{\infty}\rho^{\frac{d^{k}-1}{d-1}}. \label{Perf2}%
\end{align}

From (\ref{Perf1}) and (\ref{Perf2}), we obtain%
\begin{equation}
E\left[  T_{d}\right]  =E\left[  X\right]  E\left[  Q_{d}\right]
+\rho\left\{  E\left[  X_{R}\right]  -E\left[  X\right]  \right\}  .
\label{Rela-1}%
\end{equation}
Specifically, if $E\left[  X_{R}\right]  =E\left[  X\right]  $ (for example,
the exponential service times), then%
\begin{equation}
E\left[  T_{d}\right]  =E\left[  X\right]  E\left[  Q_{d}\right]  ,
\label{Rela-2}%
\end{equation}
which is the Little's formula in this supermarket model.

It is seen from (\ref{Perf1}) that $E\left[  Q_{d}\right]  $ only depends on
the traffic intensity $\rho=\lambda/\mu$, where $\lambda=\omega De$ and
$\mu=-\alpha T^{-1}e$; and from (\ref{Perf2}) that $E\left[  T_{d}\right]  $
depends not only on the traffic intensity $\rho$ but also on the mean
$E\left[  X_{R}\right]  $ of the residual PH service time, where $E\left[
X_{R}\right]  =\theta\left(  -T\right)  ^{-1}e$. Based on this, it is clear
that performance numerical computation of this supermarket model can be given
easily for more general MAP inputs and PH service times, although here our
numerical examples are simple.

\subsection{Numerical examples}

In this subsection, we provide some numerical examples which are used to
indicate how the performance measures of this supermarket model depend on the
non-Poisson MAP inputs and on the non-exponential PH service times.

\textbf{Example one: The Erlang service times}

In this supermarket model, the customers arrive at this system as a Poisson
process with arrival rate $N\lambda$, and the service times at each server are
an Erlang distribution E$\left[  m,\eta\right]  $. Let $\lambda=1$. Then
$\rho=m/\eta$. When $\rho<1$, we have $\eta>m$. Figure 3 shows how $E\left[
Q_{d}\right]  $ depends on the different parameter pairs $\left(  m,d\right)
=\left(  2,2\right)  ,\left(  3,2\right)  ,\left(  4,2\right)  $ and $\left(
2,10\right)  $, respectively. It is seen that $E\left[  Q_{d}\right]  $
decreases as $d$ increases or as $\eta$ increases, and it increases as $m$ increases.

\begin{figure}[ptb]
\centering                  \includegraphics[width=10cm]{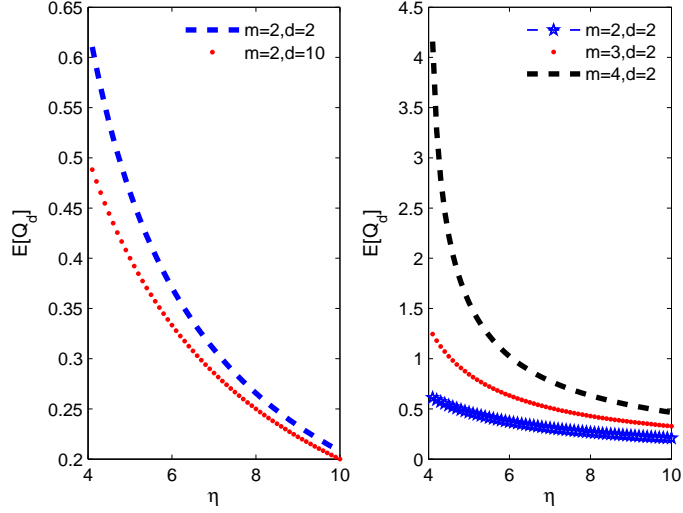}
\caption{$E[Q_{d}]$ vs $\eta$ for $(m,d)=(2,2),(3,2),(4,2)$ and $(2,10)$ }%
\label{figure:fig5-6}%
\end{figure}

\textbf{Example two: Performance comparisons between the exponential and PH
service times}

We consider two related supermarket models with Poisson inputs of arrival rate
$N\lambda$: one with exponential service times, and another with PH service
times. For the two supermarket models, our goal is to observe the influence of
different service time distributions on the performance of this supermarket
model. To that end, the parameters of this system are taken as%
\[
\mu=3.4118,\text{ }\alpha=\left(  \frac{1}{2},\frac{1}{2}\right)  ,T=\left(
\begin{array}
[c]{cc}%
-5 & 3\\
2 & -7
\end{array}
\right)  .
\]

Under the exponential and PH service times, Figure 4 depicts how $E\left[
Q_{d}\right]  $ and $E\left[  T_{d}\right]  $ depend on the arrival rate
$\lambda\in\left[  1,3\right]  $ with $\lambda<\mu$, and on the choice number
$d=1,2$. It is seen that $E\left[  Q_{d}\right]  $ and $E\left[  T_{d}\right]
$ decrease as $d$ increases, while $E\left[  Q_{d}\right]  $ and $E\left[
T_{d}\right]  $ increase as $\lambda$ increases.

\begin{figure}[ptb]
\centering                  \includegraphics[width=7cm]{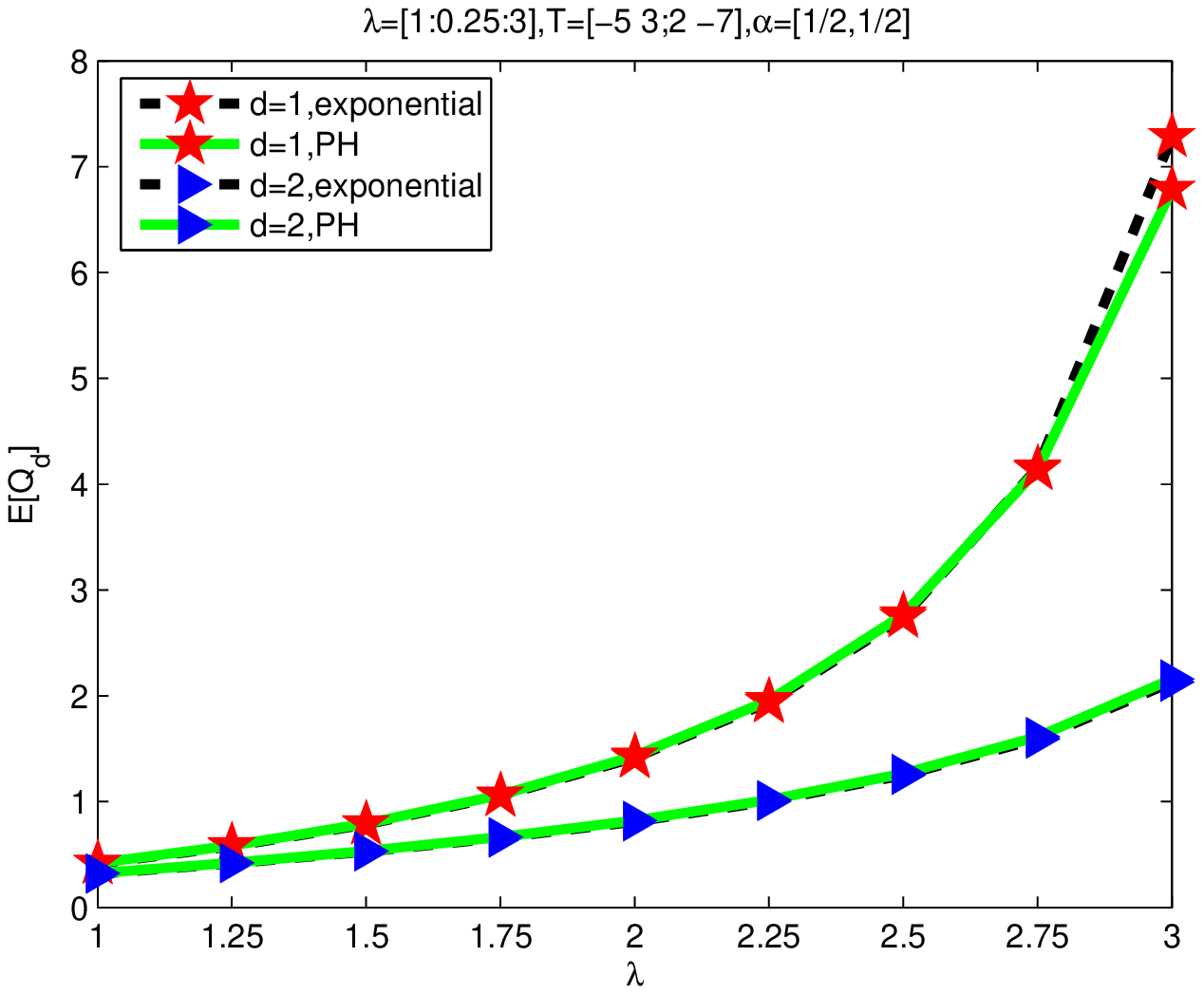}
\includegraphics[width=7cm]{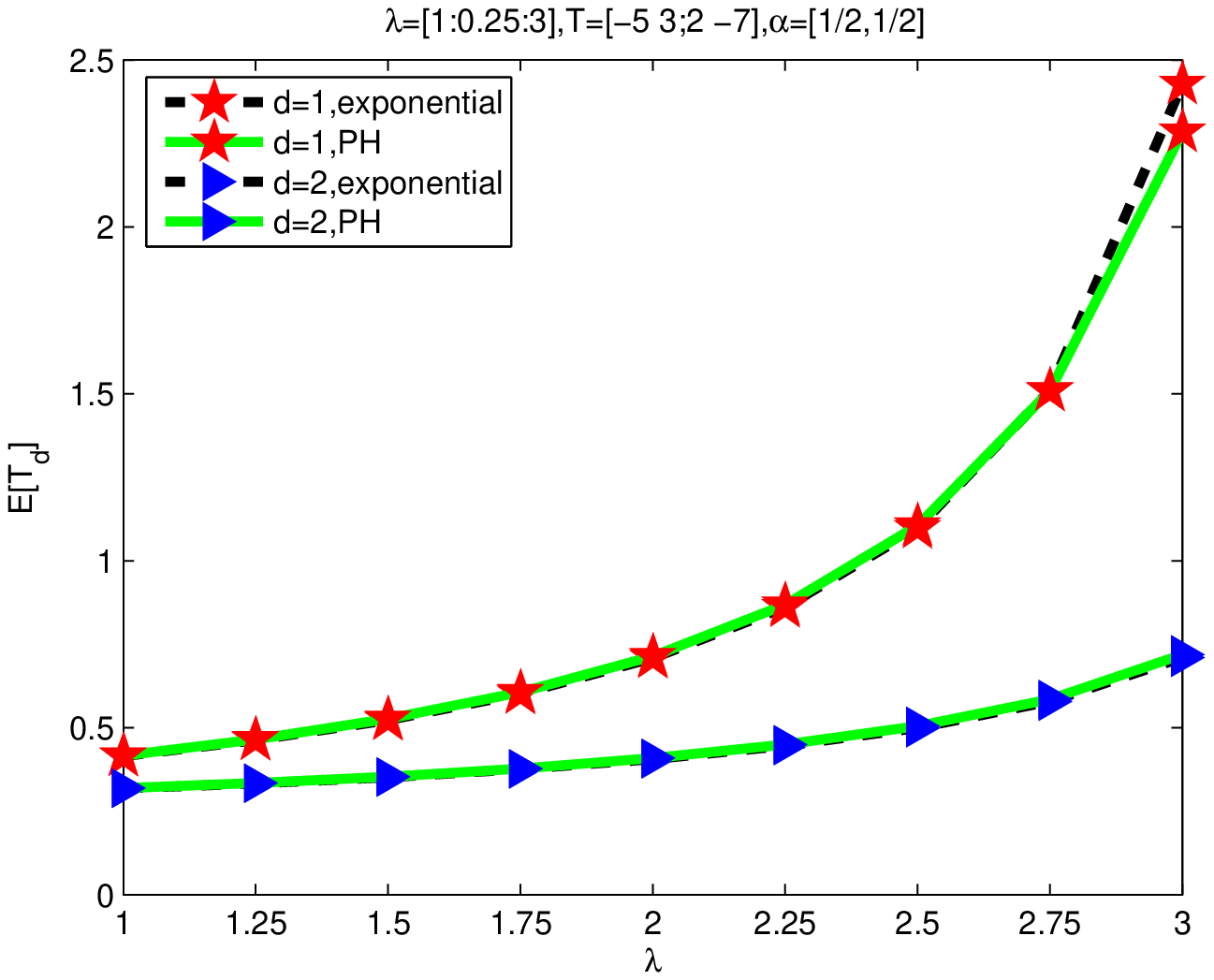}  \caption{Performance comparison
between the exponential and PH service times}%
\label{figure:fig-4}%
\end{figure}

\textbf{Example three: The role of the PH service times}

In this supermarket model with $d=2$, the customers arrive at this system as a
Poisson process with arrival rate $N\lambda$, and the service times at each
server are a PH distribution with irreducible representation $\left(
\alpha,T\left(  i\right)  \right)  $, $\alpha=\left(  1/2,1/2\right)  $,
\[
T\left(  1\right)  =\left(
\begin{array}
[c]{cc}%
-5 & 3\\
2 & -7
\end{array}
\right)  ,\text{ }T\left(  2\right)  =\left(
\begin{array}
[c]{cc}%
-4 & 3\\
2 & -7
\end{array}
\right)  ,\text{ }T\left(  3\right)  =\left(
\begin{array}
[c]{cc}%
-4 & 4\\
2 & -7
\end{array}
\right)  .
\]
It is seen that some minor changes are designed in the first rows of the
matrices $T\left(  i\right)  $ for $i=1,2,3$. Let $\lambda=1$. Then%
\[
\rho\left(  1\right)  =0.2931,\text{ }\rho\left(  2\right)  =0.3636,\text{
}\rho\left(  3\right)  =0.4250.
\]
This gives%
\[
\rho\left(  1\right)  <\rho\left(  2\right)  <\rho\left(  3\right)  .
\]
Figure 5 indicates how $E\left[  T_{d}\right]  $ depends on the different
transition rate matrices $T\left(  i\right)  $ for $i=1,2,3$, and%
\[
E\left[  T_{d}\left(  1\right)  \right]  <E\left[  T_{d}\left(  2\right)
\right]  <E\left[  T_{d}\left(  3\right)  \right]  .
\]
It is seen that $E\left[  T_{d}\right]  $ decreases as $d$ increases.

\begin{figure}[ptb]
\centering   \includegraphics[width=10cm]{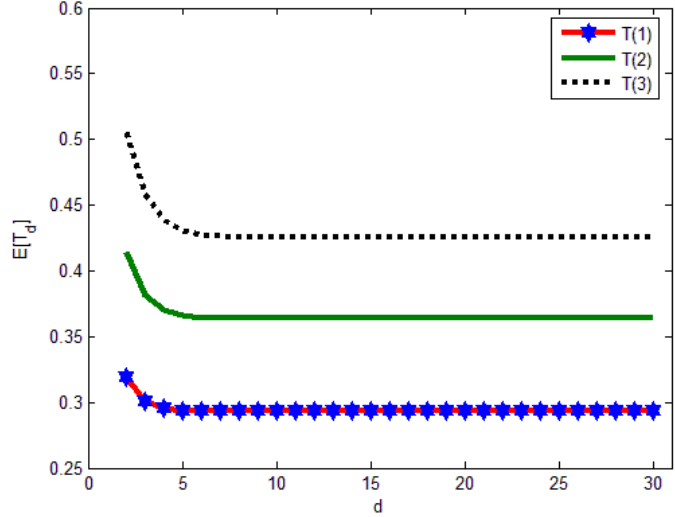}  \caption{$E\left[
T_{d}(i)\right]  $ vs the transition rate matrices $T\left(  i\right)  $ for
$i=1,2,3$}%
\label{figure:fig-5}%
\end{figure}

\textbf{Example four: The role of the MAP inputs}

In this supermarket model, the service time distribution is exponential with
service rate $\mu=1,$ and the arrival processes are the MAP of irreducible
matrix descriptor $\left(  C\left(  N\right)  ,D\left(  N\right)  \right)  $,
where%
\[
C=\left(
\begin{array}
[c]{cc}%
-5-\frac{2}{7}\lambda & 5\\
7 & -7-2\lambda
\end{array}
\right)  ,\text{ }D=\left(
\begin{array}
[c]{cc}%
\frac{2}{7}\lambda & 0\\
0 & 2\lambda
\end{array}
\right)  .
\]
It is easy to check that $\omega=\left(  7/12,5/12\right)  $, and the
stationary arrival rate $\lambda^{\ast}=\omega De=\lambda$. If $\mu=1$ and
$\rho=\lambda^{\ast}/\mu=\lambda<1$, then $\lambda\in\left(  0,1\right)  $.

Figure 6 shows how $E\left[  Q_{d}\right]  $ and $E\left[  T_{d}\right]  $
depend on the parameter $\lambda$ of the MAP under different choice numbers
$d=1,2,5,10$. It is seen that $E\left[  Q_{d}\right]  $ and $E\left[
T_{d}\right]  $ decrease as $d$ increases, while $E\left[  Q_{d}\right]  $ and
$E\left[  T_{d}\right]  $ increase as $\lambda$ increases.

\begin{figure}[ptb]
\centering                  \includegraphics[width=7cm]{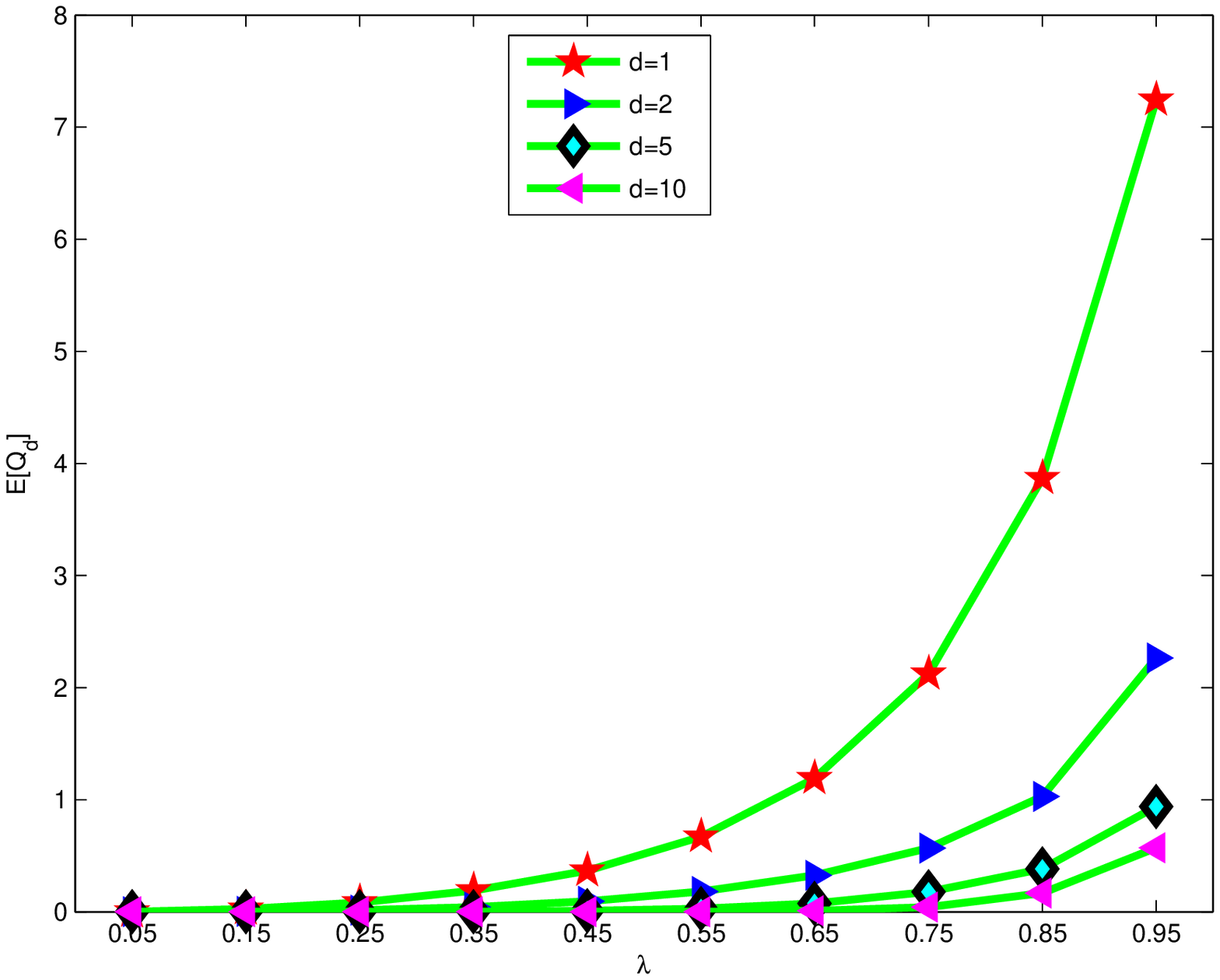}
\includegraphics[width=7cm]{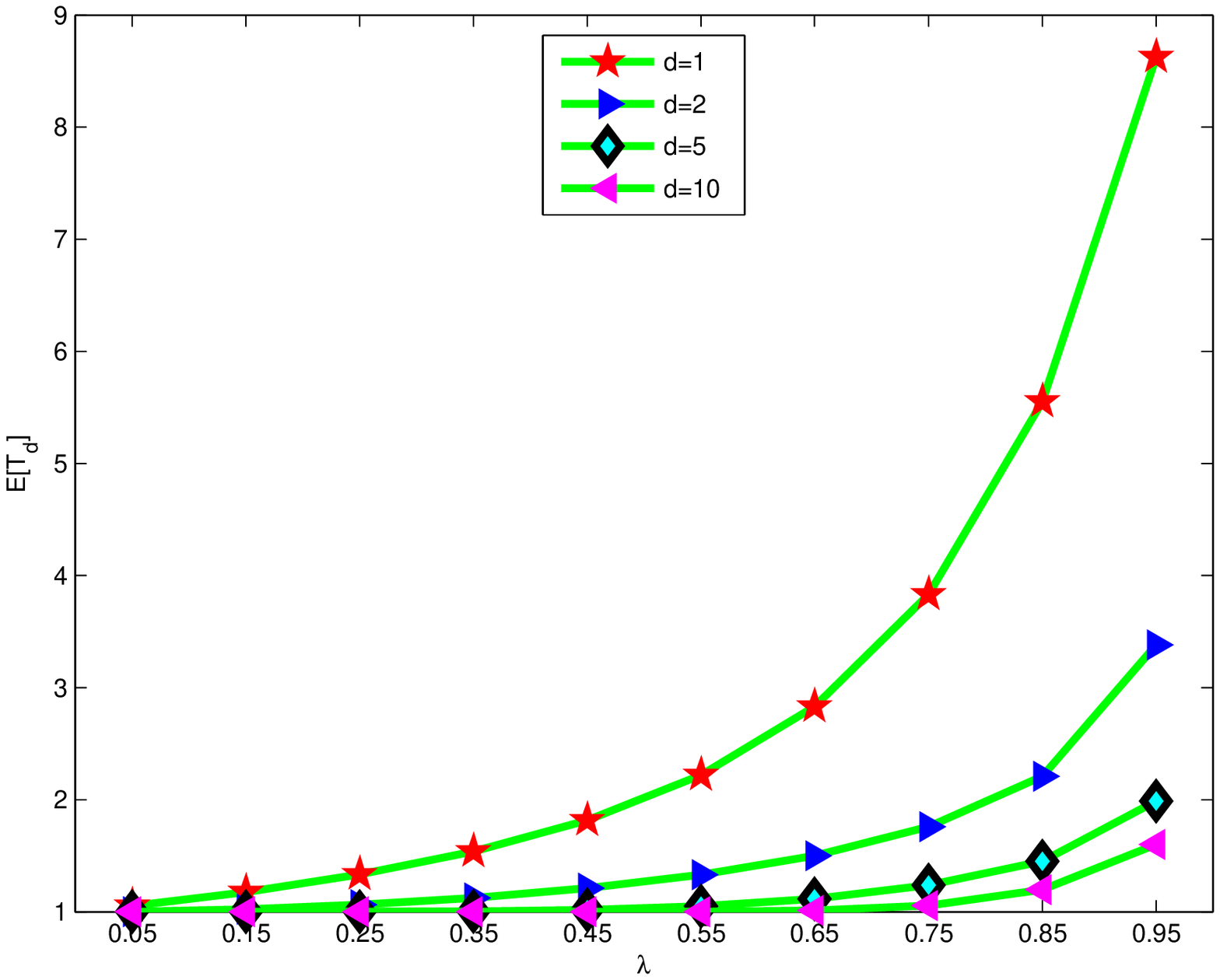}  \caption{The role of the MAP inputs}%
\label{figure:fig-6}%
\end{figure}

\section{Concluding Remarks}

In this paper, we analyze a more general block-structured supermarket model
with non-Poisson MAP inputs and with non-exponential PH service times, and set
up an infinite-dimensional system of differential vector equations satisfied
by the expected fraction vector through a detailed probability analysis, where
an important result: The invariance of environment factors is obtained. We
apply the phase-structured operator semigroup to proving the phase-structured
mean-field limit, which indicates the asymptotic independence of the
block-structured queueing processes in this supermarket model. Furthermore, we
provide an effective algorithm for computing the fixed point by means of the
matrix-analytic method. Using the fixed point, we provide two performance
measures of this supermarket model, and use some numerical examples to
illustrate how the two performance measures depend on the non-Poisson MAP
inputs and on the non-exponential PH service times. From many practical
applications, the block-structured supermarket model is an important queueing
model to analyze the relation between the system performance and the job
routing rule, and it can also help to design reasonable architecture to
improve the performance and to balance the load.

Note that this paper provide a clear picture for how to use the
phase-structured mean-field model as well as the matrix-analytic method to
analyze performance measures of more general supermarket models. We show that
this picture is organized as three key parts: (1) Setting up system of
differential equations, (2) necessary proofs of the phase-structured
mean-field limit, and (3) performance computation of this supermarket model
through the fixed point. Therefore, the results of this paper give new
highlight on understanding performance analysis and nonlinear Markov processes
for more general supermarket models with non-Poisson inputs and with
non-exponential service times. Along such a line, there are a number of
interesting directions for potential future research, for example:

\begin{itemize}
\item analyzing non-Poisson inputs such as renewal processes;

\item studying non-exponential service time distributions, for example,
general distributions, matrix-exponential distributions and heavy-tailed
distributions; and

\item discussing the bulk arrival processes, such as BMAP inputs, and the bulk
service processes, where effective algorithms for the fixed point are
necessary and interesting.
\end{itemize}

Up to now, we believe that a larger gap exists when dealing with either
renewal inputs or general service times in a supermarket model, because a more
challenging infinite-dimensional system of differential equations need be
established, a more complicated mean-field limit need be proved, and
computation of the fixed point will be more interesting, difficult and challenging.

\section*{Acknowledgements}

The authors thank the Associate Editor and two reviewers for many valuable
comments to sufficiently improve the presentation of this paper. At the same
time, the first author acknowledges that this research is partly supported by
the National Natural Science Foundation of China (No. 71271187) and the Hebei
Natural Science Foundation of China (No. A2012203125).

\section*{Three Appendices}

\subsection*{Appendix A: Proof of Theorem 1}

To prove Equations (\ref{Equ4-1}) to (\ref{Equ4-4}) in Theorem
\ref{The:InvEnF}, we need the following computational steps. Note that%
\begin{align*}
&  \sum_{m=1}^{d}C_{d}^{m}\left\{  \sum_{j=1}^{m_{B}}\left[  u_{k-1;l,j}%
^{(N)}\left(  t\right)  -u_{k;l,j}^{(N)}\left(  t\right)  \right]  \right\}
^{m-1}\left\{  \sum_{j=1}^{m_{B}}\left[  u_{k;l,j}^{(N)}\left(  t\right)
\right]  \right\}  ^{d-m}\\
&  =C_{d}^{d}\left\{  \sum_{j=1}^{m_{B}}\left[  u_{k-1;l,j}^{(N)}\left(
t\right)  -u_{k;l,j}^{(N)}\left(  t\right)  \right]  \right\}  ^{d-1}\\
&  +\sum_{m=1}^{d-1}C_{d}^{m}\left\{  \sum_{j=1}^{m_{B}}\left[  u_{k-1;l,j}%
^{(N)}\left(  t\right)  -u_{k;l,j}^{(N)}\left(  t\right)  \right]  \right\}
^{m-1}\left\{  \sum_{j=1}^{m_{B}}\left[  u_{k;l,j}^{(N)}\left(  t\right)
\right]  \right\}  ^{d-m}%
\end{align*}
and%
\begin{align*}
&  \sum_{m=1}^{d-1}C_{d}^{m}\left\{  \sum_{j=1}^{m_{B}}\left[  u_{k-1;l,j}%
^{(N)}\left(  t\right)  -u_{k;l,j}^{(N)}\left(  t\right)  \right]  \right\}
^{m-1}\left\{  \sum_{j=1}^{m_{B}}\left[  u_{k;l,j}^{(N)}\left(  t\right)
\right]  \right\}  ^{d-m}\\
&  +\sum_{m=1}^{d-1}C_{d}^{m}\left\{  \sum_{j=1}^{m_{B}}\left[  u_{k-1;l,j}%
^{(N)}\left(  t\right)  -u_{k;l,j}^{(N)}\left(  t\right)  \right]  \right\}
^{m-1}\\
&  \times\sum_{\substack{r_{1}+r_{2}+\cdots+r_{m_{A}}=d-m\\\sum_{i\neq
l}^{m_{A}}r_{i}\geq1\\0\leq r_{j}\leq d-m,1\leq j\leq m_{A}}}\left(
\begin{array}
[c]{c}%
d-m\\
r_{1},r_{2},\ldots,r_{m_{A}}%
\end{array}
\right)  \prod_{i=1}^{m_{A}}\left\{  \sum_{j=1}^{m_{B}}\left[  u_{k;i,j}%
^{(N)}\left(  t\right)  \right]  \right\}  ^{r_{i}}\\
&  =\sum_{m=1}^{d-1}C_{d}^{m}\left\{  \sum_{j=1}^{m_{B}}\left[  u_{k-1;l,j}%
^{(N)}\left(  t\right)  -u_{k;l,j}^{(N)}\left(  t\right)  \right]  \right\}
^{m-1}\left\{  \sum_{i=1}^{m_{A}}\sum_{j=1}^{m_{B}}\left[  u_{k;i,j}%
^{(N)}\left(  t\right)  \right]  \right\}  ^{d-m},
\end{align*}
since $\left\{  \sum_{j=1}^{m_{B}}\left[  u_{k;l,j}^{(N)}\left(  t\right)
\right]  \right\}  ^{d-m}$ corresponds to the case with $\sum_{i\neq l}%
^{m_{A}}r_{i}=0$ and $r_{l}=d-m$, and%
\begin{align*}
&  \left\{  \sum_{j=1}^{m_{B}}\left[  u_{k;l,j}^{(N)}\left(  t\right)
\right]  \right\}  ^{d-m}+\sum_{\substack{r_{1}+r_{2}+\cdots+r_{m_{A}%
}=d-m\\\sum_{i\neq l}^{m_{A}}r_{i}\geq1\\0\leq r_{j}\leq d-m,1\leq j\leq
m_{A}}}\left(
\begin{array}
[c]{c}%
d-m\\
r_{1},r_{2},\ldots,r_{m_{A}}%
\end{array}
\right)  \prod_{i=1}^{m_{A}}\left\{  \sum_{j=1}^{m_{B}}\left[  u_{k;i,j}%
^{(N)}\left(  t\right)  \right]  \right\}  ^{r_{i}}\\
&  =\sum_{\substack{r_{1}+r_{2}+\cdots+r_{m_{A}}=d-m\\0\leq r_{j}\leq
d-m,1\leq j\leq m_{A}}}\left(
\begin{array}
[c]{c}%
d-m\\
r_{1},r_{2},\ldots,r_{m_{A}}%
\end{array}
\right)  \prod_{i=1}^{m_{A}}\left\{  \sum_{j=1}^{m_{B}}\left[  u_{k;i,j}%
^{(N)}\left(  t\right)  \right]  \right\}  ^{r_{i}}\\
&  =\left\{  \sum_{i=1}^{m_{A}}\sum_{j=1}^{m_{B}}\left[  u_{k;i,j}%
^{(N)}\left(  t\right)  \right]  \right\}  ^{d-m}.
\end{align*}
we obtain%
\begin{align*}
&  C_{d}^{d}\left\{  \sum_{j=1}^{m_{B}}\left[  u_{k-1;l,j}^{(N)}\left(
t\right)  -u_{k;l,j}^{(N)}\left(  t\right)  \right]  \right\}  ^{d-1}\\
&  +\sum_{m=1}^{d-1}C_{d}^{m}\left\{  \sum_{j=1}^{m_{B}}\left[  u_{k-1;l,j}%
^{(N)}\left(  t\right)  -u_{k;l,j}^{(N)}\left(  t\right)  \right]  \right\}
^{m-1}\left\{  \sum_{i=1}^{m_{A}}\sum_{j=1}^{m_{B}}u_{k;l,j}^{(N)}\left(
t\right)  \right\}  ^{d-m}\\
&  =\sum_{m=1}^{d}C_{d}^{m}\left\{  \sum_{j=1}^{m_{B}}\left[  u_{k-1;l,j}%
^{(N)}\left(  t\right)  -u_{k;l,j}^{(N)}\left(  t\right)  \right]  \right\}
^{m-1}\left\{  \sum_{i=1}^{m_{A}}\sum_{j=1}^{m_{B}}\left[  u_{k;l,j}%
^{(N)}\left(  t\right)  \right]  \right\}  ^{d-m}\\
&  =C_{d}^{1}\left\{  \sum_{i=1}^{m_{A}}\sum_{j=1}^{m_{B}}\left[
u_{k;l,j}^{(N)}\left(  t\right)  \right]  \right\}  ^{d-1}+\sum_{m=2}^{d}%
C_{d}^{m}\left\{  \sum_{j=1}^{m_{B}}\left[  u_{k-1;l,j}^{(N)}\left(  t\right)
-u_{k;l,j}^{(N)}\left(  t\right)  \right]  \right\}  ^{m-1}\\
&  \times\left\{  \sum_{i=1}^{m_{A}}\sum_{j=1}^{m_{B}}\left[  u_{k;l,j}%
^{(N)}\left(  t\right)  \right]  \right\}  ^{d-m}.
\end{align*}
Using $\frac{m_{1}}{m}C_{m}^{m_{1}}=C_{m-1}^{m_{1}-1}$, we can obtain%
\begin{align*}
&  \sum_{m=2}^{d}C_{d}^{m}\left\{  \sum_{j=1}^{m_{B}}\left[  u_{k-1;l,j}%
^{(N)}\left(  t\right)  -u_{k;l,j}^{(N)}\left(  t\right)  \right]  \right\}
^{m-1}\left\{  \sum_{i=1}^{m_{A}}\sum_{j=1}^{m_{B}}\left[  u_{k;l,j}%
^{(N)}\left(  t\right)  \right]  \right\}  ^{d-m}\\
&  +\sum_{m=2}^{d}C_{d}^{m}\sum_{m_{1}=1}^{m-1}\frac{m_{1}}{m}C_{m}^{m_{1}%
}\left\{  \sum_{j=1}^{m_{B}}\left[  u_{k-1;l,j}^{(N)}\left(  t\right)
-u_{k;l,j}^{(N)}\left(  t\right)  \right]  \right\}  ^{m_{1}-1}\\
&  \times\sum_{\substack{n_{1}+n_{2}+\cdots+n_{m_{A}}=m-m_{1}\\\sum_{i\neq
l}^{m_{A}}n_{i}\geq1\\0\leq n_{j}\leq m-m_{1},1\leq j\leq m_{A}}}\left(
\begin{array}
[c]{c}%
m-m_{1}\\
n_{1},n_{2},\ldots,n_{m_{A}}%
\end{array}
\right) \\
&  \times\prod_{i=1}^{m_{A}}\left\{  \sum_{j=1}^{m_{B}}\left[  u_{k-1;i,j}%
^{(N)}\left(  t\right)  -u_{k;i,j}^{(N)}\left(  t\right)  \right]  \right\}
^{n_{i}}\\
&  \times\sum_{\substack{r_{1}+r_{2}+\cdots+r_{m_{A}}=d-m\\0\leq r_{j}\leq
d-m,1\leq j\leq m_{A}}}\left(
\begin{array}
[c]{c}%
d-m\\
r_{1},r_{2},\ldots,r_{m_{A}}%
\end{array}
\right)  \prod_{i=1}^{m_{A}}\left\{  \sum_{j=1}^{m_{B}}u_{k;i,j}^{(N)}\left(
t\right)  \right\}  ^{r_{i}}\\
&
\end{align*}%
\begin{align*}
&  =\sum_{m=2}^{d}C_{d}^{m}\sum_{m_{1}=1}^{m}C_{m-1}^{m_{1}-1}\left\{
\sum_{j=1}^{m_{B}}\left[  u_{k-1;l,j}^{(N)}\left(  t\right)  -u_{k;l,j}%
^{(N)}\left(  t\right)  \right]  \right\}  ^{m_{1}-1}\\
&  \times\sum_{\substack{n_{1}+n_{2}+\cdots+n_{m_{A}}=m-m_{1}\\0\leq n_{j}\leq
m-m_{1},1\leq j\leq m_{A}}}\left(
\begin{array}
[c]{c}%
m-m_{1}\\
n_{1},n_{2},\ldots,n_{m_{A}}%
\end{array}
\right) \\
&  \times\prod_{i=1}^{m_{A}}\left\{  \sum_{j=1}^{m_{B}}\left[  u_{k-1;i,j}%
^{(N)}\left(  t\right)  -u_{k;i,j}^{(N)}\left(  t\right)  \right]  \right\}
^{n_{i}}\left\{  \sum_{i=1}^{m_{A}}\sum_{j=1}^{m_{B}}\left[  u_{k;l,j}%
^{(N)}\left(  t\right)  \right]  \right\}  ^{d-m}\\
&  =\sum_{m=2}^{d}C_{d}^{m}\sum_{m_{1}=1}^{m}\frac{m_{1}}{m}C_{m}^{m_{1}%
}\left\{  \sum_{j=1}^{m_{B}}\left[  u_{k-1;l,j}^{(N)}\left(  t\right)
-u_{k;l,j}^{(N)}\left(  t\right)  \right]  \right\}  ^{m_{1}-1}\\
&  \times\left\{  \sum_{i\neq l}^{m_{A}}\sum_{j=1}^{m_{B}}\left[
u_{k-1;i,j}^{(N)}\left(  t\right)  -u_{k;i,j}^{(N)}\left(  t\right)  \right]
\right\}  ^{m-m_{1}}\left\{  \sum_{i=1}^{m_{A}}\sum_{j=1}^{m_{B}}\left[
u_{k;i,j}^{(N)}\left(  t\right)  \right]  \right\}  ^{d-m}\\
&  =\sum_{m=2}^{d}C_{d}^{m}\sum_{m_{1}-1=0}^{m-1}C_{m-1}^{m_{1}-1}\left\{
\sum_{j=1}^{m_{B}}\left[  u_{k-1;l,j}^{(N)}\left(  t\right)  -u_{k;l,j}%
^{(N)}\left(  t\right)  \right]  \right\}  ^{m_{1}-1}\\
&  \times\left\{  \sum_{i\neq l}^{m_{A}}\sum_{j=1}^{m_{B}}\left[
u_{k-1;i,j}^{(N)}\left(  t\right)  -u_{k;i,j}^{(N)}\left(  t\right)  \right]
\right\}  ^{m-1-\left(  m_{1}-1\right)  }\left\{  \sum_{i=1}^{m_{A}}\sum
_{j=1}^{m_{B}}\left[  u_{k;i,j}^{(N)}\left(  t\right)  \right]  \right\}
^{d-m}\\
&  =\sum_{m=2}^{d}C_{d}^{m}\left\{  \sum_{i=1}^{m_{A}}\sum_{j=1}^{m_{B}%
}\left[  u_{k-1;i,j}^{(N)}\left(  t\right)  -u_{k;i,j}^{(N)}\left(  t\right)
\right]  \right\}  ^{m-1}\left\{  \sum_{i=1}^{m_{A}}\sum_{j=1}^{m_{B}}\left[
u_{k;i,j}^{(N)}\left(  t\right)  \right]  \right\}  ^{d-m},
\end{align*}
we have%
\begin{align*}
&  C_{d}^{1}\left\{  \sum_{i=1}^{m_{A}}\sum_{j=1}^{m_{B}}\left[
u_{k;l,j}^{(N)}\left(  t\right)  \right]  \right\}  ^{d-1}\\
&  +\sum_{m=2}^{d}C_{d}^{m}\left\{  \sum_{i=1}^{m_{A}}\sum_{j=1}^{m_{B}%
}\left[  u_{k-1;i,j}^{(N)}\left(  t\right)  -u_{k;i,j}^{(N)}\left(  t\right)
\right]  \right\}  ^{m-1}\left\{  \sum_{i=1}^{m_{A}}\sum_{j=1}^{m_{B}}\left[
u_{k;i,j}^{(N)}\left(  t\right)  \right]  \right\}  ^{d-m}\\
&  =\sum_{m=1}^{d}C_{d}^{m}\left\{  \sum_{i=1}^{m_{A}}\sum_{j=1}^{m_{B}%
}\left[  u_{k-1;i,j}^{(N)}\left(  t\right)  -u_{k;i,j}^{(N)}\left(  t\right)
\right]  \right\}  ^{m-1}\left\{  \sum_{i=1}^{m_{A}}\sum_{j=1}^{m_{B}}\left[
u_{k;i,j}^{(N)}\left(  t\right)  \right]  \right\}  ^{d-m}.
\end{align*}
Thus for $k\geq1$ we obtain%
\begin{align*}
L_{k;l}^{\left(  N\right)  }\left(  u_{k-1}^{\left(  N\right)  }\left(
t\right)  ,u_{k}^{\left(  N\right)  }\left(  t\right)  \right)  =  &
\sum_{m=1}^{d}C_{d}^{m}\left\{  \sum_{i=1}^{m_{A}}\sum_{j=1}^{m_{B}}\left[
u_{k-1;i,j}^{(N)}\left(  t\right)  -u_{k;i,j}^{(N)}\left(  t\right)  \right]
\right\}  ^{m-1}\\
&  \times\left\{  \sum_{i=1}^{m_{A}}\sum_{j=1}^{m_{B}}\left[  u_{k;i,j}%
^{(N)}\left(  t\right)  \right]  \right\}  ^{d-m},
\end{align*}
which is independent of phase $l\in\left\{  1,2,\ldots,m_{A}\right\}  $. Thus
we have%
\[
L_{k}^{\left(  N\right)  }\left(  u_{k-1}^{\left(  N\right)  }\left(
t\right)  ,u_{k}^{\left(  N\right)  }\left(  t\right)  \right)  =L_{k;l}%
^{\left(  N\right)  }\left(  u_{k-1}^{\left(  N\right)  }\left(  t\right)
,u_{k}^{\left(  N\right)  }\left(  t\right)  \right)  .
\]
Similarly, for phase $l\in\left\{  1,2,\ldots,m_{A}\right\}  $, we have%
\begin{align*}
L_{1;l}^{\left(  N\right)  }\left(  \left[  u_{0}^{\left(  N\right)  }\left(
t\right)  \otimes\alpha\right]  ,u_{1}^{\left(  N\right)  }\left(  t\right)
\right)  =  &  \sum_{m=1}^{d}C_{d}^{m}\left[  \sum_{i=1}^{m_{A}}\sum
_{j=1}^{m_{B}}\left(  u_{0;i}^{\left(  N\right)  }\left(  t\right)  \alpha
_{j}-u_{1;i,j}^{\left(  N\right)  }\left(  t\right)  \right)  \right]
^{m-1}\\
&  \times\left[  \sum_{i=1}^{m_{A}}\sum_{j=1}^{m_{B}}u_{1;i,j}^{\left(
N\right)  }\left(  t\right)  \right]  ^{d-m}.
\end{align*}
This gives%
\[
L_{1}^{\left(  N\right)  }\left(  \left[  u_{0}^{\left(  N\right)  }\left(
t\right)  \otimes\alpha\right]  ,u_{1}^{\left(  N\right)  }\left(  t\right)
\right)  =L_{1;l}^{\left(  N\right)  }\left(  \left[  u_{0}^{\left(  N\right)
}\left(  t\right)  \otimes\alpha\right]  ,u_{1}^{\left(  N\right)  }\left(
t\right)  \right)
\]
This completes the proof. \textbf{{\rule{0.08in}{0.08in}}}

\subsection*{Appendix B: The Mean-Field Limit}

In this appendix, we use the operator semigroup to provide a mean-field limit
for the sequence of Markov processes $\{\mathbf{U}^{(N)}(t),t\geq0\}$, which
indicates the asymptotic independence of the block-structured queueing
processes in this supermarket model. Note that the limits of the sequences of
Markov processes can usually be discussed by the three main techniques:
Operator semigroups, martingales, and stochastic equations. Readers may refer
to Ethier and Kurtz \cite{Eth:1986} for more details.

To use the operator semigroups of Markov processes, we first need to introduce
some state spaces as follows. For the vectors $\mathbf{u}^{\left(  N\right)
}=\left(  u_{0}^{\left(  N\right)  },u_{1}^{\left(  N\right)  },u_{2}^{^{(N)}%
}\left(  t\right)  \ldots\right)  $ where $u_{0}^{\left(  N\right)  }$ is a
probability vector of size $m_{A}$ and the size of the row vector
$u_{k}^{\left(  N\right)  }$ is $m_{A}m_{B}$ for $k\geq1$, we write%
\begin{align*}
\widetilde{\Omega}_{N}=  &  \left\{  \mathbf{u}^{\left(  N\right)  }%
:u_{1}^{\left(  N\right)  }\geq u_{2}^{\left(  N\right)  }\geq u_{3}^{\left(
N\right)  }\geq\cdots\geq0,\right. \\
&  \left.  1=u_{0}^{\left(  N\right)  }e\geq u_{1}^{\left(  N\right)  }e\geq
u_{2}^{\left(  N\right)  }e\geq\cdots\geq0,\right. \\
&  \left.  Nu_{k}^{\left(  N\right)  }\text{ \ is a vector of nonnegative
integers for }k\geq0\right\}  .
\end{align*}
and%
\[
\Omega_{N}=\left\{  \mathbf{u}^{\left(  N\right)  }\in\widetilde{\Omega}%
_{N}:\mathbf{u}^{\left(  N\right)  }e<+\infty\right\}  .
\]
At the same time, for the vector $\mathbf{u}=\left(  u_{0},u_{1},u_{2}%
,\ldots\right)  $ where $u_{0}$ is a probability vector of size $m_{A}$ and
the size of the row vector $u_{k}$ is $m_{A}m_{B}$ for $k\geq1$, we set%
\[
\widetilde{\Omega}=\{\mathbf{u}:u_{1}\geq u_{2}\geq u_{3}\geq\cdots
\geq0;\ \ 1=u_{0}^{\left(  N\right)  }e\geq u_{1}^{\left(  N\right)  }e\geq
u_{2}^{\left(  N\right)  }e\geq\cdots\geq0\}
\]
and%
\[
\Omega=\left\{  \mathbf{u}\in\widetilde{\Omega}:\mathbf{u}e<+\infty\right\}
.
\]
Obviously, $\Omega_{N}\subsetneqq\Omega\subsetneqq\widetilde{\Omega}$ and
$\Omega_{N}\subsetneqq\widetilde{\Omega}_{N}\subsetneqq\widetilde{\Omega}$.

In the vector space $\widetilde{\Omega}$, we take a metric%
\begin{align}
\rho\left(  \mathbf{u},\mathbf{u}^{\prime}\right)   &  =\max\left\{
\max_{1\leq i\leq m_{A}}\left\{  |u_{0;i}-u_{0;i}^{\prime}|\right\}  ,\right.
\nonumber\\
&  \left.  \max_{\substack{0\leq i\leq m_{A}\\0\leq j\leq m_{B}}}\sup_{k\geq
1}\left\{  \dfrac{|u_{k;i,j}-u_{k;i,j}^{\prime}|}{k+1}\right\}  \right\}
\label{Eq10.0}%
\end{align}
for $\mathbf{u},\mathbf{u}^{\prime}\in\widetilde{\Omega}$. Note that under the
metric $\rho\left(  \mathbf{u},\mathbf{u}^{\prime}\right)  ,$ the vector space
$\widetilde{\Omega}$ is separable and compact.

\subsubsection*{B.1: The operator semigroup}

For $\mathbf{g}\in\Omega_{N}$, we write%
\[
L_{1}\left(  g_{0}\otimes\alpha,g_{1}\right)  =\sum_{m=1}^{d}C_{d}^{m}\left[
\sum_{l=1}^{m_{A}}\sum_{j=1}^{m_{B}}\left(  g_{0;l}\left(  t\right)
\alpha_{j}-g_{1;l,j}\right)  \right]  ^{m-1}\left(  \sum_{l=1}^{m_{A}}%
\sum_{j=1}^{m_{B}}g_{1;l,j}\right)  ^{d-m},
\]
and for $k\geq2$%
\[
L_{k}\left(  g_{k-1},g_{k}\right)  =\sum_{m=1}^{d}C_{d}^{m}\left[  \sum
_{l=1}^{m_{A}}\sum_{j=1}^{m_{B}}\left(  g_{k-1;l,j}-g_{k;l,j}\right)  \right]
^{m-1}\left[  \sum_{l=1}^{m_{A}}\sum_{j=1}^{m_{B}}g_{k;l,j}\right]  ^{d-m}.
\]

Now, we consider the infinite-dimensional Markov process $\{\mathbf{U}%
^{(N)}(t),t\geq0\}$ on state space $\Omega_{N}$ (or $\widetilde{\Omega}_{N}$
in a similar analysis) for $N=1,2,3,\ldots$. Note that the stochastic
evolution of this supermarket model of $N$ identical servers is described as
the Markov process $\left\{  \mathbf{U}^{(N)}(t),t\geq0\right\}  $, where%
\[
\frac{\text{d}}{\text{d}t}\left(  \mathbf{U}^{(N)}(t)\right)  =\mathbf{A}%
_{N}\text{ }f\left(  \mathbf{U}^{(N)}(t)\right)  ,
\]
where $\mathbf{A}_{N}$ acting on functions $f:\Omega_{N}\rightarrow
\mathbf{C}^{1}$ is the generating operator of the Markov process $\left\{
\mathbf{U}^{(N)}(t),t\geq0\right\}  $,%
\begin{equation}
\mathbf{A}_{N}=\mathbf{A}_{N}^{\text{A-In}}+\mathbf{A}_{N}%
^{\text{A-Transition}}+\mathbf{A}_{N}^{\text{S-Transition}}+\mathbf{A}%
_{N}^{\text{S-Out}}, \label{Eq10.1}%
\end{equation}
for $\mathbf{g}\in\Omega_{N}$%
\begin{align}
\mathbf{A}_{N}^{\text{A-In}}f(\mathbf{g})=  &  N\sum\limits_{k=2}^{\infty}%
\sum_{j=1}^{m_{B}}\sum_{i=1}^{m_{A}}\left[  \sum_{l=1}^{m_{A}}\left(
g_{k-1;l,j}d_{l,i}-g_{k;i,j}\sum_{q=1}^{m_{A}}d_{i,q}\right)  L_{k}\left(
g_{k-1},g_{k}\right)  \right] \nonumber\\
&  \times\left[  f\left(  \mathbf{g}+\frac{e_{k;i,j}}{N}\right)  -f\left(
\mathbf{g}\right)  \right] \nonumber\\
&  +N\sum_{j=1}^{m_{B}}\sum_{i=1}^{m_{A}}\left[  \sum_{l=1}^{m_{A}}\left(
g_{0;l}d_{l,i}\alpha_{j}-g_{1;i,j}\sum_{q=1}^{m_{A}}d_{i,q}\right)
L_{1}\left(  g_{0}\otimes\alpha,g_{1}\right)  \right] \nonumber\\
&  \times\left[  f\left(  \mathbf{g}+\frac{e_{1;i,j}}{N}\right)  -f\left(
\mathbf{g}\right)  \right]  , \label{Eq10.2}%
\end{align}%
\begin{align}
\mathbf{A}_{N}^{\text{A-Transition}}=  &  N\sum\limits_{k=1}^{\infty}%
\sum_{j=1}^{m_{B}}\sum_{i=1}^{m_{A}}\sum_{l=1}^{m_{A}}\left[  g_{k;l,j}%
c_{l,i}+g_{k;i,j}\sum_{q=1}^{m_{A}}d_{i,q}\right] \nonumber\\
&  \times\left[  f\left(  \mathbf{g}-\frac{e_{k;l,j}}{N}+\frac{e_{k;i,j}}%
{N}\right)  -f\left(  \mathbf{g}\right)  \right] \nonumber\\
&  +N\sum_{i=1}^{m_{A}}\sum_{l=1}^{m_{A}}\left[  g_{0;l}c_{l,i}+g_{0,i}%
\sum_{q=1}^{m_{A}}d_{i,q}\right] \nonumber\\
&  \times\left[  f\left(  \mathbf{g}-\frac{e_{0;l}}{N}+\frac{e_{0;i}}%
{N}\right)  -f\left(  \mathbf{g}\right)  \right]  , \label{Eq10.2-2}%
\end{align}%
\begin{align}
\mathbf{A}_{N}^{\text{S-Transition}}  &  =N\sum\limits_{k=1}^{\infty}%
\sum_{i=1}^{m_{A}}\sum_{j=1}^{m_{B}}\sum_{r=1}^{m_{B}}\left(  g_{k;i,r}%
t_{r,j}\right) \nonumber\\
&  \times\left[  f(\mathbf{g}-\dfrac{\mathbf{e}_{k;i,r}}{N}+\dfrac
{\mathbf{e}_{k;i,j}}{N})-f(\mathbf{g})\right]  \label{Eq10.2-3}%
\end{align}
and%
\begin{equation}
\mathbf{A}_{N}^{\text{S-Out}}=N\sum\limits_{k=1}^{\infty}\sum_{i=1}^{m_{A}%
}\sum_{j=1}^{m_{B}}\sum_{r=1}^{m_{B}}\left(  g_{k+1;i,r}t_{r}^{0}\alpha
_{j}\right)  \left[  f(\mathbf{g})-f(\mathbf{g-}\frac{\mathbf{e}_{k;i,j}}%
{N})\right]  , \label{Eq10.3}%
\end{equation}
where $\mathbf{e}_{k;l,j}$ is a row vector of infinite size with the $\left(
k;i,j\right)  $th entry being one and all others being zero. Thus it follows
from Equations (\ref{Eq10.1}) to (\ref{Eq10.3}) that
\begin{align}
\mathbf{A}_{N}f(\mathbf{g})=  &  N\sum\limits_{k=2}^{\infty}\sum_{j=1}^{m_{B}%
}\sum_{i=1}^{m_{A}}\left[  \sum_{l=1}^{m_{A}}\left(  g_{k-1;l,j}%
d_{l,i}-g_{k;i,j}\sum_{q=1}^{m_{A}}d_{i,q}\right)  L_{k}\left(  g_{k-1}%
,g_{k}\right)  \right] \nonumber\\
&  \times\left[  f\left(  \mathbf{g}+\frac{e_{k;i,j}}{N}\right)  -f\left(
\mathbf{g}\right)  \right] \nonumber\\
&  +N\sum_{j=1}^{m_{B}}\sum_{i=1}^{m_{A}}\left[  \sum_{l=1}^{m_{A}}\left(
g_{0;l}d_{l,i}\alpha_{j}-g_{1;i,j}\sum_{q=1}^{m_{A}}d_{i,q}\right)
L_{1}\left(  g_{1}\otimes\alpha,g_{1}\right)  \right] \nonumber\\
&  \times\left[  f\left(  \mathbf{g}+\frac{e_{1;i,j}}{N}\right)  -f\left(
\mathbf{g}\right)  \right] \nonumber\\
&  +N\sum\limits_{k=1}^{\infty}\sum_{j=1}^{m_{B}}\sum_{i=1}^{m_{A}}\sum
_{l=1}^{m_{A}}\left(  g_{k;l,j}c_{l,i}+g_{k;i,j}\sum_{q=1}^{m_{A}}%
d_{i,q}\right) \nonumber\\
&  \times\left[  f\left(  \mathbf{g}-\frac{e_{k;l,j}}{N}+\frac{e_{k;i,j}}%
{N}\right)  -f\left(  \mathbf{g}\right)  \right] \nonumber\\
&  +N\sum_{i=1}^{m_{A}}\sum_{l=1}^{m_{A}}\left(  g_{0;l}c_{l,i}+\sum
_{q=1}^{m_{A}}d_{i,q}\right)  \left[  f\left(  \mathbf{g}-\frac{e_{0;l}}%
{N}+\frac{e_{0;i}}{N}\right)  -f\left(  \mathbf{g}\right)  \right] \nonumber\\
&  +N\sum\limits_{k=1}^{\infty}\sum_{i=1}^{m_{A}}\sum_{j=1}^{m_{B}}\sum
_{r=1}^{m_{B}}\left\{  \left(  g_{k;i,r}t_{r,j}\right)  \left[  f(\mathbf{g}%
-\dfrac{\mathbf{e}_{k;i,r}}{N}+\dfrac{\mathbf{e}_{k;i,j}}{N})-f(\mathbf{g}%
)\right]  \right. \nonumber\\
&  \left.  +\left(  g_{k+1;i,r}t_{r}^{0}\alpha_{j}\right)  \left[
f(\mathbf{g})-f(\mathbf{g}-\frac{\mathbf{e}_{k;i,j}}{N})\right]  \right\}  .
\label{Eq10.4}%
\end{align}

\begin{Rem}
If the MAP is a Poisson process, then $m_{A}=1$ and $C=-\lambda$ and
$D=\lambda$; and if the PH service time distribution is exponential, then
$m_{B}=1$, $T=-\mu$ and $T^{0}\alpha=\mu$. In this case, it is easy to check
from (\ref{Eq10.4}) that%
\begin{align*}
\mathbf{A}_{N}f(\mathbf{g})=  &  \lambda N\left(  1-g_{1}^{d}\right)  \left[
f\left(  \mathbf{g}+\frac{e_{1}}{N}\right)  -f\left(  \mathbf{g}\right)
\right] \\
&  +\lambda N\sum\limits_{k=2}^{\infty}\left(  g_{k-1}^{d}-g_{k}^{d}\right)
\left[  f\left(  \mathbf{g}+\frac{e_{k}}{N}\right)  -f\left(  \mathbf{g}%
\right)  \right] \\
&  -\mu N\sum\limits_{n=1}^{\infty}\left(  g_{n}-g_{n+1}\right)  \left[
f\left(  \mathbf{g}\right)  -f\left(  \mathbf{g}-\frac{e_{n}}{N}\right)
\right]  ,
\end{align*}
which is the same as (1.5) for $d=2$ in Vvedenskaya et al \cite{Vve:1996}.
\end{Rem}

\subsubsection*{B.2: The mean-Field limit}

We compute%
\[
\lim_{N\rightarrow\infty}\frac{f\left(  \mathbf{g}+\dfrac{e_{k;i,j}}%
{N}\right)  -f\left(  \mathbf{g}\right)  }{\dfrac{1}{N}}=\frac{\partial
}{\partial g_{k;i,j}}f(\mathbf{g}),
\]%
\[
\lim_{N\rightarrow\infty}\frac{f\left(  \mathbf{g}\right)  -f\left(
\mathbf{g-}\dfrac{e_{k;i,j}}{N}\right)  }{\dfrac{1}{N}}=\frac{\partial
}{\partial g_{k;i,j}}f(\mathbf{g})
\]
and%
\[
\lim_{N\rightarrow\infty}\frac{f\left(  \mathbf{g-}\dfrac{e_{k;l,j}}{N}%
+\dfrac{e_{k;i,j}}{N}\right)  -f\left(  \mathbf{g}\right)  }{\dfrac{1}{N}%
}=\frac{\partial}{\partial g_{k;i,j}}f(\mathbf{g})-\frac{\partial}{\partial
g_{k;l,j}}f(\mathbf{g}).
\]

The operator semigroup of the Markov process $\left\{  \mathbf{U}%
^{(N)}(t),t\geq0\right\}  $ is defined as $\mathbf{T}_{N}(t)$, where if
$f:\Omega_{N}\rightarrow\mathbf{C}^{1}$, then for $\mathbf{g}\in\Omega_{N}$
and $t\geq0$%
\begin{equation}
\mathbf{T}_{N}(t)f(\mathbf{g})=E\left[  f(\mathbf{U}_{N}(t)\text{ }|\text{
}\mathbf{U}_{N}(0)=\mathbf{g}\right]  . \label{Eq10.5}%
\end{equation}
Note that $\mathbf{A}_{N}$ is the generating operator of the operator
semigroup $\mathbf{T}_{N}(t)$, it is easy to see that $\mathbf{T}_{N}%
(t)=\exp\left\{  \mathbf{A}_{N}t\right\}  $ for $t\geq0$.\qquad\qquad\qquad

\begin{Def}
A operator semigroup $\left\{  \mathbf{S}\left(  t\right)  :t\geq0\right\}  $
on the Banach space $L=C(\widetilde{\Omega})$ is said to be strongly
continuous if $\lim_{t\rightarrow0}\mathbf{S}\left(  t\right)  f=f$ for every
$f\in L$; it is said to be a contractive semigroup if $\left\|  \mathbf{S}%
\left(  t\right)  \right\|  \leq1$ for $t\geq0$.
\end{Def}

Let $L=C(\widetilde{\Omega})$ be the Banach space of continuous functions
$f:\widetilde{\Omega}\rightarrow\mathbf{R}$ with uniform metric $\left\Vert
f\right\Vert =\underset{u\in\widetilde{\Omega}}{\max}\left\vert
f(u)\right\vert $, and similarly, let $L_{N}=C(\Omega_{N})$. The inclusion
$\Omega_{N}\subset\widetilde{\Omega}$ induces a contraction mapping $\Pi
_{N}:L\rightarrow L_{N},\Pi_{N}f(u)=f(u)$ for $f\in L$ and $u\in\Omega_{N}$.

Now, we consider the limiting behavior of the sequence $\{(\mathbf{U}%
^{(N)}(t),t\geq0\}$ of Markov processes for $N=1,2,3,\ldots$. Two formal
limits for the sequence $\left\{  \mathbf{A}_{N}\right\}  $ of generating
operators and for the sequence $\left\{  \mathbf{T}_{N}(t)\right\}  $ of
semigroups are expressed as $\mathbf{A}=\lim_{N\rightarrow\infty}%
\mathbf{A}_{N}$ and $\mathbf{T}\left(  t\right)  =\lim_{N\rightarrow\infty
}\mathbf{T}_{N}(t)$ for $t\geq0$, respectively. It follows from (\ref{Eq10.4})
that as $N\rightarrow\infty$%
\begin{align}
\mathbf{A}f(\mathbf{g})=  &  \sum\limits_{k=2}^{\infty}\sum_{j=1}^{m_{B}}%
\sum_{i=1}^{m_{A}}\left[  \sum_{l=1}^{m_{A}}\left(  g_{k-1;l,j}d_{l,i}%
-g_{k;i,j}\sum_{q=1}^{m_{A}}d_{i,q}\right)  L_{k}\left(  g_{k-1},g_{k}\right)
\right]  \frac{\partial}{\partial g_{k;i,j}}f(\mathbf{g})\nonumber\\
&  +\sum_{j=1}^{m_{B}}\sum_{i=1}^{m_{A}}\left[  \sum_{l=1}^{m_{A}}\left(
g_{0;l}d_{l,i}\alpha_{j}-g_{1;i,j}\sum_{q=1}^{m_{A}}d_{i,q}\right)
L_{1}\left(  g_{0}\otimes\alpha,g_{1}\right)  \right]  \frac{\partial
}{\partial g_{1;i,j}}f(\mathbf{g})\nonumber\\
&  +\sum\limits_{k=1}^{\infty}\sum_{j=1}^{m_{B}}\sum_{i=1}^{m_{A}}\sum
_{l=1}^{m_{A}}\left(  g_{k;l,j}c_{l,i}+g_{k;i,j}\sum_{q=1}^{m_{A}}%
d_{i,q}\right)  \left[  \frac{\partial}{\partial g_{k;i,j}}f(\mathbf{g}%
)-\frac{\partial}{\partial g_{k;l,j}}f(\mathbf{g})\right] \nonumber\\
&  +\sum_{i=1}^{m_{A}}\sum_{l=1}^{m_{A}}\left(  g_{0;l}c_{l,i}+g_{0,i}%
\sum_{q=1}^{m_{A}}d_{i,q}\right)  \left[  \frac{\partial}{\partial g_{0;i}%
}f(\mathbf{g})-\frac{\partial}{\partial g_{0;l}}f(\mathbf{g})\right]
\nonumber\\
&  +\sum\limits_{k=1}^{\infty}\sum_{i=1}^{m_{A}}\sum_{j=1}^{m_{B}}\sum
_{r=1}^{m_{B}}\left\{  \left(  g_{k;i,r}t_{r,j}\right)  \left[  \frac{\partial
}{\partial g_{k;i,j}}f(\mathbf{g})-\frac{\partial}{\partial g_{k;r,r}%
}f(\mathbf{g})\right]  \right. \nonumber\\
&  +\left.  \left(  g_{k+1;i,r}t_{r}^{0}\alpha_{j}\right)  \frac{\partial
}{\partial g_{k;i,j}}f(\mathbf{g})\right\}  . \label{Eq10.5-1}%
\end{align}

We define a mapping: $\mathbf{g}\rightarrow\mathbf{u}(t,\mathbf{g})$, where
$\mathbf{u}(t,\mathbf{g)}$ is a solution to the system of differential vector
equations (\ref{EqS1}) to (\ref{EqS6}). Note that the operator semigroup
$\mathbf{T}(t)$ acts in the space $L$, thus if $f\in L$ and $\mathbf{g}%
\in\widetilde{\Omega}$, then%
\begin{equation}
\mathbf{T}(t)f(\mathbf{g})=f\left(  \mathbf{u}(t,\mathbf{g})\right)  .
\label{Eq10.6}%
\end{equation}

From (\ref{Eq10.4}) and (\ref{Eq10.5-1}), it is easy to see that the operator
semigroups $\mathbf{T}_{N}(t)$ and $\mathbf{T}(t)$ are strongly continuous and
contractive, see, for example, Section 1.1 in Chapter one of Ethier and Kurtz
\cite{Eth:1986}. We denote by $\mathcal{D}(\mathbf{A})$ the domain of the
generating operator $\mathbf{A}$. It follows from (\ref{Eq10.6}) that if $f$
is a function from $L$ and has the partial derivatives $\dfrac{\partial
}{\partial g_{k;i,j}}f\left(  \mathbf{g}\right)  $ $\in L$ for $k\geq1,1\leq
i\leq m_{A},1\leq j\leq m_{B}$, and $\sup_{k\geq1,1\leq i\leq m_{A},1\leq
j\leq m_{B}}\left\{  \left\vert \dfrac{\partial}{\partial g_{k;i,j}%
}f(\mathbf{g})\right\vert \right\}  <\infty$, then $f\in\mathcal{D}%
(\mathbf{A})$.

Let $\mathbf{D}$ be the set of all functions $f\in L$ that have the partial
derivatives $\dfrac{\partial}{\partial g_{k;i,j}}f\left(  \mathbf{g}\right)  $
and $\dfrac{\partial^{2}}{\partial g_{k_{1};m,n}\partial g_{k_{2};r;s}%
}f(\mathbf{g})$, and there exists $C=C(f)<+\infty$ such that%
\begin{equation}
\sup_{\substack{k\geq1\\1\leq i\leq m_{A},1\leq j\leq m_{B}\\\mathbf{g}%
\in\widetilde{\Omega}}}\left\{  \left\vert \dfrac{\partial}{\partial
g_{k;i,j}}f(\mathbf{g})\right\vert \right\}  <C \label{Eq10.7}%
\end{equation}
and%
\begin{equation}
\sup_{\substack{_{\substack{k_{1},k_{2}\geq1\\1\leq m,r\leq m_{A},1\leq
n,s\leq m_{B}}}\\\mathbf{g}\in\widetilde{\Omega}}}\left\{  \left\vert
\dfrac{\partial^{2}}{\partial g_{k_{1};m,n}\partial g_{k_{2};r;s}}%
f(\mathbf{g})\right\vert \right\}  <C. \label{Eq10.8}%
\end{equation}

We call that $f\in L$ depends only on the first $K$ subvectors if for
$\mathbf{g}^{\left(  1\right)  }$, $\mathbf{g}^{\left(  2\right)  }%
\in\widetilde{\Omega},$ it follows from $g_{i}^{\left(  1\right)  }%
=g_{i}^{\left(  2\right)  }$ for $1\leq i\leq K$ that $f(\mathbf{g}^{\left(
1\right)  })=f(\mathbf{g}^{\left(  2\right)  })$, where $g_{i}^{\left(
1\right)  }$ and $g_{i}^{\left(  2\right)  }$ are row vectors of size
$m_{A}m_{B}$ for $1\leq i\leq K$. A similar and simple proof of that in
Proposition 2 in Vvedenskaya et al \cite{Vve:1996} can show that the set of
functions from $L$ that depends on the first finite subvectors is dense in $L$.

The following lemma comes from Proposition 1 in Vvedenskaya et al
\cite{Vve:1996}. We restated it here for convenience of description.

\begin{Lem}
\label{Lem:Inequ}Consider an infinite-dimensional system of differential
equations: For $k\geq0,$%
\[
z_{k}\left(  0\right)  =c_{k}%
\]
and%
\[
\dfrac{\text{d}z_{k}(t)}{\text{d}t}=\sum\limits_{i=0}^{\infty}z_{i}%
(t)a_{i,k}(t)+b_{k}(t),
\]
and let $\sum\limits_{i=0}^{\infty}\left|  a_{i,k}(t)\right|  \leq a,\left|
b_{k}(t)\right|  \leq b_{0}\exp\left\{  bt\right\}  ,\left|  c_{k}\right|
\leq\varrho,b_{0}\geq0$ and $a<b$. Then%
\[
z_{k}(t)\leq\varrho\exp\left\{  at\right\}  +\dfrac{b_{0}}{b-a}\left[
\exp\left\{  bt\right\}  -\exp\left\{  at\right\}  \right]  .
\]
\end{Lem}

\begin{Def}
Let $A$ be a closed linear operator on the Banach space $L=C(\widetilde
{\Omega})$. A subspace $\mathbf{D}$ of $\mathcal{D}\left(  A\right)  $ is said
to be a core for $A$ if the closure of the restriction of $A$ to $\mathbf{D}$
is equal to $A$, i.e., $\overline{A|_{\mathbf{D}}}=A$.
\end{Def}

For any matrix $\mathbf{A}=\left(  a_{i,j}\right)  $, we define its norm as
follows:%
\[
\left\Vert \mathbf{A}\right\Vert =\max_{i}\left\{  \sum\limits_{j}\left\vert
a_{i,j}\right\vert \right\}  .
\]
It is easy to compute that%
\[
\left\Vert I\otimes\mathbf{A}\right\Vert =\left\Vert \mathbf{A}\right\Vert ,
\]%
\[
\left\Vert \mathbf{A}\otimes I\right\Vert =\left\Vert \mathbf{A}\right\Vert ,
\]%
\[
\left\Vert \text{diag}\left(  De\right)  \right\Vert =\left\Vert D\right\Vert
.
\]

We introduce some notation%
\[
M_{1}=\sum_{m=1}^{d}C_{d}^{m}=2^{d}-1,
\]%
\[
M_{2}=m_{A}m_{B}\sum_{m=1}^{d}C_{d}^{m}\left(  d+m-2\right)  ,
\]%
\[
a=\left\Vert T^{0}\alpha\right\Vert +\left\Vert \left[  C+\text{diag}\left(
De\right)  \right]  \oplus T\right\Vert +2\left\Vert D\right\Vert \left(
M_{1}+M_{2}\right)  .
\]

The following lemma is a key to prove that the set$\mathcal{\ }\mathbf{D}$ is
a core for the generating operator $\mathbf{A}$.

\begin{Lem}
\label{Lem:Pro2}Let $\mathbf{u}(t)$ be a solution to the system of
differential vector equations (\ref{EqS1}) to (\ref{EqS2}). Then%
\begin{equation}
\underset{1\leq i,m\leq m_{A},1\leq j,n\leq m_{B}}{\underset{k,k_{1}\geq
1}{\sup}}\left\{  \left|  \dfrac{\partial u_{k;i,j}(t,\mathbf{g})}{\partial
g_{k_{1};m,n}}\right|  \right\}  \leq\varrho\exp\left\{  at\right\}  ,
\label{EquT6.11}%
\end{equation}
and%
\begin{equation}
\sup_{\substack{k,k_{1},k_{2}\geq1\\1\leq i,m,r\leq m_{A}\\1\leq j,n,s\leq
m_{B}}}\left\{  \left|  \dfrac{\partial^{2}u_{k;i,j}(t,\mathbf{g})}{\partial
g_{k_{1};m,n}\partial g_{k_{2};r;s}}\right|  \right\}  \leq\widehat{\varrho
}\exp\left\{  at\right\}  +\frac{2\left\|  D\right\|  }{a}\left(  \exp\left\{
2at\right\}  -\exp\left\{  at\right\}  \right)  . \label{EquT6.12}%
\end{equation}
\end{Lem}

\textbf{Proof:} We only prove Inequalities (\ref{EquT6.11}), while
Inequalities (\ref{EquT6.12}) can be proved similarly.

Notice that $\mathbf{u}(t)$ is a solution to the system of differential vector
equations (\ref{EqS1}) to (\ref{EqS2}) and possesses the derivatives
$\dfrac{\partial u_{k;i,j}(t,\mathbf{g})}{\partial g_{k_{1};m,n}}$ and
$\dfrac{\partial^{2}u_{k;i,j}(t,\mathbf{g})}{\partial g_{k_{1};m,n}\partial
g_{k_{2};r;s}}$. For simplicity of description, we set $u_{k;i,j,k_{1}%
;m,n}^{\prime}=\dfrac{\partial u_{k;i,j}(t,\mathbf{g})}{\partial g_{k_{1}%
;m,n}}$. It follows from (\ref{EqS1}) to (\ref{EqS2}) that for $k,k_{1}\geq$
$1$, $1\leq i,m\leq m_{A}$ and $1\leq j,n\leq m_{B}$,%
\begin{align*}
\dfrac{\text{d}u_{k;i,j,k_{1};m,n}^{\prime}}{\text{d}t}=  &  \sum
\limits_{l=1}^{m_{A}}\left(  u_{k-1;l,j,k_{1};m,n}^{\prime}d_{l,i}%
-u_{k;i,j,k_{1};m,n}^{\prime}\sum\limits_{q=1}^{m_{A}}d_{i,q}\right)
L_{k}\left(  u_{k-1}\left(  t\right)  ,u_{k}\left(  t\right)  \right) \\
&  +\sum\limits_{l=1}^{m_{A}}\left(  u_{k-1;l,j}d_{l,i}-u_{k;l,j}%
\sum\limits_{q=1}^{m_{A}}d_{i,q}\right)  L_{k}^{\prime}\left(  u_{k-1}\left(
t\right)  ,u_{k}\left(  t\right)  \right) \\
&  +\sum\limits_{l=1}^{m_{A}}u_{k;l,j,k_{1};m,n}^{\prime}c_{l,i}%
+u_{k;i,j,k_{1};m,n}^{\prime}\sum_{q=1}^{m_{A}}d_{i,q}\\
&  +\sum_{s=1}^{m_{B}}u_{k;i,s,k_{1};m,n}^{\prime}t_{s,j}+\sum_{s=1}^{m_{B}%
}u_{k+1;i,s,k_{1};m,n}^{\prime}t_{s}^{0}\alpha_{j},
\end{align*}
and%
\begin{align*}
&  L_{k}^{\prime}\left(  u_{k-1}\left(  t\right)  ,u_{k}\left(  t\right)
\right)  =\sum_{m=1}^{d}C_{d}^{m}\left(  m-1\right)  \left[  \sum_{l=1}%
^{m_{A}}\sum_{j=1}^{m_{B}}\left(  u_{k-1;l,j}-u_{k;l,j}\right)  \right]
^{m-2}\\
&  \times\left[  \sum_{l=1}^{m_{A}}\sum_{j=1}^{m_{B}}u_{k;l,j}\right]
^{d-m}\left[  \sum_{l=1}^{m_{A}}\sum_{j=1}^{m_{B}}\left(  u_{k-1;l,j,k_{1}%
;m,n}^{\prime}-u_{k;l,j,k_{1};m,n}^{\prime}\right)  \right] \\
&  +\sum_{m=1}^{d}C_{d}^{m}\left(  d-m\right)  \left[  \sum_{l=1}^{m_{A}}%
\sum_{j=1}^{m_{B}}\left(  u_{k-1;l,j}-u_{k;l,j}\right)  \right]  ^{m-1}\\
&  \times\left[  \sum_{l=1}^{m_{A}}\sum_{j=1}^{m_{B}}u_{k;l,j}\right]
^{d-m-1}\left[  \sum_{l=1}^{m_{A}}\sum_{j=1}^{m_{B}}u_{k;l,j,k_{1}%
;m,n}^{\prime}\right]  .
\end{align*}
Using Lemma \ref{Lem:Inequ}, we obtain Inequalities (\ref{EquT6.11}) with%
\begin{align*}
a =  &  \left\Vert I\otimes T^{0}\alpha\right\Vert +\left\Vert \left[
C+\text{diag}\left(  De\right)  \right]  \oplus T\right\Vert \\
&  +\left[  \left\Vert D\otimes I\right\Vert +\left\Vert \text{diag}\left(
De\right)  \otimes I\right\Vert \right]  \left(  M_{1}+M_{2}\right) \\
=  &  \left\Vert T^{0}\alpha\right\Vert +\left\Vert \left[  C+\text{diag}%
\left(  De\right)  \right]  \oplus T\right\Vert +2\left\Vert D\right\Vert
\left(  M_{1}+M_{2}\right)  ,
\end{align*}%
\[
\text{\ }b_{0}=0
\]
and%
\[
\text{ }\varrho=\sup_{\substack{k,k_{1}\geq1\\1\leq i,m\leq m_{A},1\leq
j,n\leq m_{B}}}\left\{  \left\vert u_{k;i,j,k_{1};m,n}^{\prime}\left(
0\right)  \right\vert \right\}  .
\]
This completes this proof. \textbf{{\rule{0.08in}{0.08in}}}

\begin{Lem}
\label{Lem:Core}The set$\mathcal{\ }\mathbf{D}$ is a core for the operator $A$.
\end{Lem}

\textbf{Proof: }It is obvious that $\mathbf{D}$ is dense in $L$ and
$\mathbf{D}\in\mathcal{D}(A)$. Let $\mathbf{D}_{0}$ be the set of functions
from $\mathbf{D}$ which depend only on the first $K$ subvectors of size
$m_{A}m_{B}$. It is easy to see that $\mathbf{D}_{0}$ is dense in $L$.
Therefore, Using proposition 3.3 in Chapter 1 of Ethier and Kurtz
\cite{Eth:1986}, it can show that for any $t\geq0,$ the operator
$\mathbf{T}(t)$ does not bring $\mathbf{D}_{0}$ out of $\mathbf{D}$. Select an
arbitrary function $\varphi\in\mathbf{D}_{0}$ and let $f(\mathbf{g}%
)=\varphi(\mathbf{u}(t;\mathbf{g}))$, $\mathbf{g\in}\widetilde{\Omega}$. It
follows form Lemma \ref{Lem:Pro2} that $f$ has partial derivatives
$\dfrac{\partial}{\partial g_{k;i,j}}f\left(  \mathbf{g}\right)  $ and
$\dfrac{\partial^{2}}{\partial g_{k_{1};m,n}\partial g_{k_{2};r;s}%
}f(\mathbf{g})$ that satisfy conditions (\ref{Eq10.7}) and (\ref{Eq10.8}).
Therefore $f\in\mathbf{D}$. This completes the proof.
\textbf{{\rule{0.08in}{0.08in}}}

In what follows we can prove Theorem 2 given in Section 5.

\textbf{Proof of Theorem 2:} This proof is to use the convergence of operator
semigroups as well as the convergence of their corresponding generating
generators, e.g., see Theorem 6.1 in Chapter 1 of Ethier and Kurtz
\cite{Eth:1986}. Lemma \ref{Lem:Core} shows that the set$\mathcal{\ }%
\mathbf{D}$ is a core for the generating operator $\mathbf{A}$. For any
function $f\in\mathbf{D}$, we have%
\[
N\left[  f\left(  \mathbf{g}-\frac{e_{n;i,j}}{N}\right)  -f\left(
\mathbf{g}\right)  \right]  -\frac{\partial}{\partial g_{n;i,j}}f\left(
\mathbf{g}\right)  =-\frac{\gamma_{n;i,j}^{\left(  1\right)  }}{N}%
\frac{\partial^{2}f\left(  \mathbf{g}-\gamma_{n;i,j}^{\left(  2\right)
}\right)  }{\partial g_{n;i,j}^{2}},
\]
and%
\[
\left|  \left|  \frac{\gamma_{n;i,j}^{\left(  1\right)  }}{N}\frac{\partial
^{2}f\left(  \mathbf{g}-\gamma_{n;i,j}^{\left(  2\right)  }\right)  }{\partial
g_{n;i,j}^{2}}\right|  \right|  \leq\frac{\Re}{N}.
\]
Thus we obtain%
\begin{align*}
|\mathbf{A}_{N}f(\mathbf{g})-f(\mathbf{g})|\leq &  \dfrac{\Re}{N}\left\{
\sum\limits_{k=2}^{\infty}\sum_{j=1}^{m_{B}}\sum_{i=1}^{m_{A}}\left[
\sum_{l=1}^{m_{A}}\left(  g_{k-1;l,j}d_{l,i}-g_{k;i,j}\sum_{q=1}^{m_{A}%
}d_{i,q}\right)  L_{k}\left(  g_{k-1},g_{k}\right)  \right]  \right. \\
&  +\sum_{j=1}^{m_{B}}\sum_{i=1}^{m_{A}}\left[  \sum_{l=1}^{m_{A}}\left(
g_{0;l}d_{l,i}\alpha_{j}-g_{1;i,j}\sum_{q=1}^{m_{A}}d_{i,q}\right)
L_{1}\left(  g_{0}\otimes\alpha,g_{1}\right)  \right] \\
&  +\sum\limits_{k=1}^{\infty}\sum_{j=1}^{m_{B}}\sum_{i=1}^{m_{A}}\sum
_{l=1}^{m_{A}}\left(  g_{k;l,j}\left|  c_{l,i}\right|  +g_{k;i,j}\sum
_{q=1}^{m_{A}}d_{i,q}\right) \\
&  +\sum_{i=1}^{m_{A}}\sum_{l=1}^{m_{A}}\left(  g_{0;l}\left|  c_{l,i}\right|
+g_{0,i}\sum_{q=1}^{m_{A}}d_{i,q}\right) \\
&  \left.  +\sum\limits_{k=1}^{\infty}\sum_{i=1}^{m_{A}}\sum_{j=1}^{m_{B}}%
\sum_{l=1}^{m_{B}}\left(  g_{k;i,l}\left|  t_{l,j}\right|  +g_{k+1;i,l}%
t_{l}^{0}\alpha_{j}\right)  \right\}  .
\end{align*}
Note that%
\[
L_{1}\left(  g_{0}\otimes\alpha,g_{1}\right)  =\frac{\left(  g_{0}e\right)
^{d}-\left(  g_{1}e\right)  ^{d}}{g_{0}e-g_{1}e}\leq d
\]
and%
\[
L_{1}\left(  g_{k-1},g_{k}\right)  =\frac{\left(  g_{k-1}e\right)
^{d}-\left(  g_{k}e\right)  ^{d}}{g_{k-1}e-g_{k}e}\leq d,
\]
we obtain%
\begin{align*}
|\mathbf{A}_{N}f(\mathbf{g})-f(\mathbf{g})|  &  \leq\dfrac{\Re}{N}\left[
d\sum\limits_{k=2}^{\infty}\sum_{j=1}^{m_{B}}\sum_{i=1}^{m_{A}}\sum
_{l=1}^{m_{A}}g_{k-1;l,j}d_{l,i}+d\sum_{j=1}^{m_{B}}\sum_{i=1}^{m_{A}}%
\sum_{l=1}^{m_{A}}g_{0;l}d_{l,i}\alpha_{j}\right. \\
&  \left.  +\left(  \left\|  C\right\|  +m_{A}\left\|  D\right\|  +\left\|
T\right\|  +\left\|  T^{0}\alpha\right\|  \right)  \sum\limits_{k=0}^{\infty
}g_{k}e\right] \\
&  \leq\dfrac{\Re}{N}\left[  \left(  \left\|  C\right\|  +\left(
d+m_{A}\right)  \left\|  D\right\|  +\left\|  T\right\|  +\left\|  T^{0}%
\alpha\right\|  \right)  \sum\limits_{k=0}^{\infty}g_{k}e\right]  .
\end{align*}
For $\mathbf{g}\in\Omega$, it is clear that $\mathbf{g}e=\sum\limits_{k=0}%
^{\infty}g_{k}e<+\infty$. Thus we get
\[
\underset{N\rightarrow\infty}{\lim}\underset{\mathbf{g\in}\Omega}{\sup
}|\mathbf{A}_{N}f(\mathbf{g})-\mathbf{A}f(\mathbf{g})|=0\text{.}%
\]
This gives%
\[
\underset{N\rightarrow\infty}{\lim}\underset{\mathbf{g}\in\Omega}{\sup}\left|
\mathbf{T}_{N}(t)f(\mathbf{g})-f(\mathbf{u}(t;\mathbf{g}))\right|  =0.
\]
\newline This completes the proof. \textbf{{\rule{0.08in}{0.08in}}}

\subsection*{Appendix C: Proof of Theorem 3}

To prove Theorem \ref{The:Coupling}, we need to extend the coupling method
given in Turner \cite{Tur:1996} and Martin and Suhov \cite{Mar:1999} such that
this coupling method can be applied to discussing stability of more general
block-structured supermarket models.

In the two supermarket models $Q$ and $R$, they have the same parameters:
$N,d,m_{A},C,D$, $m_{B},\alpha,T$, and the same initial state at $t=0$; while
the only difference between both of them is their choice numbers: $d\left(
Q\right)  =1$ and $d\left(  R\right)  \geq2$.

To set up a coupling between the two infinite-dimensional Markov processes
$\left\{  U_{N}^{\left(  Q\right)  }\left(  t\right)  :t\geq0\right\}  $ and
$\left\{  U_{N}^{\left(  R\right)  }\left(  t\right)  :t\geq0\right\}  $, we
need introduce some notation as follows. For a supermarket model $S$ with
$k\geq1$, $1\leq i\leq m_{A}$ and $1\leq i\leq m_{B}$, we denote by
$A_{k}^{\left(  i,j\right)  }(S)$ and $D_{k}^{\left(  i,j\right)  }(S)$ the
$k$th arrival time and the $k$th departure time when the MAP environment
process is at state $i$ and the PH service environment process is at state $j$.

As discussed in Section 4 of Martin and Suhov \cite{Mar:1999}, we introduce
the notation of "shadow"\ customers to build up the coupling relation between
the two supermarket models $Q$ and $R$. For $k$ and $\left(  i,j\right)  $,
the time of the shadow customer arriving at the supermarket model $Q$ is
written as $A_{k}^{\left(  i,j\right)  }\left(  R\right)  $, and at time
$A_{k}^{\left(  i,j\right)  }\left(  Q\right)  $ the shadow customer is
replaced by the real customer immediately. The relationship between the shadow
and real customers are described by Figure 8 (a), while there will not exist a
shadow customer in Figure 8 (b).

\begin{figure}[ptb]
\centering          \includegraphics[width=12cm]{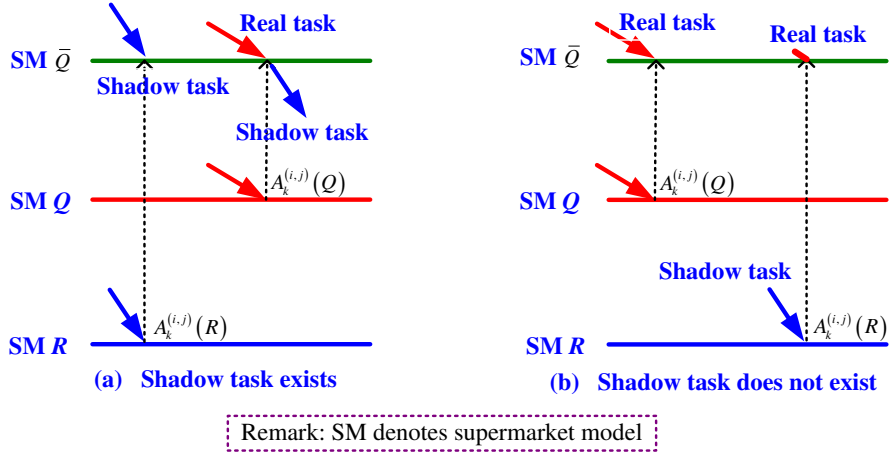}  \caption{The
shadow and real tasks }%
\label{figure:fig-7}%
\end{figure}

From the two supermarket models $Q$ and $R$, we construct a new supermarket
model $\overline{Q}$ with shadow customers such that at environment state pair
$\left(  i,j\right)  $, each arrival time in the supermarket model
$\overline{Q}$ is the same time as that in the supermarket model $R$, while
each departure time is the same time as that in supermarket model $Q$. Based
on this, we can set up a coupling between the two supermarket models $R$ and
$Q$ by means of the supermarket model $\overline{Q}$.

For a supermarket model $S$ and for $k\geq1,1\leq i\leq m_{A},1\leq i\leq
m_{B},x\geq0,$ we define%
\[
\psi_{x}^{\left(  i,j\right)  }(S,t)=\sum_{n=1}^{N}\left[  l_{n}^{\left(
i,j\right)  }(S,t)-x\right]  _{+},
\]
where $l_{n}^{\left(  i,j\right)  }(S,t)$ is the queue length of the $n$th
server with environment state pair $\left(  i,j\right)  $ at time $t$, and
$[y]_{+}=\max(y,0)$.

The following lemma gives a useful property of $\psi_{x}^{\left(  i,j\right)
}(S,t)$ for the two supermarket models $\overline{Q}$ and $R$.

\begin{Lem}
\label{pro:r-x}If $\psi_{y}^{\left(  i,j\right)  }(R,t)\leq\psi_{y}^{\left(
i,j\right)  }(\overline{Q},t)$ for all $y$ and $\psi_{x}^{\left(  i,j\right)
}(R,t)=\psi_{x}^{\left(  i,j\right)  }(\overline{Q},t)$, then%
\begin{equation}
\#\left\{  n:l_{n}^{\left(  i,j\right)  }(R,t)\leq x\right\}  \leq\#\left\{
n:l_{n}^{\left(  i,j\right)  }(\overline{Q},t)\leq x\right\}  \label{equ4.1}%
\end{equation}
and%
\begin{equation}
\#\left\{  n:l_{n}^{\left(  i,j\right)  }(R,t)\geq x\right\}  \leq\#\left\{
n:l_{n}^{\left(  i,j\right)  }(\overline{Q},t)\geq x\right\}  , \label{equ4.2}%
\end{equation}
where $\#\left\{  A\right\}  $ means the number of elements in the set $A$.
\end{Lem}

\textbf{Proof: }If $\psi_{y}^{\left(  i,j\right)  }(R,t)\leq\psi_{y}^{\left(
i,j\right)  }(\overline{Q},t)$ for all $y$ and $\psi_{x}^{\left(  i,j\right)
}(R,t)=\psi_{x}^{\left(  i,j\right)  }(\overline{Q},t)$, then for $y=x+1$%
\[
-\psi_{x+1}^{\left(  i,j\right)  }(R,t)\geq-\psi_{x+1}^{\left(  i,j\right)
}(\overline{Q},t),
\]
using $\psi_{x}^{\left(  i,j\right)  }(R,t)=\psi_{x}^{\left(  i,j\right)
}(\overline{Q},t)$ we get%
\begin{equation}
\psi_{x}^{\left(  i,j\right)  }(R,t)-\psi_{x+1}^{\left(  i,j\right)
}(R,t)\geq\psi_{x}^{\left(  i,j\right)  }(\overline{Q},t)-\psi_{x+1}^{\left(
i,j\right)  }(\overline{Q},t). \label{equ4.3}%
\end{equation}
Similarly, for $y=x-1$ we have%
\begin{equation}
\psi_{x}^{\left(  i,j\right)  }(R,t)-\psi_{x-1}^{\left(  i,j\right)
}(R,t)\leq\psi_{x}^{\left(  i,j\right)  }(\overline{Q},t)-\psi_{x-1}^{\left(
i,j\right)  }(\overline{Q},t). \label{equ4.4}%
\end{equation}
Since%
\[
\psi_{x}^{\left(  i,j\right)  }(S,t)=\sum_{n=1}^{N}\left[  l_{n}^{\left(
i,j\right)  }(S,t)-x\right]  _{+},
\]
we obtain%
\[
\psi_{x}^{\left(  i,j\right)  }(S,t)-\psi_{x+1}^{\left(  i,j\right)
}(S,t)=\sum_{n=1}^{N}\left\{  \left[  l_{n}^{\left(  i,j\right)
}(S,t)-x\right]  _{+}-\left[  l_{n}^{\left(  i,j\right)  }(S,t)-\left(
x+1\right)  \right]  _{+}\right\}  .
\]
To calculate $\psi_{x}^{\left(  i,j\right)  }(S,t)-\psi_{x+1}^{\left(
i,j\right)  }(S,t)$, we analyze the following two cases:

\textbf{Case one: }If $l_{n}^{\left(  i,j\right)  }(S,t)\leq x$, then $\left[
l_{n}^{\left(  i,j\right)  }(S,t)-x\right]  _{+}=\left[  l_{n}^{\left(
i,j\right)  }(S,t)-\left(  x+1\right)  \right]  _{+}=0$.

\textbf{Case two: }If $l_{n}^{\left(  i,j\right)  }(S,t)>x$, then $\left[
l_{n}^{\left(  i,j\right)  }(S,t)-x\right]  _{+}-\left[  l_{n}^{\left(
i,j\right)  }(S,t)-\left(  x+1\right)  \right]  _{+}=1$.

If $\sum_{n=1}^{N}\left\{  \left[  l_{n}^{\left(  i,j\right)  }(S,t)-x\right]
_{+}-\left[  l_{n}^{\left(  i,j\right)  }(S,t)-\left(  x+1\right)  \right]
_{+}\right\}  =k$, then $k$ is the number of servers whose queue length is
bigger than $x$. That is $\#\left\{  n:l_{n}^{\left(  i,j\right)
}(S,t)>x\right\}  =k$. Hence, we obtain%

\begin{align}
\psi_{x}^{\left(  i,j\right)  }(S,t)-\psi_{x+1}^{\left(  i,j\right)  }(S,t)
&  =\sum_{n=1}^{N}\left\{  \left[  l_{n}^{\left(  i,j\right)  }(S,t)-x\right]
_{+}-\left[  l_{n}^{\left(  i,j\right)  }(S,t)-\left(  x+1\right)  \right]
_{+}\right\} \nonumber\\
&  =\#\left\{  n:l_{n}^{\left(  i,j\right)  }(S,t)>x\right\}  . \label{equ4.5}%
\end{align}
It follows from (\ref{equ4.3}) to (\ref{equ4.5}) that%
\[
\#\left\{  n:l_{n}^{\left(  i,j\right)  }(R,t)>x\right\}  \geq\#\left\{
n:l_{n}^{\left(  i,j\right)  }(\overline{Q},t)>x\right\}  ,
\]
this gives%
\[
\#\left\{  n:l_{n}^{\left(  i,j\right)  }(R,t)\leq x\right\}  \leq\#\left\{
n:l_{n}^{\left(  i,j\right)  }(\overline{Q},t)\leq x\right\}  .
\]
Similarly, it follows from (\ref{equ4.4}) to (\ref{equ4.5}) that%
\[
\#\left\{  n:l_{n}^{\left(  i,j\right)  }(R,t)>x-1\right\}  \leq\#\left\{
n:l_{n}^{\left(  i,j\right)  }(\overline{Q},t)>x-1\right\}  ,
\]
which follows
\[
\#\left\{  n:l_{n}^{\left(  i,j\right)  }(R,t)\geq x\right\}  \leq\#\left\{
n:l_{n}^{\left(  i,j\right)  }(\overline{Q},t)\geq x\right\}  .
\]
This completes the proof. \textbf{{\rule{0.08in}{0.08in}}}

The following lemma sets up the coupling between the two supermarket models
$R$ and $\overline{Q}$, which is based on the arrival and departure processes.

\begin{Lem}
For the two supermarket models $R$ and $\overline{Q}$ and for $x,t\geq0,1\leq
i\leq m_{A},1\leq i\leq m_{B}$, we have
\begin{equation}
\psi_{x}^{\left(  i,j\right)  }(R,t)\leq\psi_{x}^{\left(  i,j\right)
}(\overline{Q},t). \label{equ4.6}%
\end{equation}
\end{Lem}

\textbf{Proof: }To prove (\ref{equ4.6}), we need to discuss the departure
process and the arrival process, respectively.

\textbf{(1) The departure process}

Note that the two supermarket models $R$ and $\overline{Q}$ have the same
initial state at $t=0$, thus (\ref{equ4.6}) holds at time $t=0$.

In the departing process, it is easy to see from the above coupling that at
environment state pair $\left(  i,j\right)  $, if given the server orders in
supermarket models $\overline{Q}$ and $R$ according to the queue length of
each server (including shadow tasks), then the customer departures always
occur at the same order servers. For example, if the customer departure occurs
from the server with the shortest queue length in supermarket model
$\overline{Q}$, then a customer departure must also occur from the server with
the shortest queue length in supermarket model $R$. Note that the customer
departures will be lost either from an empty server or from one containing
only shadow customers.

Let $D$ be a potential departure time at environment state pair $\left(
i,j\right)  $, and suppose that (\ref{equ4.6}) holds for $t<D$. Then we hope
to show that (\ref{equ4.6}) holds for $t=D$.

Suppose that (\ref{equ4.6}) does not hold at a departure point $D$. Then we
have $\psi_{x}^{\left(  i,j\right)  }(R,D)>\psi_{x}^{\left(  i,j\right)
}(\overline{Q},D)$.

Since (\ref{equ4.6}) holds for $t<D$, we get that $\psi_{x}^{\left(
i,j\right)  }(\overline{Q},D^{-})\leq\psi_{x}^{\left(  i,j\right)  }(R,D^{-}%
)$. Based on this, we discuss the two cases: $\psi_{x}^{\left(  i,j\right)
}(\overline{Q},D^{-})=\psi_{x}^{\left(  i,j\right)  }(R,D^{-})$ and $\psi
_{x}^{\left(  i,j\right)  }(\overline{Q},D^{-})<\psi_{x}^{\left(  i,j\right)
}(R,D^{-})$, and indicate how the two cases influence the departure process at
time $D$.

\textbf{Case one:} If $\psi_{x}^{\left(  i,j\right)  }(\overline{Q}%
,D^{-})=\psi_{x}^{\left(  i,j\right)  }(R,D^{-})$ and $\psi_{x}^{\left(
i,j\right)  }(R,D)>\psi_{x}^{\left(  i,j\right)  }(\overline{Q},D)$, then a
departure at time $D$ makes that $\psi_{x}^{\left(  i,j\right)  }(R,D)$ does
not change, while $\psi_{x}^{\left(  i,j\right)  }(\overline{Q},D)$ is
diminished. Let $a$ and $b$\ be the queue lengths at time $D$ in the two
supermarket models $\overline{Q}$ and $R$, respectively. Then for
$x=0,1,\ldots,a-1$, it is seen that
\[
\psi_{x}^{\left(  i,j\right)  }(\overline{Q},D^{-})=\sum_{n=1}^{N}\left[
l_{n}^{\left(  i,j\right)  }(\overline{Q},t)-x\right]  _{+}%
\]
reduces $1$. Similarly, for $x=0,1,\ldots,b-1$,%
\[
\psi_{x}^{\left(  i,j\right)  }(R,D^{-})=\sum_{n=1}^{N}\left[  l_{n}^{\left(
i,j\right)  }(R,t)-x\right]  _{+}%
\]
also reduce $1$. Therefore, when $x$ is $b,b+1,\ldots,a-1$ (that is $b\leq
x<a$), we have $\psi_{x}^{\left(  i,j\right)  }(R,D)>\psi_{x}^{\left(
i,j\right)  }(\overline{Q},D)$. However, when $\psi_{x}^{\left(  i,j\right)
}(\overline{Q},D^{-})=\psi_{x}^{\left(  i,j\right)  }(R,D^{-})$, both from
that (\ref{equ4.1}) holds for $t<D$ and from that the departure channels are
at a coupling, it is clear that the condition: $\psi_{x}^{\left(  i,j\right)
}(R,D)>\psi_{x}^{\left(  i,j\right)  }(\overline{Q},D)$ for $b\leq x<a$, is impossible.

\textbf{Case two: }$\psi_{x}^{\left(  i,j\right)  }(R,D^{-})<\psi_{x}^{\left(
i,j\right)  }(\overline{Q},D^{-})$. In this case, when a customer departs the
system, the two numbers $\psi_{x}^{\left(  i,j\right)  }(R,D^{-})$ and
$\psi_{x}^{\left(  i,j\right)  }(\overline{Q},D^{-})$ have only two cases:
Unchange and diminish $1$. Note that $\psi_{x}^{\left(  i,j\right)  }%
(R,D^{-})<\psi_{x}^{\left(  i,j\right)  }(\overline{Q},D^{-})$, we get that
$\psi_{x}^{\left(  i,j\right)  }(R,D^{-})+1\leq\psi_{x}^{\left(  i,j\right)
}(\overline{Q},D^{-})$. Hence, we can not obtain that $\psi_{x}^{\left(
i,j\right)  }(R,D)>\psi_{x}^{\left(  i,j\right)  }(\overline{Q},D)$.

\textbf{(2) The arrival process}

In a similar way to the above analysis in "(1) The departure process", we
discuss the coupling for the arriving process as follows.

Let $A=A_{k}^{\left(  i,j\right)  }$ be an arrival time. Then (\ref{equ4.6})
holds for $t<A$. We hope to show that (\ref{equ4.6}) holds for $t=A$.

This proof is similar to the above analysis in "(1) The departure process".
Let $a$ and $b$\ be the queue lengths at time $A$ in the two supermarket
models $\overline{Q}$ and $R$, respectively. Then $\psi_{x}^{\left(
i,j\right)  }(R,A^{-})=\psi_{x}^{\left(  i,j\right)  }(\overline{Q},A^{-})$
holds for some $x$ for $a<x\leq b$. Thus, it follows from (\ref{equ4.2}) that%
\[
\#\left\{  n:l_{n}^{\left(  i,j\right)  }(R,A^{-})\geq x\right\}
\leq\#\left\{  n:l_{n}^{\left(  i,j\right)  }(\overline{Q},A^{-})\geq
x\right\}
\]
and
\[
\#\left\{  n:l_{n}^{\left(  i,j\right)  }(R,A^{-})\geq b\right\}
\leq\#\left\{  n:l_{n}^{\left(  i,j\right)  }(\overline{Q},A^{-})\geq
a\right\}  .
\]
However, the condition: $\#\left\{  n:l_{n}^{\left(  i,j\right)  }%
(R,A^{-})\geq b\right\}  \leq\#\left\{  n:l_{n}^{\left(  i,j\right)
}(\overline{Q},A^{-})\geq a\right\}  $, is impossible, because it follows from
the above coupling that for $a<x\leq b$%
\[
\#\left\{  n:l_{n}^{\left(  i,j\right)  }(R,A^{-})\geq b\right\}  >\#\left\{
n:l_{n}^{\left(  i,j\right)  }(\overline{Q},A^{-})\geq a\right\}  .
\]
Since the queue length $a$ was chosen at the arrival time, it is seen that the
queue length $a$ must exist in the supermarket model $\overline{Q}$. In this
case, we get that $\#\left\{  n:l_{n}^{\left(  i,j\right)  }(\overline
{Q},A^{-})=a\right\}  \geq1$. Therefore, this leads to a contradiction.

Note that there are some shadow customers in supermarket model $\overline{Q}$,
the shadow customers do not affect the queue lengths in the supermarket model
$\overline{Q}$ at the arrival time $A_{k}^{\left(  i,j\right)  }(Q)$, thus
(\ref{equ4.6}) holds. This completes the proof. \textbf{{\rule{0.08in}{0.08in}%
}}

The following lemma provides the coupling between the two supermarket models
$Q$ and $R$, which is based on the arrival and departure processes.

\begin{Lem}
\label{coupling}In the two supermarket models $Q$ and $R$, for $k>0,1\leq
i\leq m_{A},1\leq j\leq m_{B}$ we have%
\begin{equation}
D_{k}^{\left(  i,j\right)  }\left(  R\right)  \leq D_{k}^{\left(  i,j\right)
}\left(  Q\right)  \label{equ4.7}%
\end{equation}
and%
\begin{equation}
A_{k}^{\left(  i,j\right)  }\left(  R\right)  \leq A_{k}^{\left(  i,j\right)
}\left(  Q\right)  . \label{equ4.8}%
\end{equation}
\end{Lem}

\textbf{Proof: }Using the above coupling, now we continue to discuss the two
supermarket models $Q$ and $R$.

Note that the two supermarket models $Q$ and $R$ have the same parameters
$N,m,c_{i,j},d_{i,j},\mu_{i}$ for $1\leq i,j\leq m$ and the same initial state
at $t=0$, the departure or arrival of the $k$th customer and the Markov
environment process in the supermarket model $Q$ correspond to those in the
supermarket model $R$. This ensures that if (\ref{equ4.7}) holds for the
departure process up to a given time, then so does (\ref{equ4.8}) for the
arrival process up to that time.

Now, we use (\ref{equ4.6}) to prove (\ref{equ4.7}).

Suppose that (\ref{equ4.7}) is false, that is, $D_{k}^{\left(  i,j\right)
}\left(  R\right)  >D_{k}^{\left(  i,j\right)  }\left(  Q\right)  $. Then the
number of customer departures before time $D$ from the supermarket model $R$
must be the same as that in the supermarket model $\overline{Q}$. Since the
arrivals in the two supermarket models $R$ and $\overline{Q}$ occur at the
same times, there must be the same total number of customers in the two
supermarket models $R$ and $\overline{Q}$. Hence, $\psi_{0}^{\left(
i,j\right)  }(R,D^{-})=\psi_{0}^{\left(  i,j\right)  }(\overline{Q},D^{-})$.
But, it is seen from (\ref{equ4.1}) that the number of servers with non-zero
queue length in the supermarket model $\overline{Q}$ is bigger than that in
the supermarket model $R$, this indicates that the number of servers with
empty server in the supermarket model $\overline{Q}$ is less than that in the
supermarket model $R$. Therefore, if a departure occurs in the supermarket
model $\overline{Q}$, then there must be a departure in the supermarket model
$R$. On the contrary, if a departure occurs in the supermarket model $R$, then
it is possible not to have a departure in the supermarket model $\overline{Q}%
$. Note that the departure time in the supermarket model $\overline{Q}$ is the
same as that in the supermarket model $Q$, hence the departure time in the
supermarket model $R$ is earlier than that in the supermarket model $Q$, that
is, $D_{k}^{\left(  i,j\right)  }\left(  R\right)  \leq D_{k}^{\left(
i,j\right)  }\left(  Q\right)  $. This leads to a contradiction of the
assumption $D_{k}^{\left(  i,j\right)  }\left(  R\right)  >D_{k}^{\left(
i,j\right)  }\left(  Q\right)  $. Hence (\ref{equ4.7}) holds. Similarly, we
can prove (\ref{equ4.8}). This completes the proof.
\textbf{{\rule{0.08in}{0.08in}}}

\textbf{Proof of Theorem \ref{The:Coupling}: }Using the lemma \ref{coupling},
we know that $D_{k}^{\left(  i,j\right)  }\left(  R\right)  \leq
D_{k}^{\left(  i,j\right)  }\left(  Q\right)  $ and $A_{k}^{\left(
i,j\right)  }\left(  R\right)  \leq A_{k}^{\left(  i,j\right)  }\left(
Q\right)  $. This indicates that for any two corresponding servers in the two
supermarket models $Q$ and $R$, the arrival and departure times in the
supermarket model $R$ are earlier than those in the supermarket model $Q$.
Hence, the queue length of any server in the supermarket model $R$ is shorter
than that of the corresponding server in the supermarket model $Q$. This shows
that the total number of customers in the supermarket model $R$ is no greater
than the total number of customers in the supermarket model $Q$ at time
$t\geq0$. Based on this, we obtain a coupling between the processes
$\{U_{Q}^{\left(  N\right)  }\left(  t\right)  \}$ and $\{U_{R}^{\left(
N\right)  }\left(  t\right)  \}$: For all $t\geq0$, the total number of
customers in the supermarket model $R$ is no greater than that in the
supermarket model $Q$. This completes the proof.
\textbf{{\rule{0.08in}{0.08in}}}

\newpage

\noindent Quan-Lin Li is Full Professor in School of Economics and
Management Sciences, Yanshan University, Qinhuangdao, China. He
received the Ph.D. degree in Institute of Applied Mathematics,
Chinese Academy of Sciences, Beijing, China in 1998. He has
published a book ({\it Constructive Computation in Stochastic Models
with Applications: The RG-Factorizations, Springer, 2010}) and over
40 research papers in a variety of journals, such as, {\it Advances
in Applied Probability, Queueing Systems, Stochastic Models,
European Journal of Operational Research, Computer Networks,
Performance Evaluation, Discrete Event Dynamic Systems, Computers \&
Operations Research, Computers \& Mathematics with Applications,
Annals of Operations Research}, and {\it International Journal of
Production Economics}. His main research interests concern with
Queueing Theory, Stochastic Models, Matrix-Analytic Methods,
Manufacturing Systems, Computer Networks, Network Security, and
Supply Chain Risk Management.

\vskip 2cm

\noindent John C.S. Lui (M¡¯93-SM¡¯02-F¡¯10) was born in Hong Kong.
He received the Ph.D. degree in computer science from the University
of California, Los Angeles, 1992. He is currently a Professor with
the Department of Computer Science and Engineering, The Chinese
University of Hong Kong (CUHK), Hong Kong. He was the chairman of
the Department from 2005 to 2011. His current research interests are
in communication networks, network/system security (e.g., cloud
security, mobile security, etc.), network economics, network
sciences (e.g., online social networks, information spreading,
etc.), cloud computing, large-scale distributed systems, and
performance evaluation theory. Professor Lui is a Fellow of the
Association for Computing Machinery (ACM), a Fellow of IEEE, a
Croucher Senior Research Fellow, and an elected member of the IFIP
WG 7.3. He serves on the Editorial Board of IEEE/ACM Transactions on
Networking, IEEE Transactions on Computers, IEEE Transactions on
Parallel and Distributed Systems, Journal of Performance Evaluation
and International Journal of Network Security. He received various
departmental teaching awards and the CUHK Vice-Chancellor¡¯s
Exemplary Teaching Award. He is also a co-recipient of the IFIP WG
7.3 Performance 2005 and IEEE/IFIP NOMS 2006 Best Student Paper
Awards.

\begin{figure}[ptb]
%Requires \usepackage{graphicx}
\centering          \includegraphics[width=12cm]{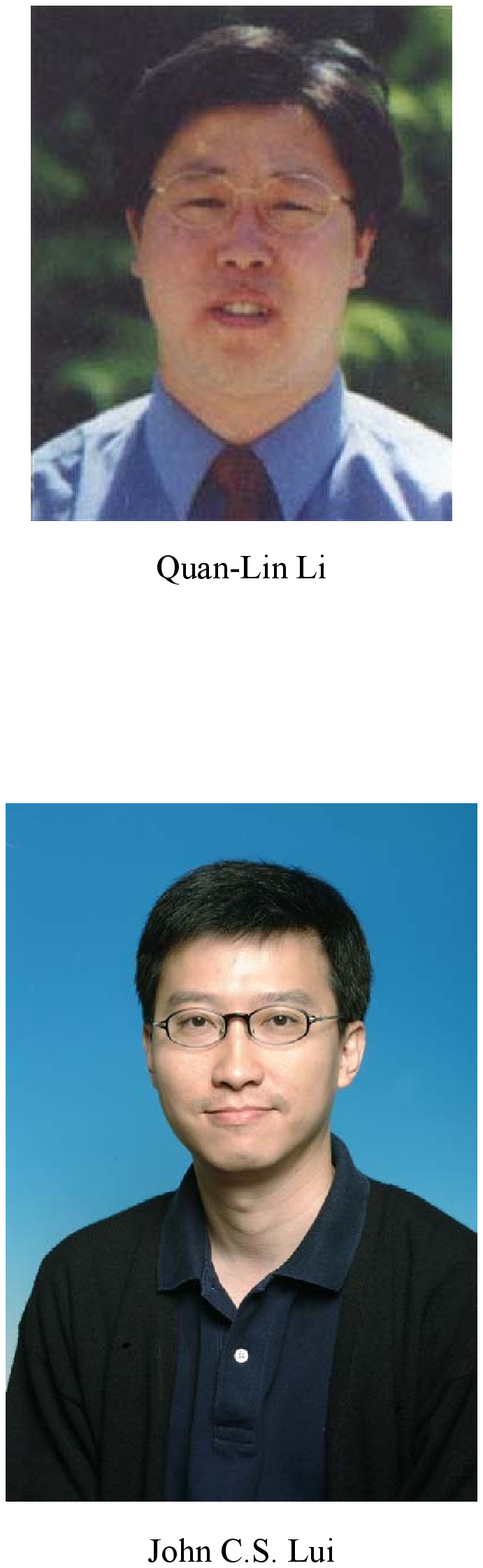}
\newline
\caption{}%
%\label{figure:fig-8}%
\end{figure}

\end{document}